\documentclass[%
reprint,
nofootinbib,
 amsmath,amssymb,
 aps,
]{revtex4-2}

\usepackage{physics}
\usepackage{romanbar}
\usepackage{graphicx}
\usepackage[caption=false]{subfig}
\usepackage{dcolumn}
\usepackage{tikz}
\usepackage{bm}
\usepackage{float}
\usepackage{xcolor}
\usepackage[symbol,hang,flushmargin]{footmisc}
\usepackage{bbold} 

\usepackage{amsthm}
\usepackage[T1]{fontenc}

\begin{document}

\preprint{APS/123-QED}

\title{Exact simulation of realistic Gottesman-Kitaev-Preskill cluster states}

\author{Milica Bani\'c$^{1,*}$}
\footnotetext[1]{Milica.Banic@nrc-cnrc.gc.ca}
\author{Valerio Crescimanna$^{1,2}$}
\author{J. Eli Bourassa$^3$}
\author{\\Carlos Gonz\'alez-Arciniegas$^3$} 
\author{Rafael N. Alexander$^3$} 
\author{Khabat Heshami$^{1,2,4}$}
\affiliation{$^1$ National Research Council of Canada, 100 Sussex Drive, Ottawa, ON, K1N 6N5, Canada\\$^2$Department of Physics, University of Ottawa, 25 Templeton Street, Ottawa, Ontario, K1N 6N5 Canada\\$^3$Xanadu, 777 Bay Street, Toronto ON, M5G 2C8, Canada\\ $^4$ Institute for Quantum Science and Technology, Department of Physics and Astronomy, University of Calgary, Alberta T2N 1N4, Canada}

\date{\today}

\begin{abstract}
{We describe a method for simulating and characterizing realistic Gottesman-Kitaev-Preskill (GKP) cluster states, rooted in the representation of resource states in terms of sums of Gaussian distributions in phase space. We apply our method to study the generation of single-mode GKP states via cat state breeding, and the formation of multimode GKP cluster states via linear optical circuits and homodyne measurements. We characterize resource states by referring to expectation values of their stabilizers, and witness operators constructed from them. Our method reproduces the results of standard Fock-basis simulations, while being more efficient, and being applicable in a broader parameter space. We also comment on the validity of the heuristic Gaussian random noise (GRN) model, through comparisons with our exact simulations: We find discrepancies in the stabilizer expectation values when homodyne measurement is involved in cluster state preparation, yet we find a close agreement between the two approaches on average.}
\end{abstract}

\maketitle


\section{Introduction}
\label{section:introduction}

{Photonic implementations of quantum information processing (QIP) are motivated by the potential for room-temperature operation, scalability of hardware, low decoherence, and ability to physically transmit quantum information over long distances via optical fibers. A number of tasks in photonic QIP---including measurement-based quantum computation, and all-optical quantum repeaters---require bosonic cluster states as resources \cite{PhysRevLett.95.010501,PhysRevA.68.022312, RevModPhys.95.045006}. One approach to constructing photonic cluster states is to encode information in the quadratures of the electromagnetic field. Particularly interesting is the case where the modes are Gottesman-Kitaev-Preskill (GKP) states \cite{GKP_PhysRevA.64.012310, PRXQuantum.2.020101, repeaters, PhysRevX.15.021003}, which are characterized by a periodic grid of peaks in phase space, with the logical information encoded in the positions of the peaks. This approach is further motivated by the relative ease with which GKP cluster states can be stitched together and processed, where Clifford gates---including entangling operations---can be implemented deterministically using Gaussian operations \cite{Bourassa2021_blueprint} and Pauli projections are implemented as homodyne measurements. Given the significance of these states and their applications, developing the ability to simulate and characterize such systems is essential and will have far reaching implications for efforts towards implementing photonic QIP. }

{It is essential when studying such architectures to account for non-idealities in the resource states. These non-idealities are inevitable because ideal GKP states are not physical}; in reality, one can only generate approximate GKP states having a finite extent in phase space, with the peaks having finite widths. Because this limits the distinguishability of the different logical states, and the state's capacity for error correction \cite{GlancyKnill_PhysRevA.73.012325, matsuura_PhysRevA.102.032408, zheng2024performanceachievableratesgottesmankitaevpreskill}, it is important to account for these features when devising and assessing the performance of CV quantum information protocols. However, accurately describing realistic GKP states can be challenging and numerically expensive; even dealing with energetic (high-quality) single-mode GKP states can be problematic, and the scaling becomes prohibitive when one treats cluster states of even a few modes. For this reason, the community typically relies on heuristic models such as the ``Gaussian random noise" (GRN) model \cite{Menicucci_PhysRevLett.112.120504, noh_PhysRevA.101.012316}, in which a noisy channel is applied to ideal GKP states, resulting in a uniform broadening of the peaks to some finite width. Although this provides a more realistic picture, the GRN model does not capture all the features of approximate GKP states that can be generated in practice \cite{Vasconcelos:10,su_PhysRevA.100.052301,PRXQuantum.4.010333, quesada_PhysRevA.100.022341, Eaton2019, takase2024generationflyinglogicalqubits}. 

{An accurate description of experimentally accessible GKP states is key to properly assessing and optimizing the performance of CV photonic architectures \cite{PhysRevA.101.032315}. Progress in this direction will enable more realistic performance estimates for CV quantum protocols. It may also enable improvements in performance by allowing for more sophisticated encoding and decoding schemes and optimizing the architecture, with the aim of mitigating logical errors due to the non-idealness of realistic GKP states. More detailed simulations will also be useful in developing methods of characterizing realistic GKP states and assessing their usefulness. Motivated by these questions, we have developed an approach for exact simulations of realistic GKP cluster state generation. 
We apply an approach in which CV states and operations are represented by sums of Gaussian distributions in phase space \cite{Eli_gaussians, PhysRevA.110.042402}. We extend earlier work---in which this method was used to characterize realistic single-mode GKP states---to simulate GKP cluster state generation by entangling these single-mode inputs states through linear unitary circuits and homodyne measurement \cite{PhysRevLett.97.110501,PhysRevA.104.012431,macronization}.}

In Section II we review the details of this formalism, with a particular emphasis on the phase-space methods required to make these calculations tractable \cite{Eli_gaussians}. In Section III we turn our focus to the computation of stabilizer expectation values (EVs), as a useful figure of merit for realistic GKP cluster states. In Section IV we implement our expression for the stabilizer EV numerically. We address single-mode GKP states up to a linear three-photon cluster state. We compare our results to those obtained with Fock-basis simulations, and with the GRN model. In Section V we summarize and conclude. 


\section{Background}

\subsection{Phase space formalism}
\label{section:phasespace}

The method presented in this manuscript makes use of the phase space formalism for quantum optics \cite{serafini2017quantum}. We represent operators in terms of their Wigner functions; the Wigner function for an ${N}$-mode operator $\mathcal{O}$ is given by 
\begin{align}
W_{\mathcal{O}}(\boldsymbol{r}) &= \frac{1}{(2\pi)^{2{N}}} \int d^{2{N}}\boldsymbol{r'} e^{-i\boldsymbol{r}^T \boldsymbol{\Omega} \boldsymbol{r'} } \chi_{\mathcal{O}}(\boldsymbol{r'}), \label{eq:Wignerdef} \\
\chi_{\mathcal{O}}(\boldsymbol{r'}) &= \text{Tr}\{\hat{D}(\boldsymbol{r}') \hat{\mathcal{O}} \}, 
\end{align}
where
\begin{align}
    \boldsymbol{r} = (x_1, p_1, ... x_N, p_N)^T
\end{align} denotes a vector of $2{N}$ quadrature variables corresponding to the ${N}$ modes, and likewise for $\boldsymbol{r}'$. By $\boldsymbol{\Omega}$ we denote the $2{N} \times 2{N}$ symplectic matrix
\begin{align}
    \boldsymbol{\Omega} = \bigoplus_{i=1}^{N} \begin{pmatrix}
        0 & 1 \\ -1 & 0
    \end{pmatrix},
\end{align}
and 
\begin{align}
    \hat{D}(\boldsymbol{r}') = e^{-i (\boldsymbol{r}')^T \boldsymbol{\Omega} \hat{\boldsymbol{r}}} \label{eq:Ddef}
\end{align}
is the usual displacement operator, with $\hat{\boldsymbol{r}}$ being the vector of $2N$ quadrature operators, $(\hat{x}_1, \hat{p}_1, ... \hat{x}_N, \hat{p}_N)^T$. 

The Wigner function as defined in Eq. \eqref{eq:Wignerdef} is normalized such that 
\begin{align}
    \int d^{2{N}}\boldsymbol{r} W_{\mathcal{O}}(\boldsymbol{r}) &=\text{Tr}\{\hat{\mathcal{O}}\},
\end{align}
resulting in the expected normalization condition when $\mathcal{\hat{O}}$ refers to the density operator of a normalized state. The expectation value of an operator $\hat{\mathcal{O}}$ with respect to the state $\hat{\rho}$ can written as
\begin{align}
    \left< \hat{\mathcal{O}} \right> &= \text{Tr}\{ \hat{\rho} \hat{\mathcal{O}} \}\\
    &= (2\pi)^{{N}} \int d\boldsymbol{r} W_{\rho}(\boldsymbol{r}) W_{\mathcal{O}}(\boldsymbol{r}), \label{eq:EV_general}
\end{align}
where $W_{\rho}(\boldsymbol{r})$ and $W_{\mathcal{O}}(\boldsymbol{r})$ are the Wigner representations of the operators $\hat{\rho}$ and $\hat{\mathcal{O}}$, respectively. 

\subsection{Sum of Gaussians formalism}
\label{section:sumGauss}

We adopt the approach described in Ref. \cite{Eli_gaussians}, in which one writes the (in general, non-Gaussian) Wigner function of a state $\hat{\rho}$ as a sum of Gaussian functions: 
\begin{align}
    W_{\rho}(\boldsymbol{r}) = \sum_m c_m G_m(\boldsymbol{r}), \label{eq:sum_gauss}
\end{align}
where the $c_m$ are complex coefficients, and
\begin{align}
    G_m(\boldsymbol{r}) = \frac{\text{exp}\left( -\frac{1}{2} \left( \boldsymbol{r} - \boldsymbol{\mu}_m \right)^T \boldsymbol{\gamma}_m^{-1} \left( \boldsymbol{r} - \boldsymbol{\mu}_m \right) \right) }{\sqrt{\text{det}(2\pi \boldsymbol{\gamma}_m)}} \label{eq:Gaussian}
\end{align}
denotes a normalized Gaussian with the mean vector $\boldsymbol{\mu}_m$ and covariance matrix $\boldsymbol{\gamma}_m$.

In principle, this expansion can be applied to any Wigner function, as long as arbitrarily many terms are allowed. Moreover, certain states of practical interest can be represented compactly in this way, despite their non-Gaussianity. For example, the Wigner function for a cat state can be written exactly in the form of Eq. \eqref{eq:sum_gauss} with four terms \cite{Eli_gaussians}. An especially attractive feature of the Gaussian expansion is that---unlike a naive Fock representation---the number of terms needed to describe a state does {not} necessarily increase for higher-energy states: For example, the expansion for a (squeezed) cat state is represented by four terms, regardless of its amplitude (and squeezing). {For a GKP state, the number of peaks does tend to increase with energy and quality; thus, the number of terms needed to represent it increases as well.}

The Wigner function for a tensor product of two states can be written as 
\begin{align}
    W_{\rho\otimes\sigma}(\boldsymbol{r}) = W_{\rho}(\boldsymbol{r}_1) W_\sigma(\boldsymbol{r}_2), \label{eq:W_tprod}
\end{align}
where $\boldsymbol{r}_1$ and $\boldsymbol{r}_2$ denote the sets of phase space variables associated the individual $\hat{\rho}$ and $\hat{\sigma}$, respectively. If both $W_{\rho}(\boldsymbol{r}_1)$ and $W_{\sigma}(\boldsymbol{r}_2)$ are expanded as in Eq. \eqref{eq:sum_gauss}, one has
\begin{align}
    W_{\rho\otimes\sigma}(\boldsymbol{r}) &= \sum_{m,n} c_m c_n G_m(\boldsymbol{r}_1) G_n(\boldsymbol{r}_2)\\
    &= \sum_{m,n} c_m c_n G_{mn}(\boldsymbol{r}),
\end{align}
where the mean and covariance matrix of $G_{mn}(\boldsymbol{r})$ are the direct products of the means and covariance matrices of $G_m(\boldsymbol{r}_1)$ and $G_n(\boldsymbol{r}_2)$:
\begin{align}
    \boldsymbol{\gamma}_{mn} &= \boldsymbol{\gamma}_m \oplus \boldsymbol{\gamma}_n\\
    \boldsymbol{\mu}_{mn} &= \boldsymbol{\mu}_m \oplus \boldsymbol{\mu}_n.\label{eq:mu_sep}
\end{align}

For GKP states, the specific form of $c_m$, $\boldsymbol{\gamma}_m$, and $\boldsymbol{\mu}_m$ depend on the details of the protocol used to generate the state; these expansions are known {for GKP states generated through cat state breeding, and for the ``Fock damped" description of finite-energy GKP states} \cite{Eli_gaussians}. In this manuscript, we focus on GKP states generated by breeding cat states \cite{Vasconcelos:10, catbreed_PhysRevA.97.022341, pizzimenti2025opticalgottesmankitaevpreskillqubitgeneration}, as indicated in Fig. \ref{fig:breeding}; in this case, {it can be shown (see Appendix \ref{appendix:catbreeddetails}, and Refs. \cite{anaelle_PhysRevA.110.012408, Vasconcelos:10}) that the state generated after $\mathcal{M}$ rounds of breeding is represented as}
\begin{align}
    W_{\rho}(\boldsymbol{r}) = |{\mathcal{N}}|^2  \sum_{k,k'=0}^{\mathcal{M}+1} {\mathcal{M}+1\choose k}{\mathcal{M}+1\choose k'}\mathcal{A}_{k,k'} G_{k,k'}(\boldsymbol{r}), \label{eq:slow_breed_wigner}
\end{align}
with $G_{k,k'}(\boldsymbol{r})$ defined as in Eq. \eqref{eq:Gaussian}, and 
\begin{align}
    \mathcal{A}_{k,k'} &= \text{exp}\left(-\frac{1}{2} e^{2\xi}(\beta_k - \beta_{k'})^2\right)\\
    \boldsymbol{\gamma}_{k,k'} &= \frac{1}{2} \begin{bmatrix}
    e^{-2\xi} & 0 \\ 0 & e^{2\xi}
    \end{bmatrix} \equiv \boldsymbol{\gamma}\\
    \boldsymbol{\mu}_{k,k'} &= \sqrt{\frac{1}{2}} \begin{bmatrix}
    \beta_k + \beta_{k'} \\ ie^{2\xi}(\beta_{k'} - \beta_k) \end{bmatrix}\\
    \beta_k &= \frac{(2k-(\mathcal{M}+1))\alpha}{2\sqrt{2^M}},
\end{align}
where $\alpha$ and $\xi$ are the amplitude and squeezing of the initial cat states, and $\mathcal{N}$ is a normalization constant. Similar expansions for other approximate GKP states can be derived \cite{Eli_gaussians}, and in situations where the exact sum-of-Gaussians expansion is more difficult to derive, one can construct approximate expansions. 
\begin{figure}
    \centering
    \includegraphics[width=0.5\textwidth]{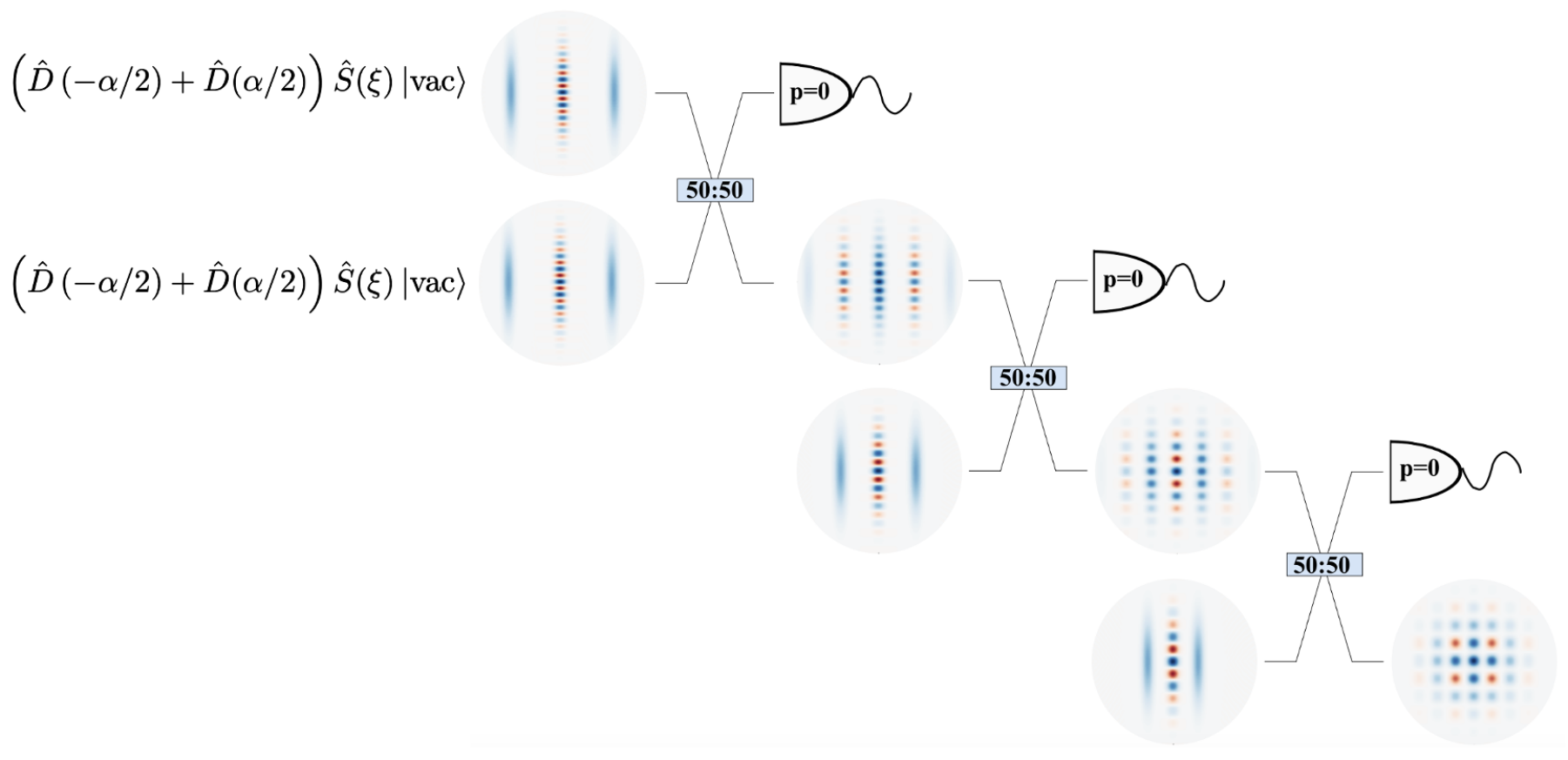}
    \caption{Sketch of $\mathcal{M}=3$ rounds of a cat breeding protocol. The scheme involves interfering squeezed cat states at a balanced beamsplitter, with one of the outputs subjected to homodyne detection (here we have taken the outcome $p=0$). The unmeasured mode becomes a closer approximation to a GKP state with each round of breeding. Details can be found in Ref. \cite{Vasconcelos:10,catbreed_PhysRevA.97.022341}.}
    \label{fig:breeding}
\end{figure}

\subsection{Describing entangling operations}
\label{section:entangling}

\begin{figure}[h]
    \centering
    \includegraphics[width=0.45\textwidth]{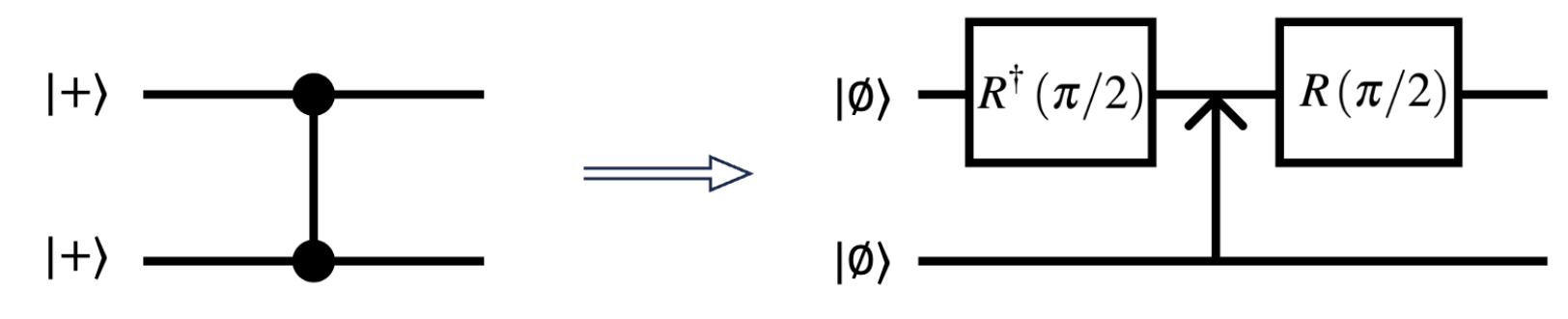}
    \caption{Implementation of a GKP CZ operation through static linear components. $\hat{R}(\theta)$ denotes a $\theta$ phase shift, and the arrow represents a beamsplitter, following the notation used in \cite{staticLO_PRXQuantum.2.040353}.}
    \label{fig:dumbbell_circuit}
\end{figure}

Single-mode GKP states can be entangled by applying Gaussian unitaries and homodyne measurement \cite{GKP_PhysRevA.64.012310,PhysRevLett.97.110501,PhysRevA.104.012431,macronization}. {For example, a GKP Bell state can be generated by applying the passive circuit shown in Fig. \ref{fig:dumbbell_circuit} to two single-mode GKP sensor states, which are defined as $\ket{\varnothing} = \hat{S}(\sqrt{2})\ket{+}$, where $\hat{S}(\sqrt{2})$ denotes a squeezing operation, and $\ket{+} = \frac{1}{\sqrt{2}} \left( \ket{0}_{\text{GKP}} + \ket{1}_{\text{GKP}} \right)$ denotes the superposition state in a GKP encoding scheme \cite{staticLO_PRXQuantum.2.040353}.}

\begin{figure}[h]
    \centering
    \includegraphics[width=0.45\textwidth]{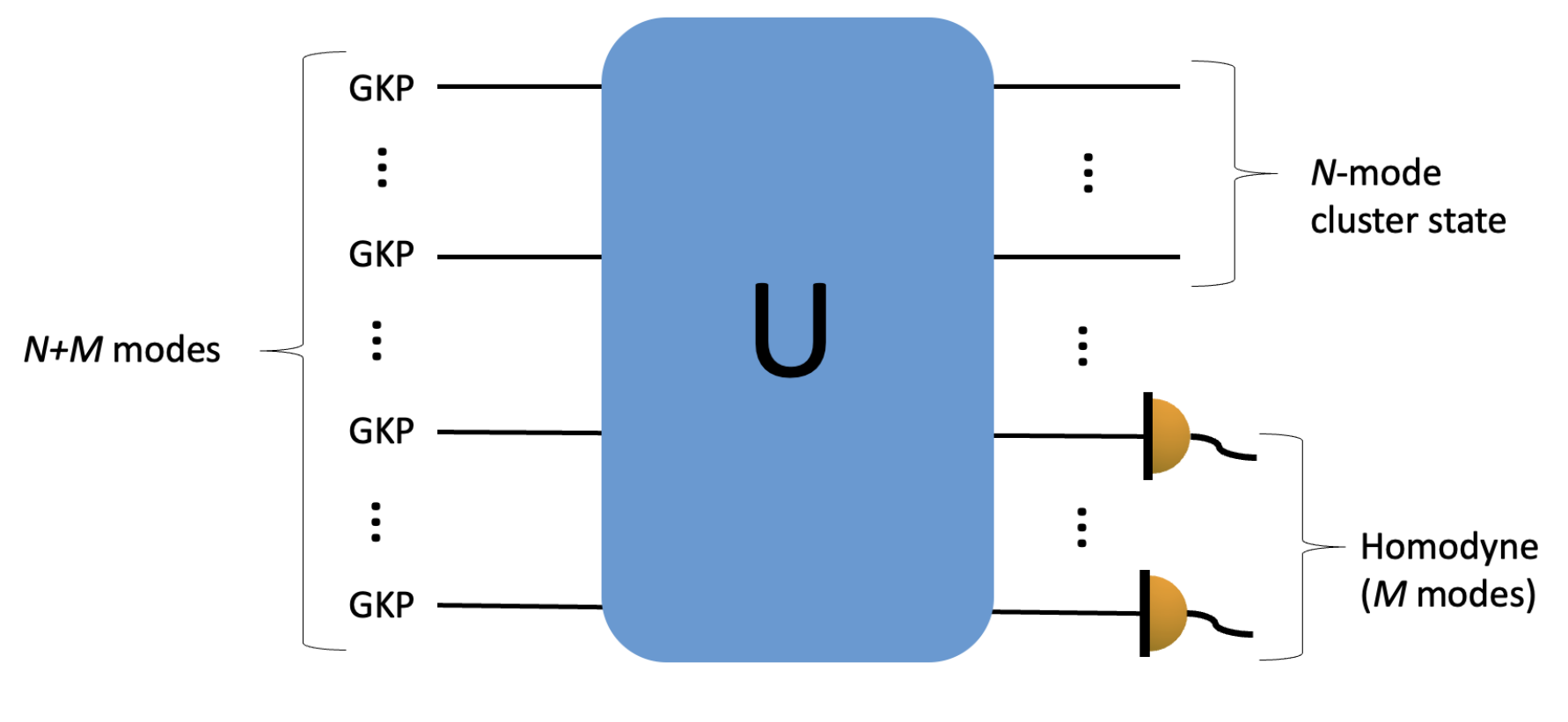}
    \caption{{Sketch of the cluster state generation circuits we consider in this manuscript: {$M+N$} single-mode GKP states are sent through a linear unitary circuit (box labelled $U$). {$M$} of the output modes are then subjected to homodyne measurement to yield a {$N$-mode} cluster state.}}
    \label{fig:circuitthing}
\end{figure}

More general cluster states can be formed by combining linear unitary circuits with homodyne measurement \cite{macronization}; the general form of these ``stitching'' circuits is sketched in Fig. \ref{fig:circuitthing}, and a specific example is shown in Fig. \ref{fig:4circuit}.

If a state $\hat{\rho}$ evolves under a Gaussian unitary ${\hat{U}}$, the Wigner function for the evolved state ${\hat{\overline{\rho}}} = {\hat{U}}
\hat{\rho}{\hat{U}}^{\dagger}$ is
\begin{align}
    W_{\overline{\rho}}(\boldsymbol{r}) = W_{{\rho}}({\boldsymbol{A}^{T}} \boldsymbol{r}), \label{eq:W_beforeandafter}
\end{align}
where $W_{\rho}(\boldsymbol{r})$ is the Wigner function of the initial density operator
(see Appendix \ref{appendix:Gaussian_unitary}). 
The {symplectic} matrix $\boldsymbol{A}$ is defined by the action of the unitary; because ${\hat{U}}$ is Gaussian, one can write the simple  input-output relation \cite{serafini2017quantum}
\begin{align}
    {\hat{\boldsymbol{r}}'} = \boldsymbol{A} \hat{\boldsymbol{r}}, \label{eq:pq_trans}
\end{align}
where $\hat{\boldsymbol{r}}$ 
is the vector of position and momentum operators corresponding to the ``input modes'', and
\begin{align}
    {\hat{\boldsymbol{r}}'} = \left( {\hat{U}^{\dagger}}\hat{x}_1{\hat{U}}, {\hat{U}^{\dagger}}\hat{p}_1{\hat{U}},... {\hat{U}^{\dagger}}\hat{x}_N{\hat{U}}, {\hat{U}^{\dagger}}\hat{p}_N{\hat{U}} \right)^T \label{eq:r_tilde}.
\end{align}

If the initial $\hat{\rho}$ is represented by a Wigner function of the sum-of-Gaussians form in Eq. \eqref{eq:sum_gauss}, the evolved Wigner function is 
\begin{align}
    W_{\overline{\rho}}(\boldsymbol{r}) &= \sum_{m} c_m G_{m}({\boldsymbol{A}^{T}} \boldsymbol{r}), \label{eq:W_rhobar}\\
    &= \sum_{m} c_m \overline{G}_{m}( \boldsymbol{r}),
\end{align}
where $\overline{G}_{m}( \boldsymbol{r})$ is a normalized Gaussian with the mean and covariance matrix 
\begin{align}
    \overline{\boldsymbol{\mu}} &= \boldsymbol{A} \boldsymbol{\mu} \label{eq:mubar}, \\
    \overline{\boldsymbol{\gamma}} &= \boldsymbol{A} \boldsymbol{\gamma} \boldsymbol{A}^T. \label{eq:gammabar}
\end{align}

\subsection{Measurement}
\label{section:homodyne}
We describe the homodyne detection and postselection as follows. Assume that an arbitrary quadrature operator
\begin{align}
    \hat{\eta}_{\theta} =  \hat{p} \text{cos}(\theta) - \hat{x} \text{sin}(\theta)
\end{align}
is being measured, with $\tilde{\eta}$ denoting the measurement outcome. We write the  state $\hat{\rho}'$ after the measurement as 
\begin{align}
    \hat{\rho}' &= \frac{\hat{K}^{\dagger}_{\tilde{\eta}} {\hat{\overline{\rho}}} \hat{K}_{\tilde{\eta}}}{\text{Tr}\{\hat{K}^{\dagger}_{\tilde{\eta}} {\hat{\overline{\rho}}} \hat{K}_{\tilde{\eta}}\}}, \label{eq:rho_post}
\end{align}
where {$\hat{\overline{\rho}}$} is the state prior to measurement, and \cite{oudot2024realisticbelltestshomodyne}
\begin{align}
    \hat{K}_{\tilde{\eta}} &= \int d\eta \Theta_{\tilde{\eta}}(\eta) \ketbra{\eta}{\eta}.
\end{align}
The function $\Theta_{\tilde{\eta}}(\eta)$ defines a window of homodyne outcomes centered at $\tilde{\eta}$. {Here we will take the limit $|\Theta_{\tilde{\eta}}(\eta)|^2 \rightarrow \delta(\eta-\tilde{\eta})$ corresponding to an ideal postselection; non-ideal postselection can be considered by adjusting the shape of $|\Theta_{\tilde{\eta}}(\eta)|^2$ -- for example, to be a Gaussian function.} The Wigner representation of $\hat{F}_{\tilde{\eta}} = \hat{K}^{\dagger}_{\tilde{\eta}} \hat{K}_{\tilde{\eta}}$ is (see Appendix \ref{appendix:homodyne}) 
\begin{align}
    W_{F_{\tilde{\eta}}}(\boldsymbol{r}) &= \frac{1}{(2\pi)}|\Theta_{\tilde{\eta}}(\eta_{\theta})|^2, 
\end{align}
where $\eta_{\theta} = p \text{cos} (\theta) - x \text{sin} (\theta)$ is a phase-space variable corresponding to $\hat{\eta}_{\theta}$. Similarly, homodyne detection in $M$ modes can be represented as
\begin{align}
    W_{F_{\boldsymbol{\tilde{\eta}}}}(\boldsymbol{r}) &= \frac{1}{(2\pi)^M}\prod_{m=1}^{M}|\Theta_{{\tilde{\eta}}_m}(\eta_{\theta m})|^2,
\end{align}
where $\boldsymbol{\tilde{\eta}}$ denotes a vector of homodyne outcomes. 

{The formalism we have summarized enables a compact representation of multimode non-Gaussian states---like GKP cluster states---given compact Gaussian representations for the single-mode input states. With this, one can obtain the density operator of the multi-mode state, from which one can compute particular figures of merit.}

\subsection{{Loss}}

{Photon loss is considered to be the dominant source of noise in photonic approaches to quantum information. Its inclusion is essential in a realistic description of photonic devices. A conventional approach to modeling loss is to introduce a fictitious beamsplitter that couples the physical mode to a ``loss mode", with the beamsplitter angle $\theta_i$ chosen to produce the correct transmittance of the physical mode \cite{RevModPhys.84.621,PhysRevA.97.032346,Helt2020, Oszmaniec2018}. Here we take the input to the loss mode to be vacuum, but the generalization to thermal states is easily made \cite{Eli_gaussians}.}

{Applying loss to $\hat{\overline{\rho}}$ (as defined in Eqs. \eqref{eq:W_rhobar} -- \eqref{eq:gammabar}) amounts to making the substitution
\begin{align}
    \overline{\boldsymbol{\mu}} &\rightarrow \boldsymbol{T} \overline{\boldsymbol{\mu}} \label{eq:mu_loss}\\
    \overline{\boldsymbol{\gamma}} &\rightarrow \boldsymbol{T}  \overline{\boldsymbol{\gamma}} \boldsymbol{T}^T + \frac{1}{2} \boldsymbol{R} \boldsymbol{R}^T \label{eq:gamma_loss},  
\end{align}
with 
\begin{align}
    \boldsymbol{T} &= \bigoplus_{i=1}^N \cos \theta_i \mathbb{1},\\
    \boldsymbol{R} &= \bigoplus_{i=1}^N \sin \theta_i \mathbb{1},\\
    \theta_i &= \cos^{-1}(\tau_i).
\end{align}
By $\theta_i$ we denote the effective beamsplitter angle for each lossy mode, and $\tau_i$ is the transmittance of the channel. This result is quoted in Ref. \cite{Eli_gaussians}, and its derivation is reviewed in Appendix \ref{appendix:loss}}. The inclusion of loss does not significantly alter the numerical cost of simulations, because it does not increase the number of terms required for the Gaussian expansion of the Wigner function; one just needs to update the means and covariance matrices according to Eqs. \eqref{eq:mu_loss} and \eqref{eq:gamma_loss}.

\section{Stabilizer expectation values}
\label{sec:stabilizers}

A relevant set of parameters in quantum information processing applications is the expectation values (EVs) of the cluster state's stabilizers. {The stabilizer EVs can be used directly as a metric for the quality of a state, or they can be used to infer other relevant metrics such as effective squeezing or entanglement witnesses \cite{sciara_witnesses, Toth_PhysRevA.72.022340}}. The operator $\hat{\mathcal{S}}$ is a stabilizer for a state $\ket{\psi}$ if it satisfies 
\begin{align}
    \hat{\mathcal{S}} \ket{\psi} = \ket{\psi}. \label{eq:stab_def}
\end{align}
An ideal $N$-mode cluster state is stabilized (and uniquely characterized) by products of single-mode Pauli operators \cite{Toth_PhysRevA.72.022340}
\begin{align}
    \hat{\mathcal{S}}_k = \hat{X}^{(k)} \prod_{\text{Neighbours $l$ of $k$}} \hat{Z}^{(l)} \label{eq:stab_def2},
\end{align}
where $k$ labels each of the $N$ vertices of the cluster state, such that the cluster state is specified by $N$ distinct stabilizers with the form given in Eq. \eqref{eq:stab_def2}.

In a square lattice GKP encoding scheme, {the generators of the Pauli group are $\sqrt{\pi}$ displacements in position and momentum.} The logical Pauli $\hat{X}$ and $\hat{Z}$ operations correspond respectively to position and momentum displacements by odd integer multiples of $\sqrt{\pi}$ \cite{GKP_PhysRevA.64.012310, staticLO_PRXQuantum.2.040353}. 
An ideal GKP qubit is stabilized by discrete displacements in phase space (by $2\sqrt{\pi}$ in position and momentum, corresponding to $\hat{X}^2$ and $\hat{Z}^2$), {so the expectation value (EV) of these operators is unity-- that is,
\begin{align}
    \text{Tr}\{ \hat{\rho}_{{\text{ideal}}} \hat{D}(\boldsymbol{\overline{r}}) \} = 1
\end{align}
for $\boldsymbol{\overline{r}} = (2n\sqrt{\pi},2m\sqrt{\pi})^T$, with $n,m \in \mathbb{Z}$.} 

For finite-energy approximations to GKP states, the EVs of the same operators are necessarily less than unity, due to the state's finite extent in phase space, {and due to the non-zero width of its peaks in phase space}.  Generally, the higher the state's quality, the closer to unity its stabilizer EVs. Hence stabilizer EVs can be used to gain insight into the quality of approximate GKP states, and its dependence on various parameters involved in the state generation protocol. {Although the elements of the Pauli group can be defined in terms of displacements by any integer multiple of $\sqrt{\pi}$, we will refer to 
stabilizer sets composed of displacements of minimal length -- namely, displacements by $\sqrt{\pi}$.} Implementations of stabilizers using larger displacements will result in a lower EV, again due to the realistic states' finite extent in phase space.

Using Eqs. \eqref{eq:rho_post} and \eqref{eq:EV_general}, the stabilizer EV for a cluster state generated as described above can be written as
\begin{align}
    \langle {\hat{\mathcal{S}}} \rangle = \text{Tr}\{ \hat{\rho}' {\hat{\mathcal{S}}} \} &= \frac{\text{Tr}\{{\hat{\overline{\rho}}} {\hat{\mathcal{S}}} \hat{K}_{\tilde{\eta}}^{\dagger} \hat{K}_{\tilde{\eta}}\}}{\text{Tr}\{{\hat{\overline{\rho}}} \hat{K}_{\tilde{\eta}}^{\dagger} \hat{K}_{\tilde{\eta}}\}}, \label{eq:S_EV} \\
    &= (2\pi)^{({N}-{M})}\frac{\int d\boldsymbol{r} W_{{\overline{\rho}}}(\boldsymbol{r}) W_{F_{\tilde{\eta}}}(\boldsymbol{r}) W_{{\mathcal{S}}}(\boldsymbol{r})}{\int d\boldsymbol{r} W_{{\overline{\rho}}}(\boldsymbol{r}) W_{F_{\tilde{\eta}}}(\boldsymbol{r})} \label{eq:numerator}
\end{align}
where ${\hat{\rho}'}$ is a ${N}-$ mode density operator, ${\hat{K}^{\dagger}_{\tilde{\eta}}}$ (or $\hat{F}_{\tilde{\eta}} = {\hat{K}^{\dagger}_{\tilde{\eta}}} {\hat{K}_{\tilde{\eta}}}$) is a ${M}-$ mode operator describing the homodyne measurement, and $\hat{\mathcal{S}}$ is a stabilizer defined over the unmeasured ${N}-{M}$ modes. The factor of $2\pi^{({N}-{M})}$ appears due to the implicit identity operator in the denominator of Eq. \eqref{eq:S_EV} (see Appendix \ref{appendix:S_details}). In a GKP encoding scheme, $W_{\mathcal{S}}(\boldsymbol{r})$ is the Wigner representation of a displacement operator, which has the form 
\begin{align}
    W_{D_{\boldsymbol{\overline{r}}}} (\boldsymbol{r}) &= \frac{1}{(2\pi)^{N}} e^{-i \boldsymbol{r}^T \boldsymbol{\Omega} \overline{\boldsymbol{r}}} \label{eq:W_D},
\end{align}
with the displacement $\overline{\boldsymbol{r}}$ set to the relevant value for the stabilizer in question; for example, for a single-mode Pauli $\hat{X}$ operator, which is implemented by a position displacement by {$m\sqrt{\pi}$ ($m \in \mathbb{Z}$) one would put $\overline{\boldsymbol{r}} = (m \sqrt{\pi},0)^T$.} 

Using a Gaussian expansion for $W_{{\overline{\rho}}}(\boldsymbol{r})$ in Eq. \eqref{eq:numerator} results in Gaussian integrals that---due to the simple form of the stabilizer's Wigner representation---can easily be evaluated analytically. {We neglect losses here for simplicity; the expressions including loss (and details of the derivation) are given in Appendix \ref{appendix:S_details}.} We obtain
\begin{widetext}
\begin{align}
    \langle {{\mathcal{S}}} \rangle &= \frac{\sum_{\boldsymbol{m}} c_{\boldsymbol{m}} g_{\boldsymbol{m}}(\tilde{\boldsymbol{\eta}};\boldsymbol{J})\text{exp}\left(i\boldsymbol{J}^T \boldsymbol{P}_C \boldsymbol{A}{\boldsymbol{\mu}}_{\boldsymbol{m}} \right) \text{exp}\left( -\frac{1}{2} \boldsymbol{J}^T \boldsymbol{P}_C \boldsymbol{A}{\boldsymbol{\gamma}}_{\boldsymbol{m}} \boldsymbol{A}^T \boldsymbol{P}_C^T \boldsymbol{J} \right)}{\sum_{\boldsymbol{m}} c_{\boldsymbol{m}} {g}_{\boldsymbol{m}}(\tilde{\boldsymbol{\eta}};\boldsymbol{0})}. \label{eq:S_result}
\end{align}
\end{widetext}
By $\boldsymbol{m}$ we denote a vector of indices $\boldsymbol{m} = (m_1, m_2,...,m_N)^T$, where the index $m_i$ refers to the Gaussian expansion---recall Eq. \eqref{eq:sum_gauss}---for one of the $N$ inputs to the {stitching circuit}; in this work, all the inputs are identical single-mode states, but this can easily be generalized.  The Wigner function of the $N$-mode input (separable) state is specified by 
\begin{align}
    \boldsymbol{\gamma}_{\boldsymbol{m}} &= \bigoplus_{i=1}^{N} \boldsymbol{\gamma}_{mi} \label{eq:oplus1}\\
    \boldsymbol{\mu}_{\boldsymbol{m}} &= \bigoplus_{i=1}^{N} \boldsymbol{\mu}_{mi}\\
    c_{\boldsymbol{m}} &= \prod_{i=1}^{N} c_{mi}, \label{eq:oplus2}
\end{align}
where the $\boldsymbol{\gamma}_{mi}$, $\boldsymbol{\mu}_{mi}$, $c_{mi}$ refer to the Gaussian expansion parameters for the $i$-th input mode (recall the discussion around Eq. \eqref{eq:W_tprod}). The matrix $\boldsymbol{A}$ is defined by the Gaussian unitary (recall Eq. \eqref{eq:pq_trans}), and the vector $\boldsymbol{J} = -\boldsymbol{\Omega}\overline{\boldsymbol{r}}$ defines the stabilizer operator (recall Eq. \eqref{eq:W_D}). We also introduce projection matrices $\boldsymbol{P}_H$ and $\boldsymbol{P}_C$: These project an object defined for the entire $N-$mode phase space into the subspace associated with the $M$ measured quadratures and the $N-M$ unmeasured modes, respectively. Finally, $g_{\boldsymbol{m}}(\boldsymbol{\tilde{\eta}};\boldsymbol{J})$ is a Gaussian function in the postselected homodyne outcomes $\boldsymbol{\tilde{\eta}}$, with the mean and covariance matrix
\begin{align}
    {\boldsymbol{\mu}}'_{\boldsymbol{m}} &= \boldsymbol{P}_H \boldsymbol{A}{\boldsymbol{\mu}}_{\boldsymbol{m}} + i \boldsymbol{P}_H \boldsymbol{A}{\boldsymbol{\gamma}}_{\boldsymbol{m}} \boldsymbol{A}^T \boldsymbol{P}_C^T \boldsymbol{J}, \label{eq:gauss_X_mu}\\
    {\boldsymbol{\gamma}}'_{\boldsymbol{m}} &= \boldsymbol{P}_H \boldsymbol{A}{\boldsymbol{\gamma}}_{\boldsymbol{m}} \boldsymbol{A}^T \boldsymbol{P}_H^T.
\end{align}

Stabilizer EVs for realistic GKP cluster states can be numerically computed with Eq. \eqref{eq:S_result}: Gaussian expansions for the single-mode input GKP states are obtained and used to construct $\boldsymbol{\gamma_m}$, $\boldsymbol{\mu_m}$, and $c_{\boldsymbol{m}}$ according to Eqs. \eqref{eq:oplus1} - \eqref{eq:oplus2}; the symplectic matrix $\boldsymbol{A}$ is derived for the Gaussian unitary circuit; the projectors $\boldsymbol{P}_H$ and $\boldsymbol{P}_C$ are defined according to the labeling of the modes to identify those modes that are measured and unmeasured, respectively (see Appendix \ref{appendix:S_details}); the vector $\tilde{\boldsymbol{\eta}}$ specifies the homodyne measurement outcomes; and the vector $\boldsymbol{J}$ is defined according to the stabilizer of interest. Eq. \eqref{eq:S_result} can be used to explore the effect of various state preparation settings on the quality of the generated cluster states; for example, changes in the unitary circuit or homodyne outcomes are reflected by modifying $\boldsymbol{A}$ and $\tilde{\boldsymbol{\eta}}$ respectively, whereas changes in the protocol used to generate the single-mode GKP inputs are reflected in $\boldsymbol{\gamma_m}$, $\boldsymbol{\mu_m}$, and $c_{\boldsymbol{m}}$. 

{\subsection{Gaussian Random Noise model}}
\label{section:GRN_intro}

Exact simulations are not tractable for cluster states of an arbitrary size. A standard, scalable approach for modelling non-ideal GKP states is to apply a Gaussian random noise (GRN) channel to ideal GKP states, resulting in a mixed state $\hat{\tilde{\rho}}$~\cite{repeaters,staticLO_PRXQuantum.2.040353,raveendran2022finite,macronization}. The effect of the GRN channel can be expressed as ~\cite{cerf2007quantum}  
\begin{equation}
    \hat{\tilde{\rho}} = \int_\mathbb{C}\mathrm{d}^2\alpha G_{\bm{\Delta}}(\alpha)\hat{D}(\alpha)\ketbra{\psi}{\psi}\hat{D}^\dagger(\alpha),
\end{equation}
where
\begin{equation}
    G_{\bm{\Delta}}(\alpha)=\frac{\exp{-\frac{\Re{\alpha}^2}{\Delta_x^2/2}}\exp{-\frac{\Im{\alpha}^2}{\Delta_p^2/2}}}{\pi\Delta_x\Delta_p/2},
\end{equation}
{and
\begin{align}
    \hat{D}(\alpha) &= \text{exp}\left( \alpha a^{\dagger} - \alpha^* a \right)\\
    &= \text{exp}\left( \sqrt{2}i \text{Im}\{\alpha\} \hat{x} -\sqrt{2}i \text{Re}\{\alpha\} \hat{p} \right).
\end{align}}
The GRN state is characterized by a Wigner function with Gaussian (rather than delta function) peaks. The $p$ and $x$ quadrature variance of the peaks is set by the ``effective squeezing'' parameters {$\Delta_p$ and $\Delta_x$}, {which are in turn often related to the performance of the GKP qubit} \cite{fukui_PhysRevLett.119.180507,noh_PhysRevA.101.012316,PhysRevA.111.052445,repeaters}. The stabilizer EVs for a GRN state are related to the effective squeezing parameters as follows ~\cite{PhysRevA.95.012305}:
\begin{align}
    \abs{\Tr\{\hat{\mathcal{S}}_x \hat{\tilde{\rho}}\}} &= \text{exp}\left(-\pi \Delta_{x}^2/2 \right) ,\label{eq:GRN_q} \\
    \abs{\Tr\{\hat{\mathcal{S}}_p\hat{\tilde{\rho}}\}} &= \text{exp}\left(-\pi \Delta_{p}^2/2 \right),
\end{align}
{where $\hat{\mathcal{S}}_x =\exp(is\hat{x})$ and $\hat{\mathcal{S}}_p=\exp(-is'\hat{p})$ denote the GKP stabilizers; the displacements $s$ and $s'$ depend on the lattice spacing of the GKP state.}

{The GRN model can be used for scalable simulations of cluster state formation by taking GRN states as the single-mode input states, with the effective squeezing parameters chosen to reproduce the stabilizer EVs of the approximate GKP states one could actually generate. The performance of a cluster state generated by stitching the input GRN states can be estimated following the methods described in Ref. \cite{macronization}, for example. However, this does not necessarily result in an accurate description of the inputs nor of the cluster state---approximate GKP states generated through realistic protocols cannot be fully characterized by effective squeezing parameters---and it is unclear how accurate GRN results are in different scenarios.}

\section{Numerical results}

\begin{figure}[h]
    \centering
     \subfloat[]{
        \includegraphics[width=0.45\textwidth]{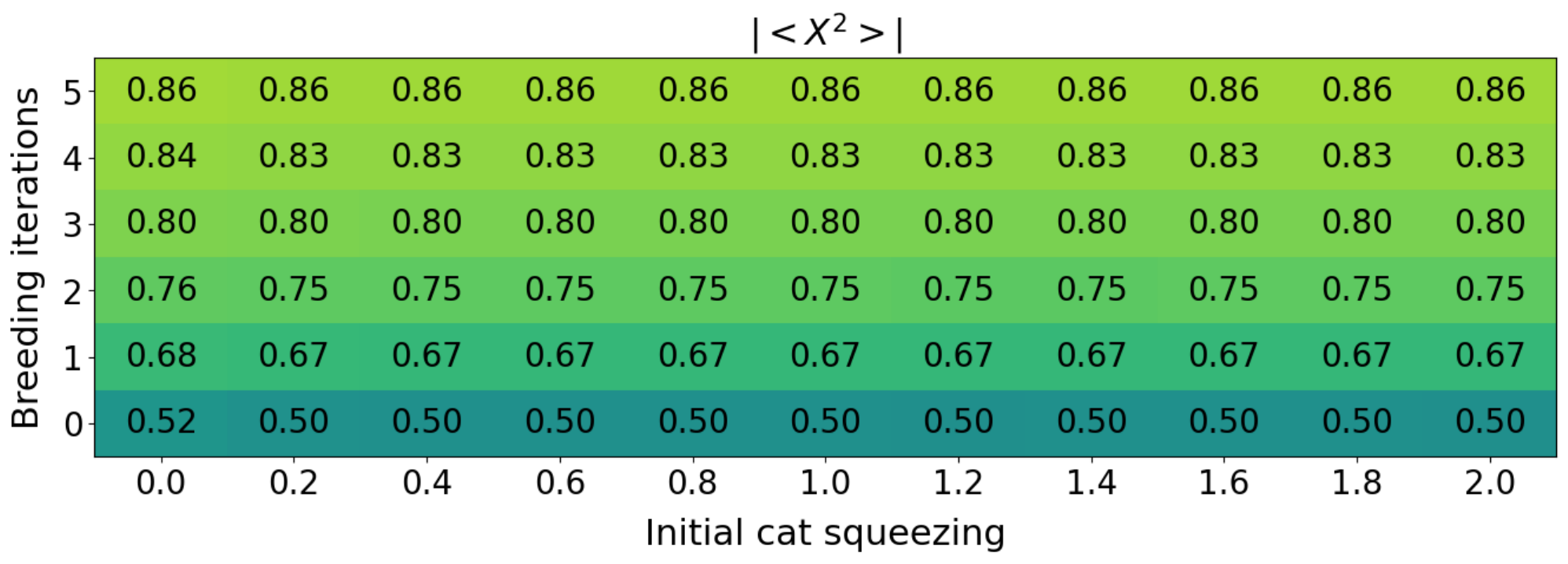}
    }

    \subfloat[]{
        \includegraphics[width=0.45\textwidth]{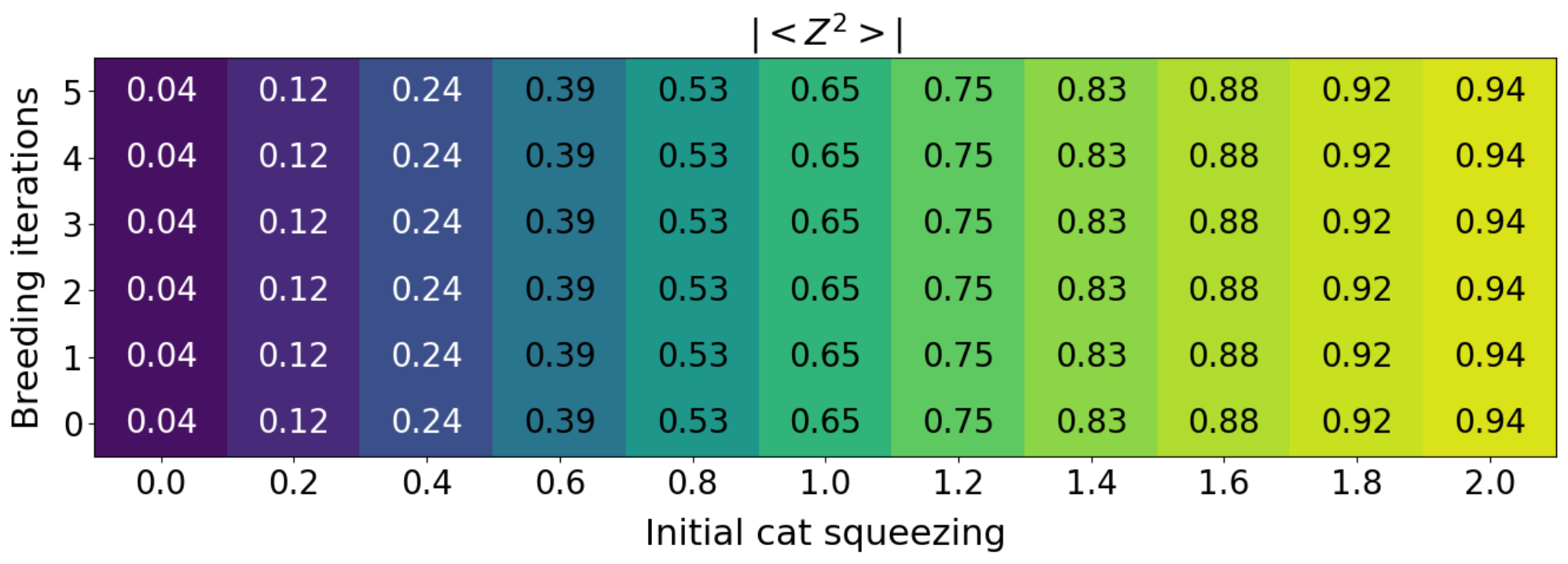}
    }
    \caption{Single-qubit stabilizer EVs for a single-mode GKP state, as a function of the parameters of the cat breeding protocol used to generate it. The stabilizer EVs increase as the squeezing of the input cat states and/or the number of iterations of cat state breeding are increased, reflecting the increasing quality of the GKP state.}
    \label{fig:GKP_breeding}
\end{figure}

In this section we compute stabilizer expectation values for single-mode and multimode GKP states. We first consider single-mode approximate GKP states generated by breeding cat states (recall Fig. \ref{fig:breeding}). In Fig.~\ref{fig:GKP_breeding}, we plot the stabilizer EVs for the bred GKP state, as a function of the squeezing of the cat states, and the number of rounds in the breeding protocol. The increasing stabilizer EVs reflect the increasing quality of the state as the key parameters of the breeding protocol are improved. 

\begin{figure}[H]
    \centering
    \includegraphics[width=0.5\textwidth]{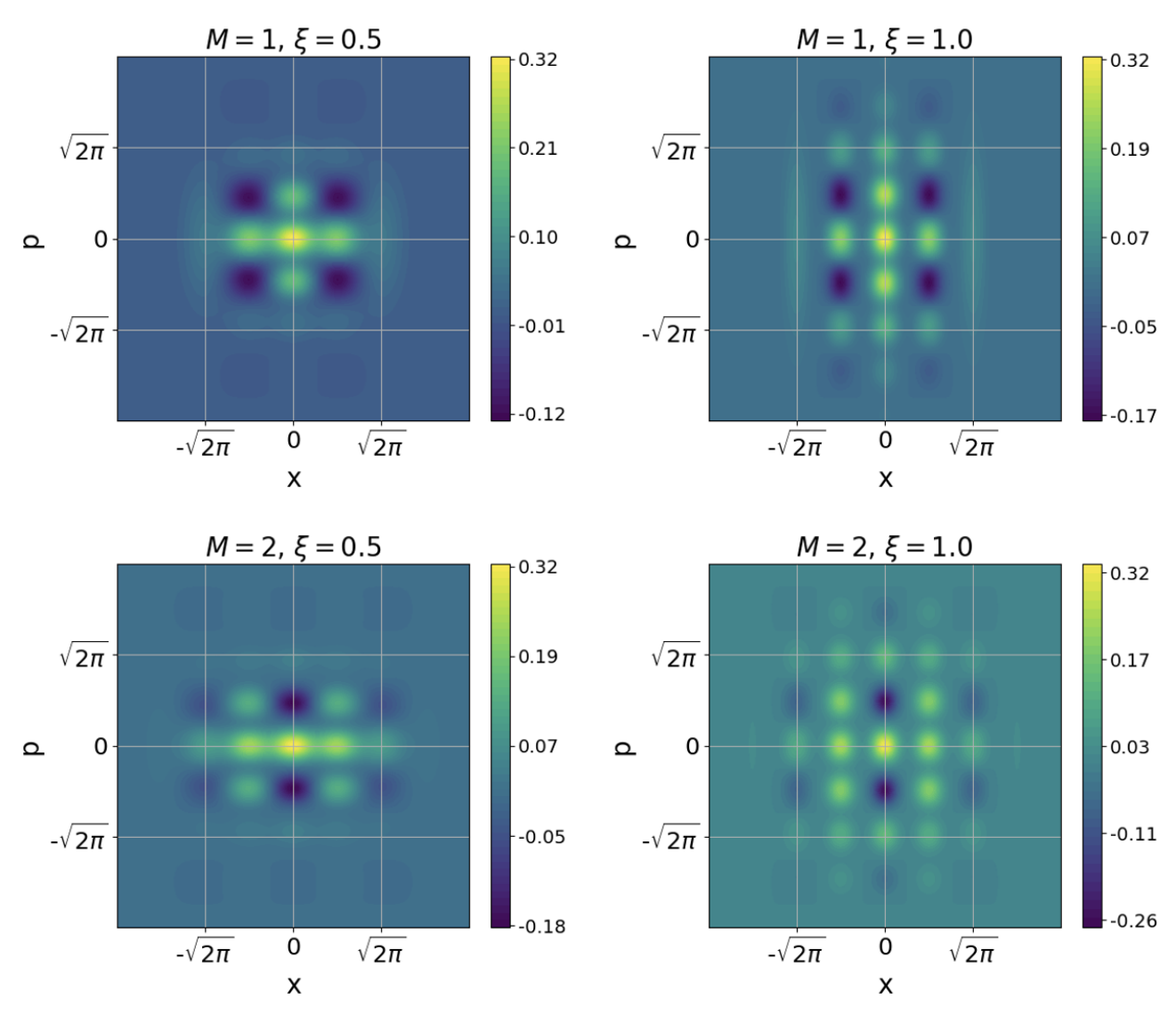}
    \caption{Wigner functions of approximate GKP states generated by the breeding protocol described in Section \ref{section:sumGauss}, for various $\mathcal{M}$ and $\xi$ (number of breeding iterations and cat squeezing, respectively).}
    \label{fig:wigner}
\end{figure}

In particular, we see that $\langle \hat{X}^2 \rangle$ depends mainly on the number of breeding iterations, while $\langle \hat{Z}^2 \rangle$ is mainly determined by the initial cat squeezing. This is because each iteration of breeding increases the number of peaks in the $x$ quadrature of the output state (see Fig. \ref{fig:wigner}), which increases the periodicity of the state along the $x$ quadrature, leaving the $p$ quadrature unaffected \cite{Vasconcelos:10, catbreed_PhysRevA.97.022341, anaelle_PhysRevA.110.012408}. {On the other hand, the squeezing of the initial cat state sets the state's periodicity in the $p$ quadrature, and the sharpness of its peaks in the $x$ quadrature; these features are reflected in $\langle \hat{Z}^2 \rangle$, which by its definition relates to the periodicity of the state in the $p$ quadrature, and the sharpness of its peaks in the $x$ quadrature.} 

{Next we address GKP Bell pairs: 
These can be generated by the circuit sketched in Fig. \ref{fig:dumbbell_circuit}, starting with approximate single-mode sensor states that can be generated, for example, through cat state breeding. Because the stitching circuit is unitary, and because the stabilizers are displacement operators, the Bell state stabilizers 
can be obtained directly from the stabilizers of the input sensor states. It is easily shown that 
\begin{align}
    \text{Tr} \{ \hat{\overline{\rho}} \hat{D}(\boldsymbol{r})\} = \text{Tr} \{ \hat{\rho} \hat{D}(\boldsymbol{r}') \}, \label{eq:stabs_unitary_evol}
\end{align}
with
\begin{align}
    \boldsymbol{r}' = \boldsymbol{A}^T {\boldsymbol{r}},
\end{align}
where {$\boldsymbol{r}'$ and ${\boldsymbol{r}}$ refer to quadrature variables (unlike in Eq. \eqref{eq:pq_trans}),} $\hat{\overline{\rho}} = \hat{U} \hat{\rho} \hat{U}^{\dagger}$ is the state after the Gaussian unitary, and $\boldsymbol{A}$ is the symplectic matrix defined in Eq. \eqref{eq:pq_trans}. Because the input $\rho$ is separable, the expectation value on the right hand side of Eq. \eqref{eq:stabs_unitary_evol} can be written as a product of single-mode expectation values: In this way, the stabilizer EVs of the output state can be inferred from EVs of displacements on the input states. This is true regardless of the input state. The stabilizer EVs obtained using the GRN model are therefore guaranteed to match the results of exact simulations, provided the effective squeezing for the input GRN states is chosen to reproduce the stabilizer EVs of the realistic input states used in the full simulation (recall Section \ref{section:GRN_intro}).} 

\begin{figure}[h]
    \centering
    \includegraphics[width=0.8\linewidth]{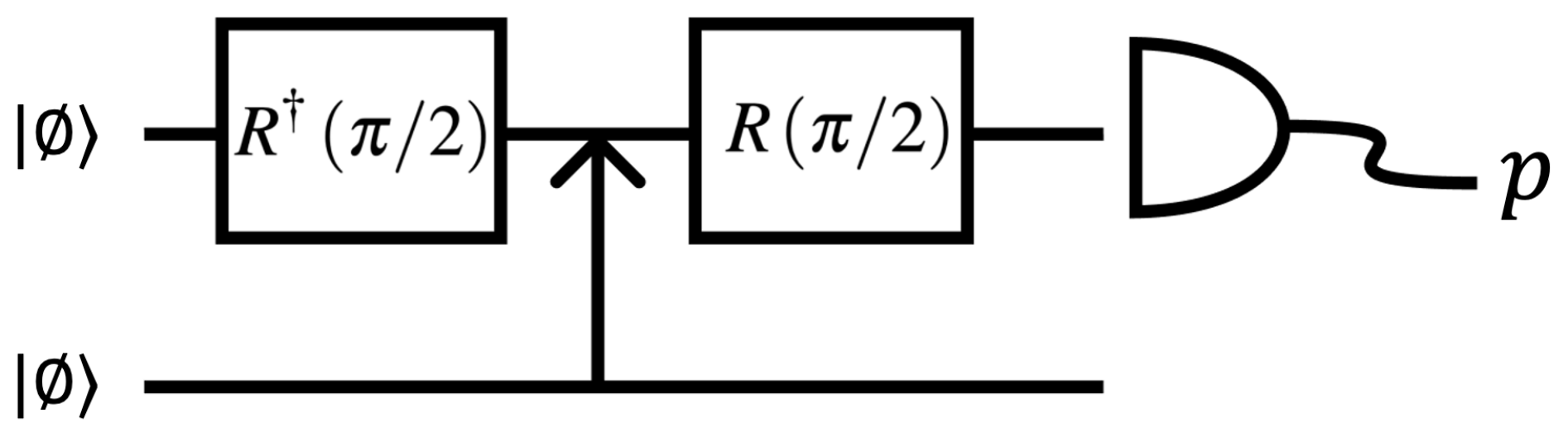}
    \caption{Sketch of the scenario we consider in Figs. \ref{fig:DB_homo} and \ref{fig:GRN_comp}: we envision creating a GKP Bell state through the unitary circuit in Fig. \ref{fig:dumbbell_circuit}, postselecting on a homodyne measurement outcome $p$ in one of the modes, and characterizing the unmeasured mode.}
    \label{fig:DBhomocircuit}
\end{figure}

\begin{figure}[h]
    \centering
     \subfloat[]{
        \includegraphics[width=0.49\textwidth]{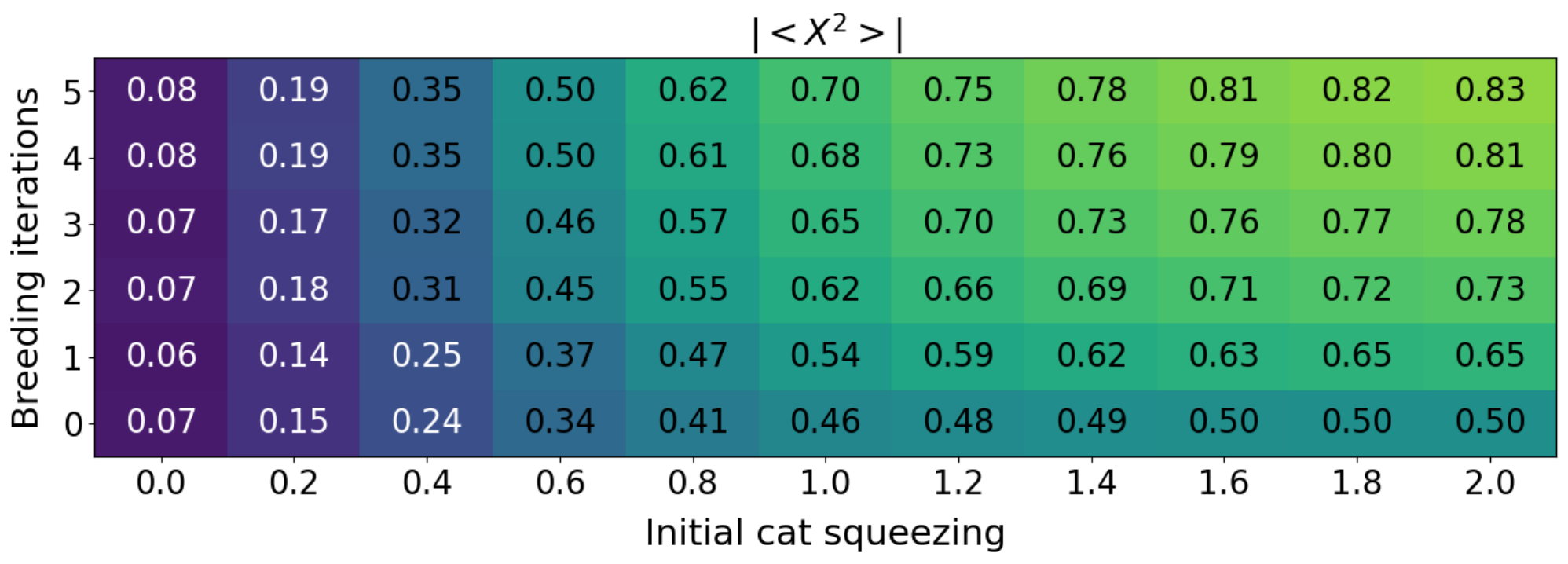}
    }

    \subfloat[]{
        \includegraphics[width=0.49\textwidth]{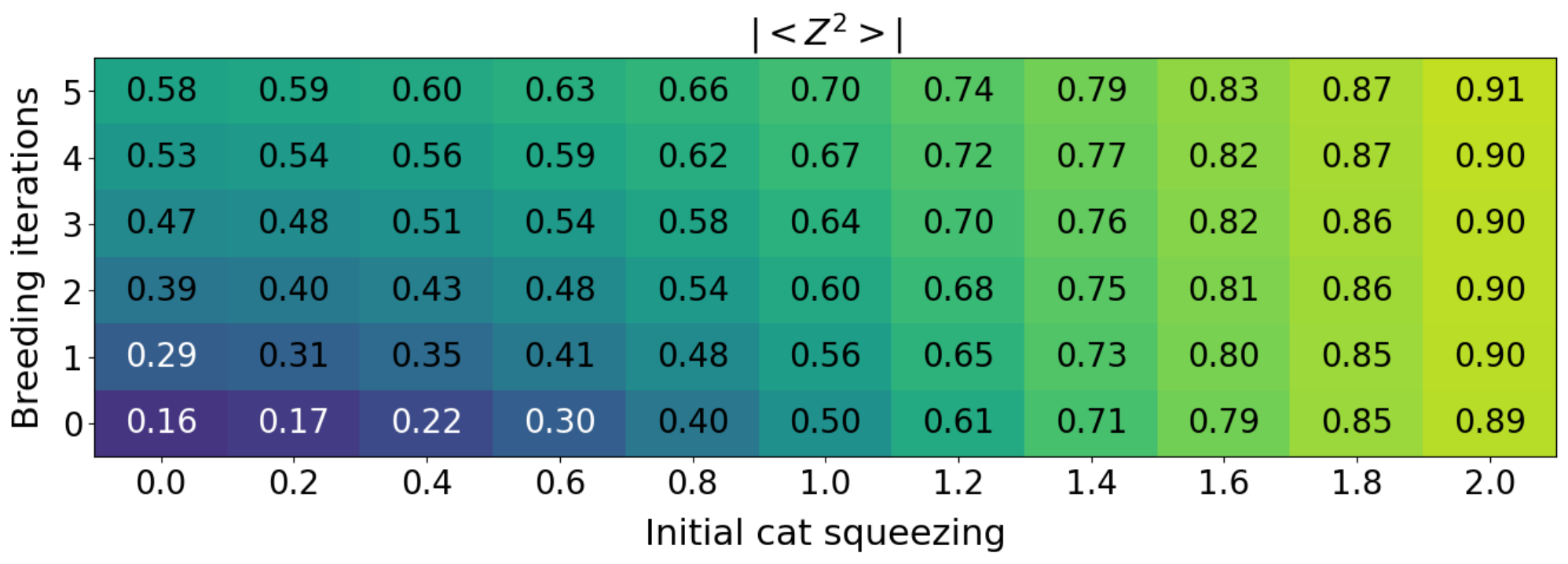}
    }
    \caption{EVs of single-qubit stabilizers evaluated for the unmeasured mode in the scheme sketched in Fig. \ref{fig:DBhomocircuit}. The results shown take the homodyne measurement outcome to be $p=0$ for odd breeding iterations, and $p = \sqrt{\pi}/2$ for even breeding iterations.}
    \label{fig:DB_homo}
\end{figure}

\begin{figure}[h]
    \centering
    \includegraphics[trim=50 5 40 20, clip,width=0.5\textwidth]
        {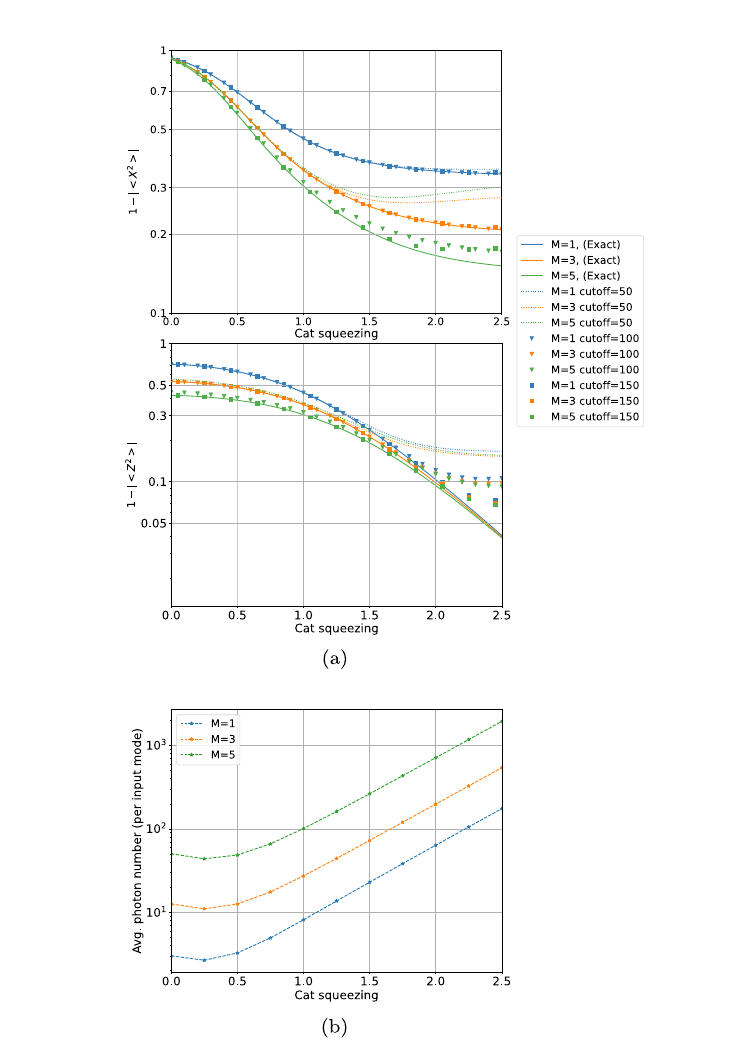}

    \caption{(a) Stabilizer EVs computed through simulations in the Fock basis (dotted lines), compared to the exact Gaussian expansion (solid lines). Single-qubit stabilizer EVs are evaluated for the unmeasured mode in Fig. \ref{fig:DBhomocircuit}. {(b) Average photon number for each input cat state for the GKP state preparation protocols considered in panel (a). The Fock basis simulations deviate from the exact result when the average photon number for the input states exceeds the photon number cutoff used in the simulation.}} 
    \label{fig:DB_homo_comp}
\end{figure}

{The situation becomes more interesting when homodyne measurements are introduced: Because the evolution is no longer unitary, direct simulation is required to determine the stabilizers of the output state following measurement.} {We consider the scenario sketched in Fig. \ref{fig:DBhomocircuit}, in one generates a GKP Bell state and measures the $p$ quadrature of one mode, leaving one unmeasured mode that we characterize in terms of stabilizer EVs; if the GKP states were ideal, the unmeasured mode would be stabilized by $\hat{X}^2$ and $\hat{Z}^2$. }
In Fig. \ref{fig:DB_homo} we plot the stabilizer EVs for the unmeasured mode, focusing on the homodyne outcome $p=0$ for odd breeding iterations, and $p=\sqrt{\pi}/2$ for even breeding iterations; the different homodyne outcome for even $\mathcal{M}$ is chosen to compensate for the fact that the cat breeding protocol produces displaced sensor states for even $\mathcal{M}$ (recall Fig. \ref{fig:wigner}, {and see Appendix \ref{appendix:displace_homo}}). {These results (and those summarized in Fig. \ref{fig:GKP_breeding}) were verified by comparisons to Fock-basis simulations. The two approaches agree closely, provided a sufficiently high Fock-basis cutoff is used.}
In Fig. \ref{fig:DB_homo_comp} we directly compare the results of Fock-basis simulations \cite{note} and our exact sum-of-Gaussians description. The discrepancies between the two grow more significant as the breeding parameters are increased to produce energetic states, with the Fock-basis results tending to underestimate the stabilizer EVs. Increasing the photon number cutoff for the Fock-basis simulations results in a closer approximation to the exact results, but even the cutoffs shown in Fig. \ref{fig:DB_homo_comp}---which are still not sufficient for certain breeding parameters---result in a costly computation that cannot be scaled to describe few-mode cluster states.

\begin{figure}[h]
    \centering
     \subfloat[]{
        \includegraphics[width=0.45\textwidth]{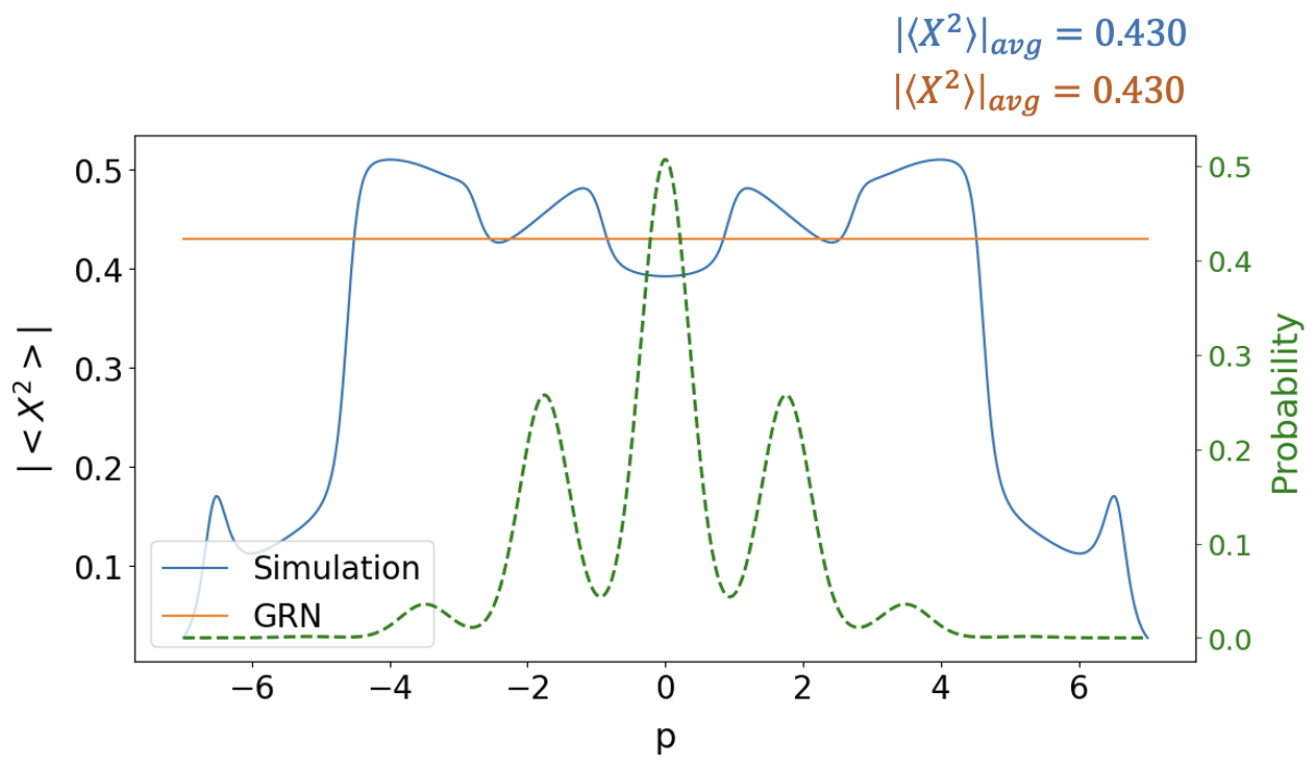}
    }

    \vspace{-0.3cm}
    \subfloat[]{
        \includegraphics[width=0.45\textwidth]{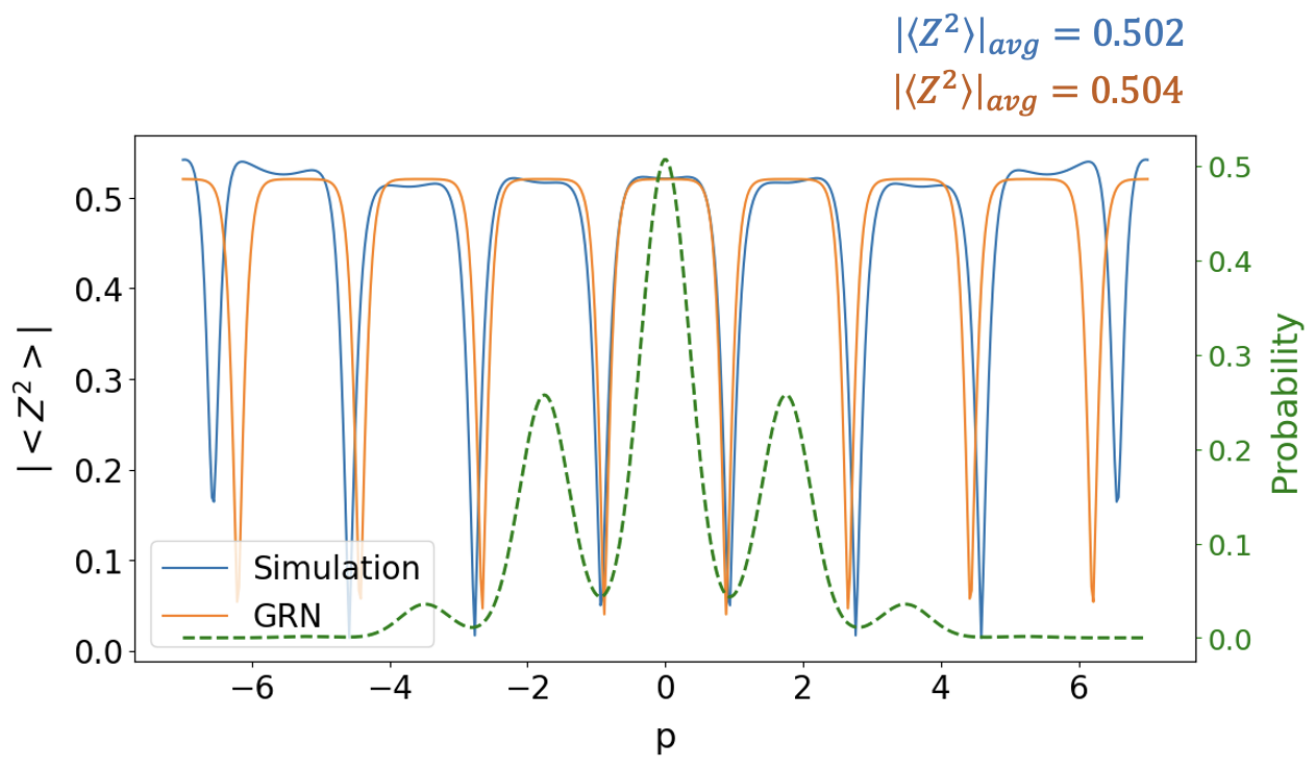}
    }
    \caption{Single-qubit stabilizer EVs evaluated for the unmeasured mode of a GKP Bell pair (see Fig. \ref{fig:DBhomocircuit}). We take the single-mode GKP states to be generated by $\mathcal{M}=3$ rounds of breeding, with an initial cat squeezing of $\xi = 0.5$. We vary the measured $p$, and plot the stabilizer EV for the unmeasured mode as calculated through direct simulation (blue line) and using the GRN model (orange line). The dashed green line represents the probability of measuring the various homodyne outcomes. {The average EVs are indicated in the top right corner.}}
    \label{fig:GRN_comp}
\end{figure}

In Fig. \ref{fig:GRN_comp} we fix a set of breeding parameters, and we plot the simulated stabilizer EVs for the unmeasured mode of the Bell state as a function of the $p$ measurement outcome. We compare this to the stabilizer EVs predicted by a GRN treatment, with the effective squeezing for the single-mode GKP states chosen to reproduce the stabilizer EVs of the bred states {(see Section \ref{section:GRN_intro} and Appendix \ref{app:ExpValsGRNS})}. The predictions of the GRN model differ from the simulation results for most $p$ measurement outcomes. There is also a clear qualitative difference, with the simulated results lacking the GRN results' periodicity over the $p$ measurement outcomes. This reflects the fact that the peaks in the bred state are not all identical, unlike in the GRN state. {In Fig. \ref{fig:GRN_comp} we also indicate the \textit{average} stabilizer EVs, weighted by the probabilities of the $p$ measurement outcomes. We find a good agreement between the average EVs: The values for $|\langle \hat{X}^2 \rangle|$ obtained using the two models deviate at the seventh significant digit, and for $|\langle \hat{Z}^2 \rangle|$ the two approaches deviate at the third significant digit.}

\begin{figure}[H]
    \centering
    \includegraphics[width=0.4\textwidth]{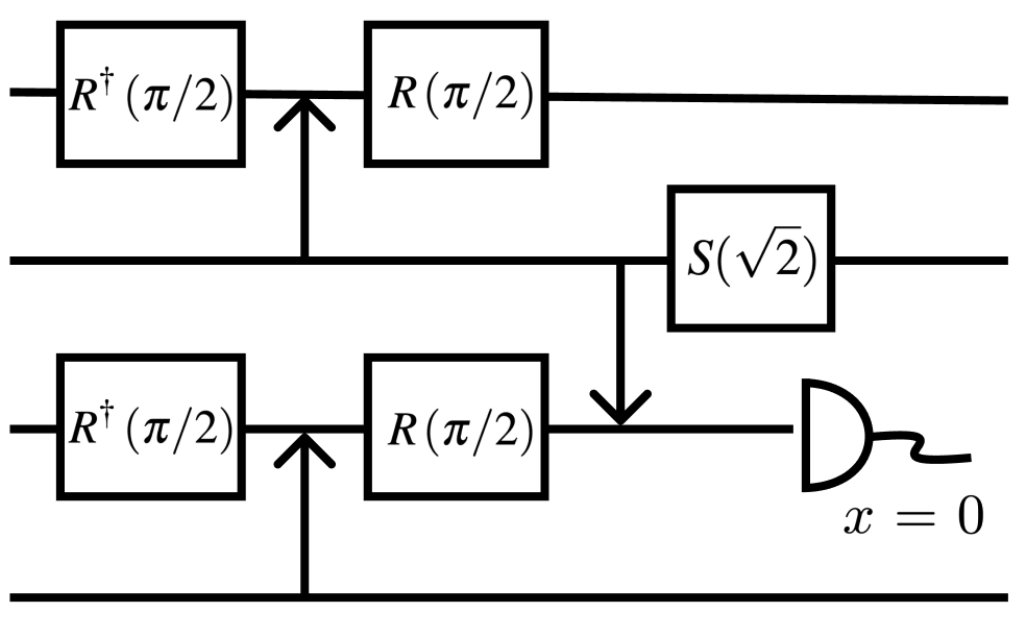}
    \caption{The circuit used to create a three-mode linear cluster state, beginning with four single-mode approximate GKP states. Inline squeezing in the second mode is included for clarity, but in practice it can be eliminated, as long as the squeezing in this mode is accounted for in the postprocessing of measurement results (see Ref. \cite{macronization}).}
    \label{fig:4circuit}
\end{figure}

\begin{figure}[h]
    \centering
     \subfloat[]{
        \includegraphics[width=0.49\textwidth]{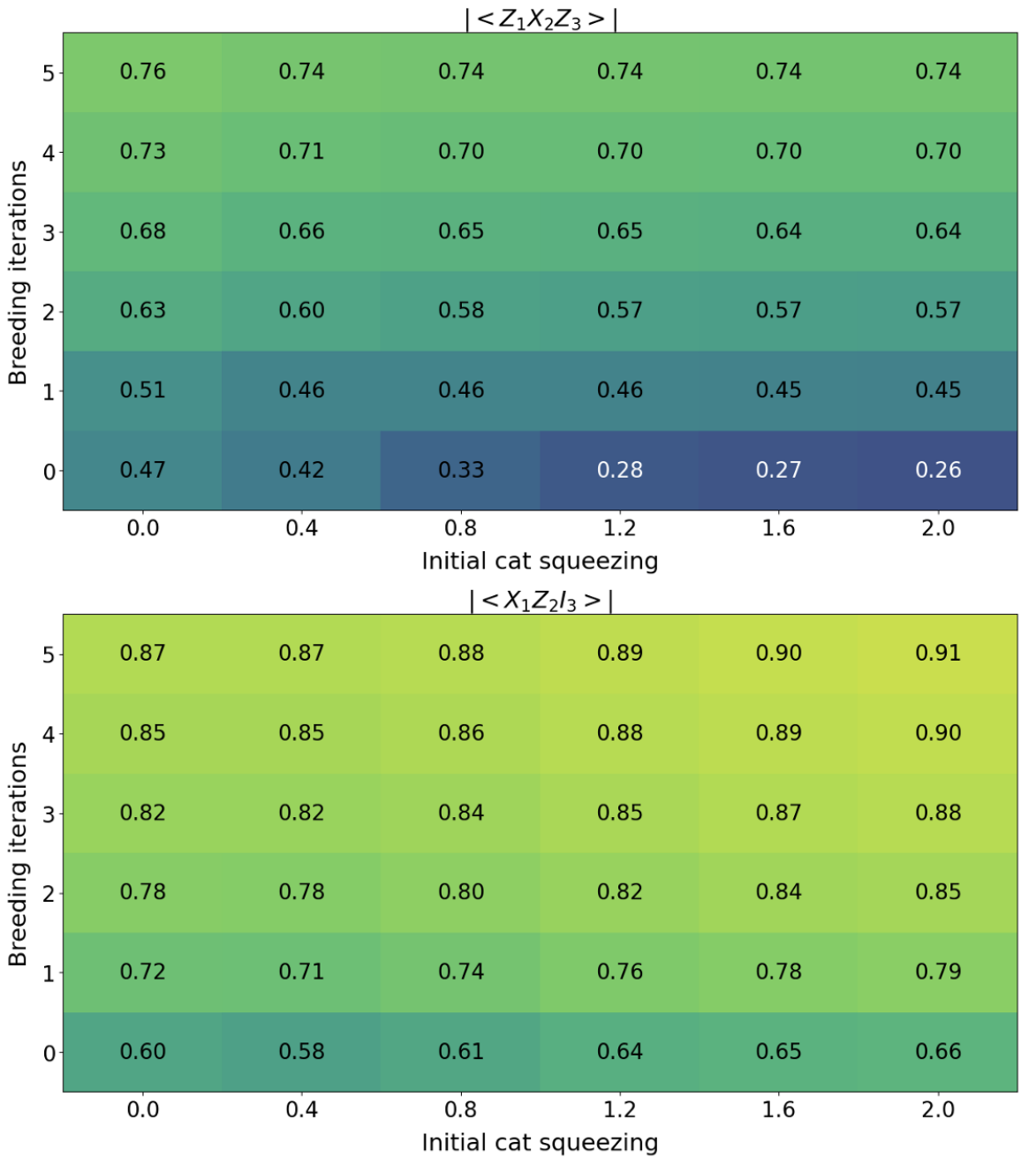}
    }

    \vspace{-0.3cm}
    \subfloat[]{
        \includegraphics[width=0.49\textwidth]{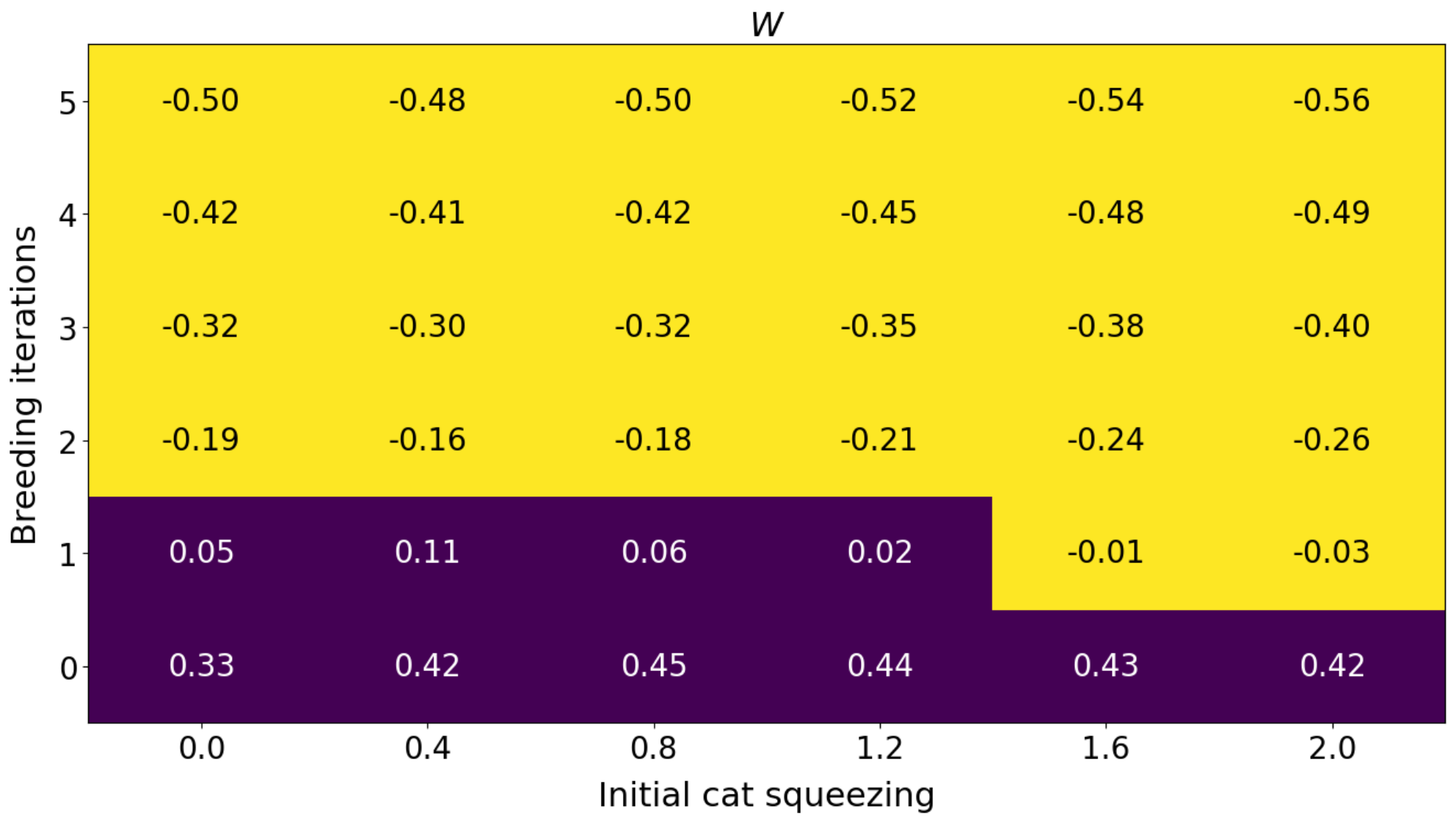}
    }
    \caption{(a) The stabilizer EVs for the three-mode linear cluster state (omitting $|\langle \hat{I}_1\hat{Z}_2\hat{X}_3 \rangle| = |\langle \hat{X}_1\hat{Z}_2\hat{I}_3 \rangle |$), and (b) the expectation value of an entanglement witness derived from these stabilizers \cite{Toth_PhysRevA.72.022340} .}
    \label{fig:3mode}
\end{figure}

\begin{figure}[h]
    \centering
    \includegraphics[width=\linewidth]{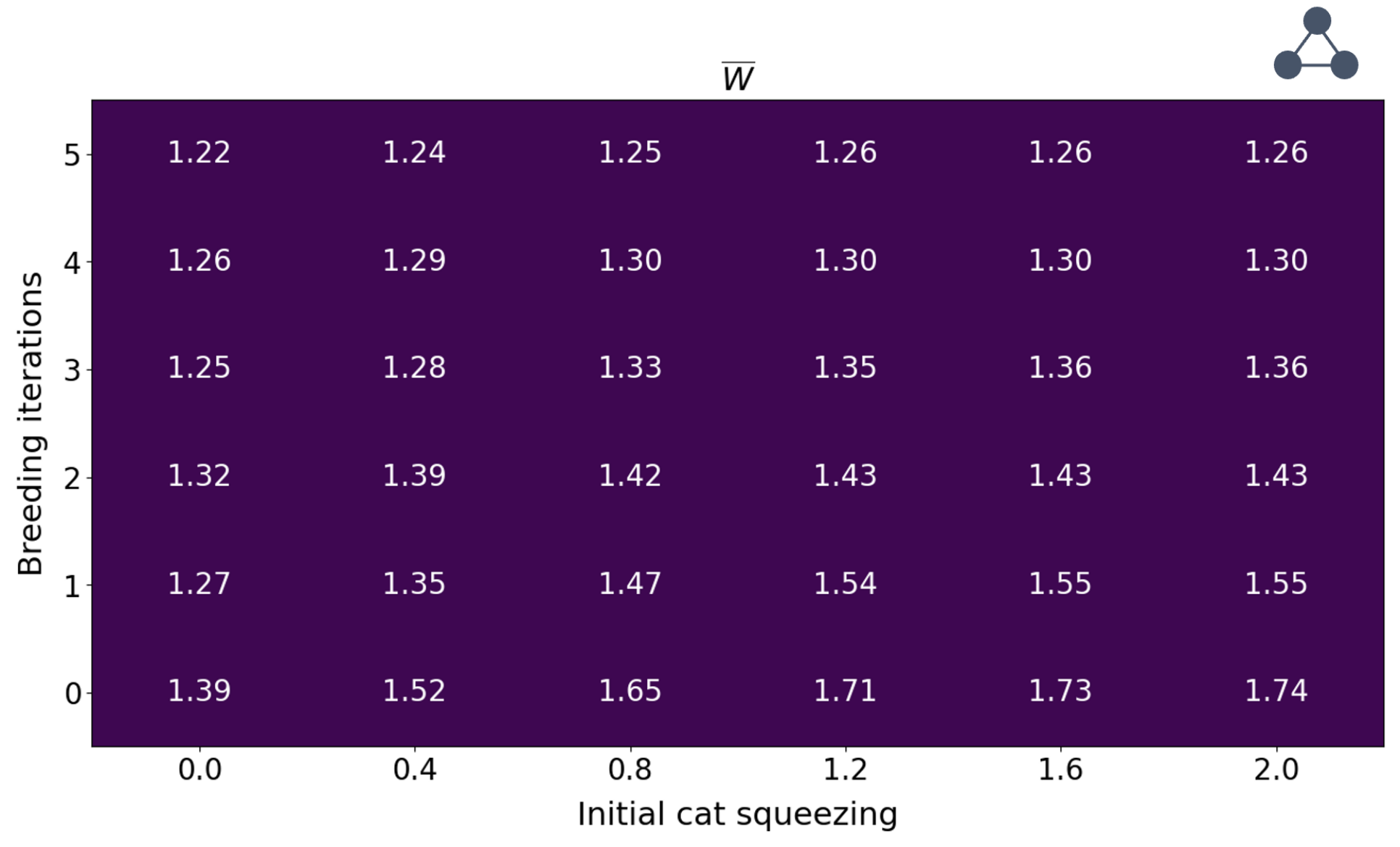}
    \caption{EVs of the witness $\hat{\overline{W}}$ (designed to witness the tripartite graph state sketched in the top right), evaluated the linear GKP cluster state generated through the circuit in Fig. \ref{fig:4circuit}. The positive EVs indicate that as expected, the distinct type of tripartite entanglement witnessed by $\hat{\overline{W}}$ is exhibited by the linear cluster state, even for parameters with which ``linear'' tripartite entanglement is witnessed in Fig. \ref{fig:3mode}b.}
    \label{fig:WGHZ}
\end{figure}

We can apply our implementation of Eq. \eqref{eq:S_result} to address the generation of more complex cluster states, requiring the use of unitary linear components and homodyne measurement. We first address the generation of a linear three-mode cluster state, through the circuit sketched in Fig. \ref{fig:4circuit}.
The stabilizer EVs for the linear cluster state are plotted in Fig. \ref{fig:3mode}(a).
The stabilizer EVs (see Fig. \ref{fig:3mode}(a)) indicate the increasing quality of the state {as the number of breeding iterations is increased. Interestingly, the squeezing of the cat states has relatively little -- and sometimes negative -- impact on the stabilizer EVs}. In Fig. \ref{fig:3mode}(b) we plot the EVs of a witness operator constructed from the cluster state's stabilizers, as suggested by T\'oth and G\"uhne \cite{Toth_PhysRevA.72.022340}: We use
\begin{align}
    \hat{\mathcal{W}} = 
    2\left(\mathbb{1}\right) - \hat{I}_1 \hat{Z}_2 \hat{X}_3 -  \hat{X}_1 \hat{Z}_2\hat{I}_3 -  \hat{Z}_1 \hat{X}_2 \hat{Z}_3.
\end{align}
A negative EV of the witness operator certifies genuine multipartite entanglement; interestingly, this threshold is crossed even with few rounds of breeding and low cat state squeezing. To confirm the significance of this witness EV's negativity, we compute the EV for the witness for a different tripartite cluster state
\begin{align}
    \hat{\overline{\mathcal{W}}} = 
    2\left(\mathbb{1}\right) -  \hat{Z}_1 \hat{Z}_2 \hat{X}_3 -  \hat{X}_1 \hat{Z}_2 \hat{Z}_3 -  \hat{Z}_1 \hat{X}_2 \hat{Z}_3,
\end{align}
which is constructed to ``witness'' the cluster state sketched in Fig. \ref{fig:WGHZ}. If the negative EVs of the witness in Fig. \ref{fig:3mode}(b) carry any significance, the EVs of a different witness should remain positive to indicate that the linear GKP cluster state does not exhibit the ``wrong" type of tripartite entanglement. This is confirmed by the results plotted in {Fig. \ref{fig:WGHZ}}. 
An interesting direction for follow-up work will be to explore whether these witness operators are a relevant figure of merit for these types of resources, and what can be inferred about the usefulness of the approximate GKP cluster state based on the witness EV. 

\section{Conclusion}

We have addressed the problem of studying realistic GKP cluster states, demonstrating an approach for simulating these states by using Gaussian expansions in phase space \cite{Eli_gaussians}, and referring to stabilizer EVs as figures of merit. We apply our approach to the generation of single-mode GKP states through ``breeding" of Schrodinger cat states, and to the generation of GKP cluster states by entangling these single-mode inputs through linear unitary components and homodyne measurement. {We demonstrate that our method is suitable for studying a broader parameter space (and higher number of modes) compared to a typical Fock-basis representation. We explore the effect of state preparation parameters on the quality of the final cluster state, and we compare our exact computations to predictions made using the GRN model.}

{Our comparisons to the GRN model open up a few questions for future work. We find deviations between the two models when homodyne measurement is introduced. These discrepancies may indicate an opportunity to improve the decoding of information encoded in GKP cluster states---for example, by adjusting the binning of the homodyne measurement outcome—using the more detailed picture of the states given by exact simulation. At the same time, the discrepancies between the GRN model and exact calculations are much smaller when one averages over measurement outcomes. More work is needed to understand what these small discrepancies mean on an operational level, and how they scale with the complexity of the cluster state.}

{It will also be interesting to assess in more detail the usefulness of stabilizers---and witness operators constructed from them---in characterizing GKP cluster states; for example, by comparing stabilizer EVs to logical error rates for realistic states using exact simulations. Due to its efficiency, this method is also a possible tool for numerically optimizing state generation protocols, perhaps with reference to figures of merit based on stabilizers, rather than fidelity to a particular state \cite{valerio_PhysRevA.109.023717}. {Finally, we envision a number of generalizations to the formalism described here, so that it can be applied to a more general class of state preparation protocols \cite{su_PhysRevA.100.052301, quesada_PhysRevA.100.022341}, and to other physical platforms \cite{travaglione_PhysRevA.66.052322,Flhmann2019,CampagneIbarcq2020,royer_PhysRevLett.125.260509}. For example, this will include describing photon-number-resolving measurement \cite{stellar_PhysRevLett.124.063605,quesada_PhysRevA.100.022341,PhysRevA.101.032315}, and addressing different input states like imperfect cat states \cite{PhysRevLett.88.250401, Ourjoumtsev2006}, or other relevant approximations to GKP states.}

\begin{acknowledgments}
    The NRC headquarters is located on the traditional unceded territory of the Algonquin Anishinaabe and Mohawk people. {The authors thank Noah Lupu-Gladstein, Anaelle Hertz, and Aaron Goldberg for helpful discussions.} M.B. and K.H. acknowledge support from the Quantum Research and Development Initiative, led by the National Research Council Canada, under the National Quantum Strategy. K.H. acknowledges funding from the NSERC Discovery Grant and Alliance programs.
\end{acknowledgments}
\bibliography{apssamp}

\onecolumngrid
\appendix
\pagebreak
\section{Review of some useful results in phase-space formalism}
\subsection{Wigner function of an operator under (Gaussian) unitary evolution}
\label{appendix:Gaussian_unitary}

We first review the derivation of Eq. \ref{eq:W_beforeandafter}, which relates the Wigner functions of a state before ($\hat{\rho}$) and after ($\hat{\overline{\rho}} = {\hat{U}} \hat{\rho} {\hat{U}^{\dagger}}$) evolution by a Gaussian unitary. The characteristic function of $\hat{\overline{\rho}}$ is
\begin{align}
    \chi_{\overline{\rho}}(\boldsymbol{r'}) &= \text{Tr}\{ \hat{D}(\boldsymbol{r}') \hat{\overline{\rho}} \}\\
    &= \text{Tr}\{ \hat{D}(\boldsymbol{r}') {\hat{U}}\hat{\rho}{\hat{U}}^{\dagger} \}\\
    &= \text{Tr}\{ {\hat{U}}^{\dagger} \hat{D}(\boldsymbol{r}') {\hat{U}}\hat{\rho} \}.
\end{align}
We have
\begin{align}
    \hat{D}(\boldsymbol{r}') &= e^{-i (\boldsymbol{r}')^T \boldsymbol{\Omega} \hat{\boldsymbol{r}}},
\end{align}
and
\begin{align}
    {\hat{U}}^{\dagger} \hat{D}(\boldsymbol{r}') {\hat{U}} &=  \text{exp} \left(-i (\boldsymbol{r}')^T \boldsymbol{\Omega} \hat{\boldsymbol{r}}'\right),
\end{align}
with $\hat{\boldsymbol{r}}'$ as defined in Eq. \eqref{eq:r_tilde}. Using Eq. \eqref{eq:pq_trans} (which holds for any Gaussian unitary), we have
\begin{align}
    {\hat{U}}^{\dagger}\hat{D}(\boldsymbol{r}') {\hat{U}} &= \text{exp} \left(-i (\boldsymbol{r}')^T \boldsymbol{\Omega} \boldsymbol{A}\hat{\boldsymbol{r}}\right)\\
    &= \text{exp} \left(-i (\boldsymbol{r}')^T \boldsymbol{\Omega} \boldsymbol{A}\boldsymbol{\Omega}^{-1}  \boldsymbol{\Omega} \hat{\boldsymbol{r}}\right)\\
    &= \text{exp} \left(-i (\boldsymbol{r}'')^T  \boldsymbol{\Omega} \hat{\boldsymbol{r}}\right) = \hat{D}(\boldsymbol{r}''),
\end{align}
where 
\begin{align}
    (\boldsymbol{r}'')^T = (\boldsymbol{r}')^T \boldsymbol{\Omega} \boldsymbol{A}\boldsymbol{\Omega}^{-1}, \\
    \boldsymbol{r}'' = \boldsymbol{\Omega}^{-1} \boldsymbol{A}^T\boldsymbol{\Omega} \boldsymbol{r}' \label{eq:r''}.
\end{align}
This gives
\begin{align}
    \chi_{\overline{\rho}}(\boldsymbol{r'}) &= \text{Tr}\{ \hat{D}(\boldsymbol{r}')\hat{\overline{\rho}} \} \\
    &=\text{Tr}\{ \hat{D}(\boldsymbol{r}'') \hat{\rho} \}\\ &= \chi_{{\rho}}(\boldsymbol{r''}).
\end{align}

The Wigner function of $\hat{\overline{\rho}}$ is 
\begin{align}
    W_{\overline{\rho}}(\boldsymbol{r}) &= \frac{1}{(2\pi)^{2N}} \int d\boldsymbol{r}' e^{-i (\boldsymbol{r})^T \boldsymbol{\Omega} \boldsymbol{r}'} \chi_{\overline{\rho}}(\boldsymbol{r}')\\
    &= \frac{1}{(2\pi)^{2N}} \int d\boldsymbol{r}'' e^{-i (\boldsymbol{r})^T \boldsymbol{\Omega} \left( \boldsymbol{\Omega}^{-1} \boldsymbol{A}^T \boldsymbol{\Omega} \right)^{-1} \boldsymbol{r}''} \chi_{{\rho}}(\boldsymbol{r''}).
\end{align}
In the second line we have changed variables using Eq. \eqref{eq:r''}, and recognizing that the Jacobian is unity. We insert a factor of $\mathbf{1} = \boldsymbol{\Omega}^{-1} \boldsymbol{\Omega}$ in the argument of the exponential, and we define
\begin{align}
    \boldsymbol{R}^T &= \boldsymbol{r}^T \boldsymbol{\Omega} \left( \boldsymbol{\Omega}^{-1} \boldsymbol{A}^T \boldsymbol{\Omega} \right)^{-1} \boldsymbol{\Omega}^{-1} \\
    &= \boldsymbol{r}^T \boldsymbol{\Omega}  \boldsymbol{\Omega}^{-1} (\boldsymbol{A}^T)^{-1} \boldsymbol{\Omega} \boldsymbol{\Omega}^{-1} \\
    &= \boldsymbol{r}^T (\boldsymbol{A}^{-1})^T.
\end{align}
Then we have 
\begin{align}
    W_{\overline{\rho}}(\boldsymbol{r}) 
    &= \frac{1}{(2\pi)^{2N}} \int d\boldsymbol{r}'' e^{-i \boldsymbol{R}^T \boldsymbol{\Omega} \boldsymbol{r}''} \chi_{{\rho}}(\boldsymbol{r''}),\\
    &= W_{\rho}(\boldsymbol{R}).
\end{align}

\subsection{Wigner representation of our homodyne measurement operator}
\label{appendix:homodyne}

We define 
\begin{align}
    \hat{K}_{\tilde{\eta}} &= \int d\eta \Theta_{\tilde{\eta}}(\eta) \ketbra{\eta}{\eta},\\
    \hat{F}_{\tilde{\eta}} &= \hat{K}^{\dagger}_{\tilde{\eta}} \hat{K}_{\tilde{\eta}}\\ & = \int d\eta \int d\eta' \Theta^*_{\tilde{\eta}}(\eta) \Theta_{\tilde{\eta}}(\eta')  \ketbra{\eta}{\eta} \eta'\rangle \langle \eta'|\\
    &= \int d\eta |\Theta_{\tilde{\eta}}(\eta)|^2  \ketbra{\eta}{\eta}.
\end{align}
due to the linearity of the Wigner transform, we have 
\begin{align}
    W_{F_{\tilde{\eta}}}(\boldsymbol{r}) = \int d\eta |\Theta_{\tilde{\eta}}(\eta)|^2 W_{\ketbra{\eta}{\eta}}(\boldsymbol{r}),
\end{align}
where $W_{\ketbra{\eta}{\eta}}(\boldsymbol{r})$ is the Wigner function of the generalized quadrature eigenstate associated with 
\begin{align}
    \hat{\eta}_{\theta} = \hat{p} \text{cos}(\theta) - \hat{x}\text{sin}(\theta),
\end{align}
which can be written in terms of the usual momentum eigenstate as
\begin{align}
\ket{\eta} = \hat{R}(\theta) \ket{P},\\
\hat{R}(\theta) = \text{exp}\left( -i\theta a^{\dagger}a \right), \label{eq:x_intermsof_p}
\end{align}

With Eq. \eqref{eq:x_intermsof_p} and Eq. \eqref{eq:W_beforeandafter} we have 
\begin{align}
    W_{\ketbra{\eta}{\eta}}(\boldsymbol{r}) = W_{\ketbra{P}{P}}(\boldsymbol{A}^{T}\boldsymbol{r})
\end{align}
where 
\begin{align}
    \boldsymbol{A} = \begin{bmatrix}
        \text{cos} (\theta) & - \text{sin} (\theta) \\ \text{sin}(\theta) & \text{cos} (\theta) \label{eq:A_rot}
    \end{bmatrix}.
\end{align}
It is easy to show that 
\begin{align}
    W_{\ketbra{P}{P}}(\boldsymbol{r}) = \frac{1}{2\pi} \delta(p-P),
\end{align}
where $p$ is a quadrature variable, and $P$ is the corresponding eigenvalue. Inverting Eq. \eqref{eq:A_rot}, we get
\begin{align}
    W_{\ketbra{\eta}{\eta}}(\boldsymbol{r}) = \frac{1}{2\pi} \delta(-x\text{sin}(\theta) + p\text{cos}(\theta)-\eta),
\end{align}
where we have used the fact that the eigenvalues $\eta$ and $P$ are equal. One can also work in terms of rotated phase space coordinates $\{\eta_{\theta}, \eta^{\perp}_{\theta}\}$ to write
\begin{align}
    W_{\ketbra{\eta}{\eta}}(\boldsymbol{r}) = \frac{1}{2\pi} \delta(\eta_{\theta}-\eta).
\end{align}
Using this notation, we have
\begin{align}
        W_{F_{\tilde{\eta}}}(\boldsymbol{r}) = \frac{1}{2\pi} \int d\eta |\Theta_{\tilde{\eta}}(\eta)|^2 \delta(\eta_{\theta}-\eta)
\end{align}
as written in Section \ref{section:homodyne}.

\subsection{Accounting for displacements in input states}\label{appendix:displace_homo}
The breeding protocol described in Section \ref{section:phasespace} produces the sensor state defined in Section \ref{section:entangling} for even breeding iterations, and a displaced sensor state for odd breeding rounds. The relative displacement needs to be taken into account when comparing the results of a stitching protocol for a particular homodyne outcome.

Consider a state $\rho$ characterized by quadrature operators $\hat{\boldsymbol{r}}$. Evolution through a Gaussian unitary can be understood as taking 
\begin{align}
    \hat{\boldsymbol{r}} \rightarrow \boldsymbol{A}\hat{\boldsymbol{r}},
\end{align}
with $\boldsymbol{A}$ being a symplectic matrix defined in Eq. \eqref{eq:pq_trans}. Now we consider the displaced state evolving through the same unitary circuit. The displacement takes 
\begin{align}
    \hat{\boldsymbol{r}} \rightarrow \hat{\boldsymbol{r}} + \overline{\boldsymbol{r}},
\end{align}
and the unitary circuit does 
\begin{align}
    \hat{\boldsymbol{r}} + \overline{\boldsymbol{r}} \rightarrow \boldsymbol{A} \hat{\boldsymbol{r}} + \boldsymbol{A} \overline{\boldsymbol{r}}.
\end{align}
The output in this case is displaced by $\boldsymbol{A} \overline{\boldsymbol{r}}$ relative to the scenario with undisplaced initial states, which can be accounted for by adjusting homodyne measurement outcomes accordingly. For example, the GKP state generated by odd $\mathcal{M}$ iterations of breeding is an approximate sensor state, even $\mathcal{M}$ produces an approximation of a sensor state displaced by $\overline{\boldsymbol{r}} = (\sqrt{\pi/2},0)^T$ (see Fig. \ref{fig:wigner}). The dumbbell stitching circuit sketched in Fig. \ref{fig:dumbbell_circuit} is characterized by 
\begin{align}
    \boldsymbol{A} = \frac{1}{\sqrt{2}} \begin{bmatrix}
        1 & 0 & 0 & -1\\ 
        0 & 1 & 1 & 0 \\
        0 & -1 & 1 & 0\\
        1 & 0 & 0 & 1
    \end{bmatrix}.
\end{align}
We have 
\begin{align}
\boldsymbol{A} \overline{\boldsymbol{r}} =
\frac{1}{\sqrt{2}} \begin{bmatrix}
        1 & 0 & 0 & -1\\ 
        0 & 1 & 1 & 0 \\
        0 & -1 & 1 & 0\\
        1 & 0 & 0 & 1
    \end{bmatrix} \begin{bmatrix}
        \sqrt{\pi/2}\\ 
        0 \\
        \sqrt{\pi/2}\\
        0
    \end{bmatrix} = \frac{\sqrt{\pi}}{2}\begin{bmatrix}
        1\\ 
        1\\
        1\\
        1
    \end{bmatrix},
\end{align}
hence our choice of $p=0$ and $p=\sqrt{\pi}/{2}$ as the homodyne measurement outcomes in Fig. \ref{fig:DB_homo} (for odd and even $\mathcal{M}$, respectively).
\newpage

\subsection{Modeling loss}
\label{appendix:loss}

We consider a scenario in which all the modes have the same transmissivity, such that the loss in each mode can be ``aggregated" and treated at one single point in the circuit, as shown in Fig. \ref{fig:loss}a \cite{Oszmaniec2018}; more general scenarios could be treated by applying loss at multiple points in the circuit. We model each ``instance'' of loss as a coupling of the physical channel to a fictitious loss channel, through a beamsplitter for which the angle $\theta_i$ is chosen to produce the correct transmission through the physical channel (see Fig. \ref{fig:loss}b.) 

\begin{figure}[h]
    \centering
     \subfloat[]{
        \includegraphics[width=0.45\textwidth]{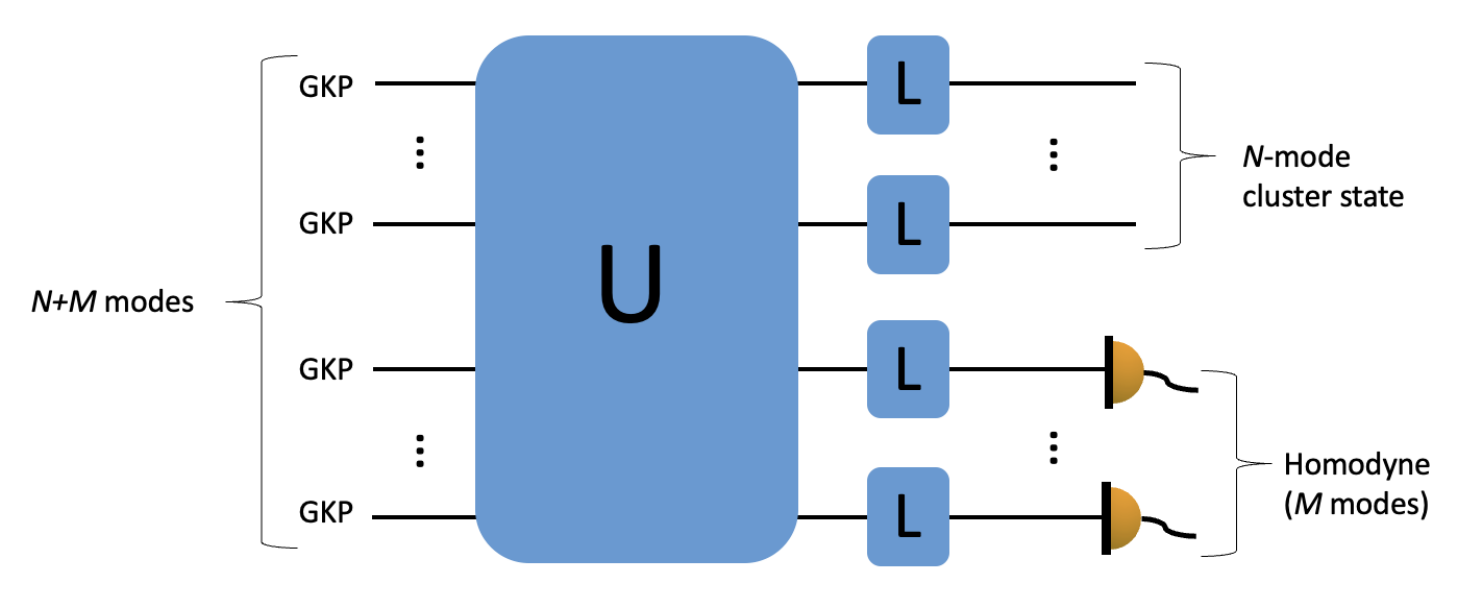}
    }
    \subfloat[]{
        \includegraphics[width=0.45\textwidth]{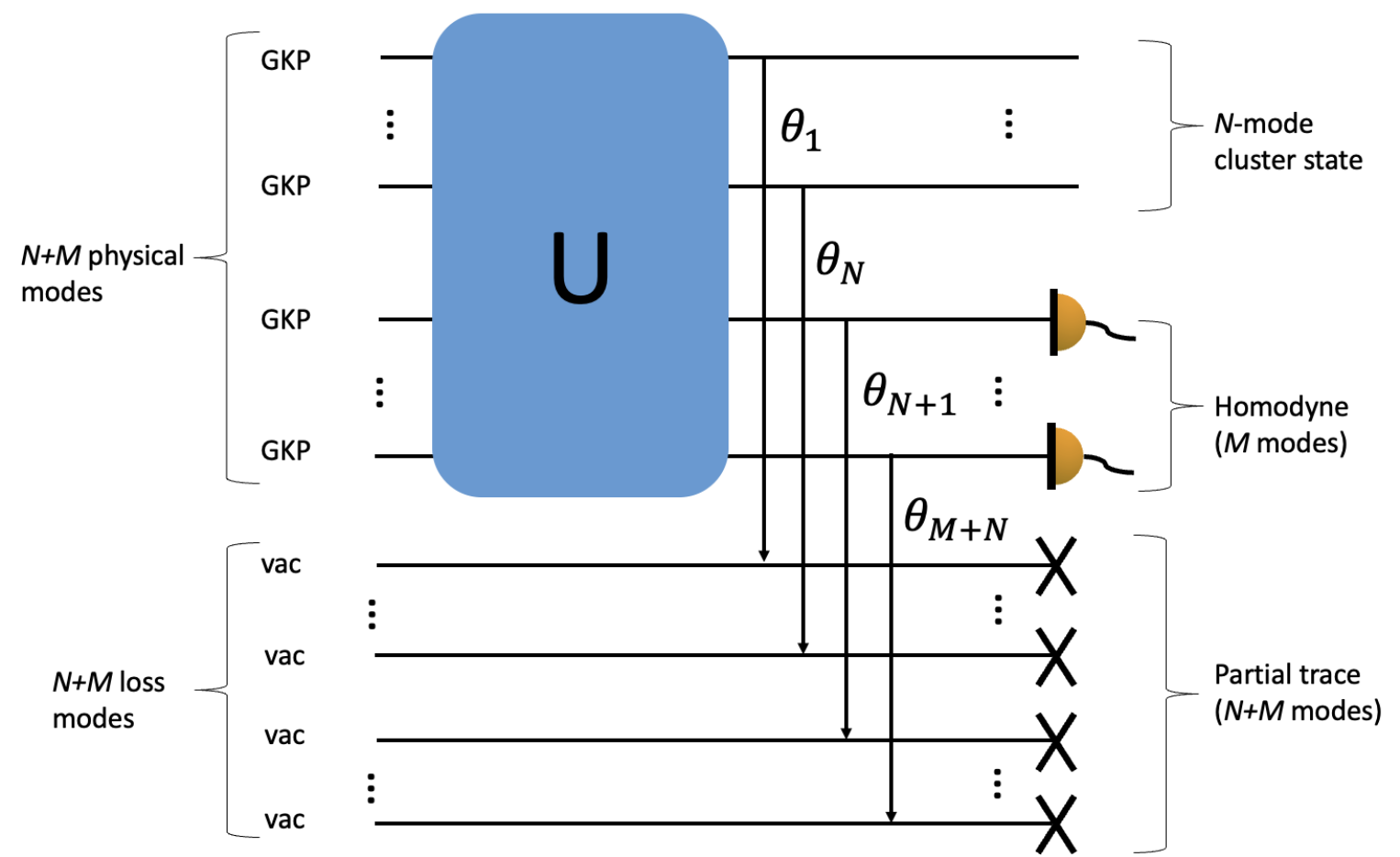}
    }
    \caption{Sketch of the treatment of loss in physical modes (panel (a)) as a beamsplitter coupling physical modes to ``loss modes" (panel (b)).}
    \label{fig:loss}
\end{figure}

Here we show that applying loss to the state $\hat{\overline{\rho}}$ characterized by 
\begin{align}
    W_{\overline{\rho}}(\boldsymbol{r}) = \sum_m c_m \overline{G}_m(\boldsymbol{r})
\end{align}can be described by making the substitution 
\begin{align}
    \boldsymbol{\overline{\mu}} &\rightarrow \boldsymbol{T} \overline{\boldsymbol{\mu}} \boldsymbol{T}^T \\
    \boldsymbol{\overline{\gamma}} &\rightarrow \boldsymbol{T} \overline{\boldsymbol{\gamma}} \boldsymbol{T}^T + \frac{1}{2}\boldsymbol{R}\boldsymbol{R}^T,
\end{align}
with 
\begin{align}
    \boldsymbol{T} &= \bigoplus_{i=1}^{N} \cos \theta_i \mathbb{1}, \label{eq:C_def}\\
    \boldsymbol{R} &= \bigoplus_{i=1}^{N} \sin \theta_i \mathbb{1}, \label{eq:S_def}.
\end{align}

\subsubsection*{Details}

Using the notation of Section \ref{section:entangling}, the state following evolution via $\hat{U}$ is represented by 
\begin{align}
    W_{\overline{\rho}} = \sum_m c_m \overline{G}_m(\boldsymbol{r}),
\end{align}
with $\overline{G}_m(\boldsymbol{r})$ being a $2(N+M)$-variable Gaussian with 
\begin{align}
    \overline{\boldsymbol{\mu}} &= \boldsymbol{A} \boldsymbol{\mu}\\
    \overline{\boldsymbol{\gamma}} &= \boldsymbol{A} \boldsymbol{\gamma} \boldsymbol{A}^T.
\end{align}

We define $\hat{\tau}$ to be the composite density operator describing the physical modes and the loss modes. Then we have (c.f. Eqs. \eqref{eq:W_tprod}--\eqref{eq:mu_sep}, and Eqs. \eqref{eq:W_beforeandafter}--\eqref{eq:gammabar})
\begin{align}
    W_{\tau}(\boldsymbol{r}) &= W_{\overline{\rho}}(\boldsymbol{r}_1) W_{\text{vac}}(\boldsymbol{r}_2)\\
    &= \sum_m c_m G'_m(\boldsymbol{r}),
\end{align}
where $W_{\text{vac}}$ is the (Gaussian) Wigner function for vacuum in the $N+M$ loss modes, with
\begin{align}
    \boldsymbol{\mu}_{\text{vac}} &= \bigoplus_{i=1}^{N} \boldsymbol{0}_{2 \times 1}, \label{eq:mu_vac}\\
    \boldsymbol{\gamma}_{\text{vac}} &= \bigoplus_{i=1}^{N} \frac{1}{2}\mathbb{1}\label{eq:gamma_vac}
\end{align}
being its mean and covariance matrices. Then $G'_m(\boldsymbol{r})$ is a $4(M+N)$--variable Gaussian with 
\begin{align}
    \boldsymbol{\mu}' &= \overline{\boldsymbol{\mu}}\oplus\boldsymbol{\mu}_{\text{vac}}\\
    \boldsymbol{\gamma}' &= \overline{\boldsymbol{\gamma}}\oplus\boldsymbol{\gamma}_{\text{vac}}.
\end{align}
We point out that the the number of terms in the Gaussian expansion does not grow with the inclusion of the loss modes, because the inputs to these are Gaussian states; namely, vacuum or thermal states. 

Applying the beamsplitters that are used to implement loss on the physical channels yields
\begin{align}
    W_{\overline{\tau}}(\boldsymbol{r})
    &= \sum_m c_m \overline{G}'_m(\boldsymbol{r}), \\
    \boldsymbol{\overline{\mu}}' &= \boldsymbol{B} \left( \overline{\boldsymbol{\mu}}\oplus\boldsymbol{\mu}_{\text{vac}}\right) \label{eq:mu_prime}\\
    \boldsymbol{\overline{\gamma}}' &= \boldsymbol{B} \left(\overline{\boldsymbol{\gamma}}\oplus\boldsymbol{\gamma}_{\text{vac}}\right) \boldsymbol{B}^T. \label{eq:gamma_prime}
\end{align}

Here $\boldsymbol{B}$ denotes the $4(N+M) \times 4(N+M)$ symplectic matrix corresponding to the beamsplitters in Fig. \ref{fig:loss}b.
A single beamsplitter coupling modes $i$ and $j$ has the symplectic matrix \cite{Eli_gaussians}
\begin{align}
    \boldsymbol{B}_{ij} &= \begin{bmatrix}
        \cos\theta \mathbb{1} & -\sin\theta \mathbb{1}\\
        \sin\theta\mathbb{1} & \cos\theta\mathbb{1}
    \end{bmatrix},
\end{align}
with our usual ordering for the quadrature variables $\boldsymbol{r} = (x_i, p_i, x_j, p_j)^T$. Then the symplectic matrix for our scenario with $N+M$ beamsplitters -- in which each beamsplitter couples a physical mode $(i)$ to a loss mode $(i+N+M)$ -- is
\begin{align}
    \boldsymbol{B} = \begin{bmatrix}
        \boldsymbol{T} & -\boldsymbol{R}\\
        \boldsymbol{R} & \boldsymbol{T}
    \end{bmatrix}, \label{eq:Bdef}
\end{align} 
with $\boldsymbol{T}$ and $\boldsymbol{R}$ defined in Eqs. \eqref{eq:C_def} and \eqref{eq:S_def}.

Tracing over the loss modes in $\hat{\overline{\tau}}$ yields the state $\hat{\tilde{\rho}}$ of the lossy physical modes; we have
\begin{align}
    W_{\tilde{\rho}}(\boldsymbol{r_1}) &= \int d\boldsymbol{r}_2 W_{\overline{\tau}}(\boldsymbol{r}), 
\end{align}
where $\boldsymbol{r}_{1(2)}$ denotes the set of $2(N+M)$ quadrature variables associated with the physical (loss) modes. Integrating over $\boldsymbol{r}_2$ just amounts to a Gaussian marginalization \cite{matrix_cookbook}: 
\begin{align}
    \int d\boldsymbol{r}_2 \overline{G}'_m(\boldsymbol{r}) &= \tilde{G}_m(\boldsymbol{r}), \\
    \tilde{\boldsymbol{\mu}} &= \boldsymbol{P} \boldsymbol{\overline{\mu}}' \label{eq:mu_tilde}\\
    \tilde{\boldsymbol{\gamma}} &= \boldsymbol{P} \boldsymbol{\overline{\gamma}}' \boldsymbol{P}^T, \label{eq:gamma_tilde}
\end{align}
where 
\begin{align}
    \boldsymbol{P} = \begin{bmatrix}
        \boldsymbol{\mathbb{1}}_{2(N+M)\times 2(N+M)} & \boldsymbol{0}_{2(N+M)\times 2(N+M)}
    \end{bmatrix}
\end{align}
is a projection matrix that selects the components of $\boldsymbol{\overline{\mu}}'$ and $\boldsymbol{\overline{\gamma}}'$ associated with $\boldsymbol{r}_1$. 

Using Eq. \eqref{eq:Bdef}, where the blocks $\boldsymbol{T}$ and $\boldsymbol{R}$ are $2(N+M)\times2(N+M)$ matrices, 
we have 
\begin{align}
    \boldsymbol{P} \boldsymbol{B} &= \begin{bmatrix}
        \boldsymbol{T} & -\boldsymbol{R}
    \end{bmatrix} \label{eq:PB}.
\end{align}
Then using Eqs. \eqref{eq:mu_prime}, \eqref{eq:gamma_prime}, and \eqref{eq:PB} in Eqs. \eqref{eq:mu_tilde} and \eqref{eq:gamma_tilde}, we have 
\begin{align}
    \tilde{\boldsymbol{\mu}} &= \boldsymbol{T} \overline{\boldsymbol{\mu}} - \boldsymbol{R} \boldsymbol{\mu}_{\text{vac}}\\
    \tilde{\boldsymbol{\gamma}} &= \boldsymbol{T} \overline{\boldsymbol{\gamma}} \boldsymbol{T}^T + \boldsymbol{R} \boldsymbol{\gamma}_{\text{vac}} \boldsymbol{R}^T
\end{align}
Using the definitions of $\boldsymbol{\mu}_{\text{vac}}$ and $\boldsymbol{\gamma}_{\text{vac}}$, we have 
\begin{align}
    \tilde{\boldsymbol{\mu}} &= \boldsymbol{T} \overline{\boldsymbol{\mu}}\\
    \tilde{\boldsymbol{\gamma}} &= \boldsymbol{T} \overline{\boldsymbol{\gamma}} \boldsymbol{T}^T + \frac{1}{2} \boldsymbol{R} \boldsymbol{R}^T.
\end{align}
This now specifies the Wigner function of $\hat{\tilde{\rho}}$ including the effects of loss, and one can describe the effects of homodyne measurement and compute its stabilizers as described in Sections \ref{section:homodyne} and \ref{sec:stabilizers}. Coupling to thermal loss modes can be dealt with in the same way, by substituting
\begin{align}
    \boldsymbol{\mu}_{\text{therm}} &= \bigoplus_{i=1}^{N} \boldsymbol{0}_{2 \times 1}, \label{eq:mu_ther}\\
    \boldsymbol{\gamma}_{\text{therm}} &= \bigoplus_{i=1}^{N} \left(\frac{1}{2} + \overline{n}_i\right)\mathbb{1},\label{eq:gamma_therm}
\end{align}
where $\boldsymbol{\mu}_{\text{vac}}$ and $\boldsymbol{\gamma}_{\text{vac}}$ appear above. 
\newpage

\section{Gaussian expansion of GKP states prepared by cat state breeding} 
\label{appendix:catbreeddetails}

In this work, we focus on GKP state generated by breeding squeezed cat states; here we lay out the derivation of the Gaussian expansion of the bred state. Our analysis is similar to those presented in Refs. \cite{catbreed_PhysRevA.97.022341} and \cite{anaelle_PhysRevA.110.012408}, but because we take a displaced initial state compared to theirs (to obtain bred states centered at the origin). We include a derivation here for clarity. We begin with ideal squeezed cat states of the form 
\begin{align}
    \left(\hat{D}\left( -\alpha/2 \right) + \hat{D}(\alpha/2) \right) \hat{S}(\xi) \ket{\text{vac}},
    \label{eq:sq_cat}
\end{align} 
where 
\begin{align}
    \hat{D}(\alpha) &= \text{exp}\left( \alpha \hat{a}^{\dagger} - \alpha^* \hat{a} \right)\\
    &= \text{exp}\left( \sqrt{2}i \text{Im}\{\alpha\} \hat{q} -\sqrt{2}i \text{Re}\{\alpha\} \hat{p} \right),
\end{align}
and we take $\text{Im} \{ \alpha \} = 0$ for simplicity. 

The input state to the breeding protocol is
\begin{align}
    \ket{\psi} = \left(\hat{D}_1\left( -\alpha/2 \right) + \hat{D}_1(\alpha/2) \right) \left(\hat{D}_2\left( -\alpha/2 \right) + \hat{D}_2(\alpha/2) \right) \hat{S}_1(\xi) \hat{S}_2(\xi) \ket{\text{vac}}, 
\end{align}
and the state following the first 50:50 beamsplitter is 
\begin{align}
    \hat{\mathcal{B}} \ket{\psi} = \hat{\mathcal{B}} \left(\hat{D}_1\left( -\alpha/2 \right) + \hat{D}_1(\alpha/2) \right) \left(\hat{D}_2\left( -\alpha/2 \right) + \hat{D}_2(\alpha/2) \right) \hat{\mathcal{B}}^{\dagger} \hat{\mathcal{B}} \hat{S}_1(\xi) \hat{S}_2(\xi) \hat{\mathcal{B}}^{\dagger} \ket{\text{vac}}, 
\end{align}
where
\begin{align}
    \hat{\mathcal{B}} &= \text{exp}\left( -i \frac{\pi}{4} \left( \hat{q}_1 \hat{p}_2 - \hat{p}_2 \hat{q}_1 \right) \right)\\
    &= \text{exp} \left( \frac{\pi}{4} \left( \hat{a}_1 \hat{a}^{\dagger}_2 - \hat{a}_2 \hat{a}^{\dagger}_1 \right) \right)
\end{align}
and we have used the fact that $\hat{\mathcal{B}} \ket{\text{vac}} = \ket{\text{vac}}$.
Recalling that
\begin{align}
    \hat{S}_1(\xi) \hat{S}_2(\xi) &= \text{exp}\left( \frac{1}{2} \left(\xi^* \hat{a}_1 ^2 - \xi \hat{a}^{\dagger 2}_1\right) \right) \text{exp}\left( \frac{1}{2} \left(\xi^* \hat{a}_2 ^2 - \xi \hat{a}^{\dagger 2}_2\right) \right) \\
    &= \text{exp}\left( \frac{1}{2} \left(\xi^* (\hat{a}_1 ^2 + \hat{a}_2 ^2) - \xi (\hat{a}^{\dagger 2}_1 + \hat{a}^{\dagger 2}_2) \right) \right),
\end{align}
and 
\begin{align}
    \hat{\mathcal{B}} \hat{a}_1 \hat{\mathcal{B}}^{\dagger
    } &= \frac{\hat{a}_1 + \hat{a}_2}{\sqrt{2}} \\
    \hat{\mathcal{B}} \hat{a}_2 \hat{\mathcal{B}}^{\dagger
    } &= \frac{\hat{a}_2 - \hat{a}_1}{\sqrt{2}},
\end{align}
we see that 
\begin{align}
    \hat{\mathcal{B}} \hat{S}_1(\xi) \hat{S}_2(\xi) \hat{\mathcal{B}}^{\dagger} = \hat{S}_1(\xi) \hat{S}_2(\xi).  
\end{align}
Similarly, we have
\begin{align}
    \hat{\mathcal{B}} \hat{D}_1(\pm \frac{\alpha}{2}) \hat{\mathcal{B}}^{\dagger} =  \hat{D}_1(\pm \frac{\alpha}{2\sqrt{2}}) \hat{D}_2(\pm \frac{\alpha}{2 \sqrt{2}})\\
    \hat{\mathcal{B}} \hat{D}_2(\pm \frac{\alpha}{2}) \hat{\mathcal{B}}^{\dagger} =  \hat{D}_1(\mp \frac{\alpha}{2\sqrt{2}}) \hat{D}_2(\pm \frac{\alpha}{2 \sqrt{2}}).
\end{align}
After the beamsplitter, the protocol involves measuring the $p$ quadrature in mode $2$ (see Fig. \ref{fig:GKP_breeding}, and recall Section \ref{section:homodyne}). Recalling that 
\begin{align}
    f(\hat{p})\ket{\tilde{p}} = f(\tilde{p}) \ket{\tilde{p}},
\end{align}
and taking the homodyne outcome to be $\tilde{p} = 0$, we have (see Ref. \cite{catbreed_PhysRevA.97.022341} for a similar analysis)
\begin{align}
    \ket{\psi}_1 = \mathcal{N} \left(\hat{D}_1\left(\frac{\alpha}{2\sqrt{2}}\right) + \hat{D}_1\left(-\frac{\alpha}{2\sqrt{2}}\right) \right)^2 \hat{S}(\xi) \ket{\text{vac}},
\end{align}
where $\ket{\psi}_1$ represents the postselected state after one iteration of breeding, and $\mathcal{N}$ denotes a normalization factor. The second round of breeding involves adding a second squeezed cat state input 
\begin{align}
    \ket{\psi''}= \left(\hat{D}_2\left(\frac{\alpha}{2\sqrt{2}}\right) + \hat{D}_2\left(-\frac{\alpha}{2\sqrt{2}}\right) \right) \hat{S}_2(\xi) \ket{\text{vac}}.
\end{align}
Interfering this with $\ket{\psi}_1$ at a 50:50 beamsplitter and again doing homodyne postselection on one output mode, this second iteration of breeding yields 
\begin{align}
    \ket{\psi}_2 = \mathcal{N} \left(\hat{D}_1\left(\frac{\alpha}{2\sqrt{2}^2}\right) + \hat{D}_1\left(-\frac{\alpha}{2\sqrt{2}^2}\right) \right)^3 \hat{S}(\xi) \ket{\text{vac}}.
\end{align}
After $\mathcal{M}$ iterations of breeding, one obtains
\begin{align}
    \ket{\psi}_{\mathcal{M}} &= \mathcal{N} \left(\hat{D}_1\left(\frac{\alpha}{2\sqrt{2}^{\mathcal{M}}}\right) + \hat{D}_1\left(-\frac{\alpha}{2\sqrt{2}^{\mathcal{M}}}\right) \right)^{\mathcal{M}+1} \hat{S}(\xi) \ket{\text{vac}} \\
    &= \mathcal{N}\sum_{k = 0}^{\mathcal{M}+1} \binom{\mathcal{M}+1}{k} \hat{D}_1\left(\frac{\alpha}{2\sqrt{2}^{\mathcal{M}}}\right)^{k} \hat{D}_1\left(-\frac{\alpha}{2\sqrt{2}^{\mathcal{M}}}\right)^{{\mathcal{M}}+1-k} \hat{S}_1 (\xi) \ket{\text{vac}}\\
    &= \mathcal{N}\sum_{k = 0}^{{\mathcal{M}}+1} \binom{{\mathcal{M}}+1}{k} \hat{D}_1\left(\frac{k\alpha}{2\sqrt{2}^{\mathcal{M}}}\right) \hat{D}_1\left(-\frac{\alpha ({\mathcal{M}}+1-k)}{2\sqrt{2}^{\mathcal{M}}}\right) \hat{S}_1 (\xi) \ket{\text{vac}}\\
    &= \mathcal{N}\sum_{k = 0}^{{\mathcal{M}}+1} \binom{{\mathcal{M}}+1}{k} \hat{D}_1\left(\frac{\alpha(2k - ({\mathcal{M}}+1))}{2\sqrt{2}^{\mathcal{M}}}\right) \hat{S}_1 (\xi) \ket{\text{vac}}.
\end{align}
We write 
\begin{align}
    \hat{\rho}_{\mathcal{M}} = \ketbra{\psi}{\psi}_{\mathcal{M}} &= |\mathcal{N}|^2  \sum_{k=0}^{{\mathcal{M}}+1} \sum_{k'=0}^{{\mathcal{M}}+1} {{\mathcal{M}}+1\choose k}{{\mathcal{M}}+1\choose k'} \hat{\mathcal{O}}_{k,k'} \label{eq:slow_breed_rho},\\
    \hat{\mathcal{O}}_{k,k'} &= \hat{D}\left(\beta_k\right)  \hat{S}(\xi) \ket{\text{vac}}\bra{\text{vac}}\hat{S}^{\dagger}(\xi
    ) \hat{D}^{\dagger}\left(\beta_{k'}\right),
\end{align}
with 
\begin{align}
    \beta_k = \frac{\alpha(2k - ({\mathcal{M}}+1))}{2\sqrt{2}^{\mathcal{M}}},
\end{align}
and the Wigner function for Eq. \eqref{eq:slow_breed_rho} is
\begin{align}
    W_{\rho}(\boldsymbol{r}) = |\mathcal{N}|^2  \sum_{k=0}^{{\mathcal{M}}+1} \sum_{k'=0}^{{\mathcal{M}}+1} {{\mathcal{M}}+1\choose k}{{\mathcal{M}}+1\choose k'} W_{k,k'}(\boldsymbol{r}),
\end{align}
where $W_{k,k'}$ is the Wigner representation of $\mathcal{O}_{k,k'}$. It can be shown that \cite{catbreed_PhysRevA.97.022341,anaelle_PhysRevA.110.012408} 
\begin{align}
    W_{k,k'}(\boldsymbol{r}) = \mathcal{A}_{k,k'} G_{k,k'}(\boldsymbol{r}), \label{eq:Wkk}
\end{align} 
with $G_{k,k'}(\boldsymbol{r})$ defined as in Eq. \eqref{eq:Gaussian}, and 
\begin{align}
    \mathcal{A}_{k,k'} &= \text{exp}\left(-\frac{1}{2} e^{2\xi}(\beta_k - \beta_{k'})^2\right)\\
    \boldsymbol{\gamma}_{k,k'} &= \frac{1}{2} \begin{bmatrix}
    e^{-2\xi} & 0 \\ 0 & e^{2\xi}
    \end{bmatrix} \equiv \boldsymbol{\gamma}\\
    \boldsymbol{\mu}_{k,k'} &= \sqrt{\frac{1}{2}} \begin{bmatrix}
    \beta_k + \beta_{k'} \\ ie^{2\xi}(\beta_{k'} - \beta_k) \end{bmatrix}.
\end{align}

\newpage

\section{Stabilizer expectation value: details of the derivation}
\label{appendix:S_details}

\subsection{Integrals}

We first address the numerator in Eq. \eqref{eq:numerator}. We have
\begin{align}
    \mathcal{I} &= \int d\boldsymbol{r} W_{\rho}(\boldsymbol{r}) W_{F_i}(\boldsymbol{r}) W_{\mathcal{O}}(\boldsymbol{r})\\
    &= \sum_{\boldsymbol{m}} c_{\boldsymbol{m}} \int d\boldsymbol{r} G_{\boldsymbol{m}}(\boldsymbol{r}) W_{F_i}(\boldsymbol{r}) W_{\mathcal{O}}(\boldsymbol{r})
\end{align}
Recognize that $\boldsymbol{r}$ refers to the quadrature variables for all the ${N}$ modes. We denote by $\boldsymbol{r}_C$ the $2({N} - {M})$  quadrature variables for the unmeasured modes. The quadrature variables that are measured are denoted by $\boldsymbol{r}_H$ (there are ${M}$ of these), and the conjugate variables to these are $\boldsymbol{r}_H^{\perp}$ (again ${M}$ of these). We now write
\begin{align}
    \mathcal{I} &= \sum_{\boldsymbol{m}} c_{\boldsymbol{m}} \int d\boldsymbol{r}_C d\boldsymbol{r}_H d\boldsymbol{r}_H^{\perp} G_{\boldsymbol{m}}(\boldsymbol{r}_C, \boldsymbol{r}_H, \boldsymbol{r}_H^{\perp}) W_{F}(\boldsymbol{r}_{H}) W_{\mathcal{O}}(\boldsymbol{r}_C).
\end{align}
We do the integral over $\boldsymbol{r}_H^{\perp}$ by using the fact that \cite{matrix_cookbook}
\begin{align}
    \int d\boldsymbol{r}_2 G(\boldsymbol{r}_1, \boldsymbol{r}_2) &= \tilde{G}(\boldsymbol{r}_1),
\end{align}
where $\tilde{G}(r_1)$ is a normalized Gaussian defined by the mean vector and covariance matrix
\begin{align}
    \tilde{\boldsymbol{\mu}} &= \boldsymbol{P} \boldsymbol{\mu}\\
    \tilde{\boldsymbol{\gamma}} &= \boldsymbol{P} \boldsymbol{\gamma} \boldsymbol{P}^{T}.\label{eq:P_def}
\end{align}
By $\boldsymbol{P}$ we denote a rectangular matrix that selects the components of $\boldsymbol{\gamma}$ and $\boldsymbol{\mu}$ that are associated with a subset of quadrature variables that we associate with $\boldsymbol{r}_1$; effectively, $\tilde{\boldsymbol{\mu}}$ and $\tilde{\boldsymbol{\gamma}}$ are obtained by dropping all the elements of ${\boldsymbol{\mu}}, {\boldsymbol{\gamma}}$ associated with $\boldsymbol{r}_2$. We have
\begin{align}
    \mathcal{I} &= \sum_{\boldsymbol{m}} c_{\boldsymbol{m}} \int d\boldsymbol{r}_C d\boldsymbol{r}_H \tilde{G}_{\boldsymbol{m}}(\boldsymbol{r}_C, \boldsymbol{r}_H) W_{F}(\boldsymbol{r}_{H}) W_{\mathcal{O}}(\boldsymbol{r}_C),
\end{align}
with $\tilde{\boldsymbol{\mu}}_m$ and $\tilde{\boldsymbol{\gamma}}_m$ defined as above. Next we use the delta functions in $W_{F}(\boldsymbol{r}_{H})$ to get
\begin{align}
    \mathcal{I} &= \frac{1}{(2\pi)^{{M}} }\sum_{\boldsymbol{m}} c_{\boldsymbol{m}} \int d\boldsymbol{r}_C  \tilde{G}_{\boldsymbol{m}}(\boldsymbol{r}_C, \tilde{\boldsymbol{R}}_H) W_{\mathcal{O}}(\boldsymbol{r}_C), \label{eq:with_tilde}
\end{align}
where $\tilde{\boldsymbol{R}}_H$ denotes the set of postselected homodyne outcomes. 

We now use the Schur decomposition to write 
\begin{align}
    \int d\boldsymbol{r}_C & G(\boldsymbol{r}_C, \tilde{\boldsymbol{R}}_H) W_{\mathcal{O}}(\boldsymbol{r}_C) = \frac{1}{\sqrt{\text{det}(2\pi \boldsymbol{\gamma})}} \text{exp}\left( -\frac{1}{2} (\tilde{\boldsymbol{R}}_H-\boldsymbol{\mu}_{H})^T \boldsymbol{\gamma
    }_{HH}^{-1} (\tilde{\boldsymbol{R}}_H-\boldsymbol{\mu}_{H})\right) \\ \times \int &d\boldsymbol{r}_C \nonumber \text{exp}\left( -\frac{1}{2} \left(\boldsymbol{r}_C-\boldsymbol{\mu}_C - \boldsymbol{\gamma}_{CH} \boldsymbol{\gamma}_{HH}^{-1} \left(\tilde{\boldsymbol{R}}_H-\boldsymbol{\mu}_H\right)\right)^T \boldsymbol{C
    }^{-1} \left(\boldsymbol{r}_C-\boldsymbol{\mu}_C - \boldsymbol{\gamma}_{CH} \boldsymbol{\gamma}_{HH}^{-1} \left(\tilde{\boldsymbol{R}}_H-\boldsymbol{\mu}_H\right)\right)\right) W_{\mathcal{O}}(\boldsymbol{r}_C), \\ &= \frac{1}{\sqrt{\text{det}(2\pi \boldsymbol{\gamma})}} \text{exp}\left( -\frac{1}{2} (\tilde{\boldsymbol{R}}_H-\boldsymbol{\mu}_{H})^T \boldsymbol{\gamma
    }_{HH}^{-1} (\tilde{\boldsymbol{R}}_H-\boldsymbol{\mu}_{H})\right) \int d\boldsymbol{r}_C \text{exp}\left( -\frac{1}{2} \left(\boldsymbol{r}_C-\boldsymbol{\xi}\right)^T \boldsymbol{C
    }^{-1} \left(\boldsymbol{r}_C-\boldsymbol{\xi}\right)\right) W_{\mathcal{O}}(\boldsymbol{r}_C),
\end{align}
where
\begin{align}
    \boldsymbol{C} = \boldsymbol{\gamma}_{CC} - \boldsymbol{\gamma}_{CH} \boldsymbol{\gamma}_{HH}^{-1} \boldsymbol{\gamma}_{HC}
\end{align}
where $\boldsymbol{C}$ is the Schur complement of the lower-right block of $\boldsymbol{\gamma}$ {\cite{matrix_cookbook}},  
and we have introduced
\begin{align}
    \boldsymbol{\xi} &= \boldsymbol{\mu}_C + \boldsymbol{\gamma}_{CH} \boldsymbol{\gamma}_{HH}^{-1} \left(\tilde{\boldsymbol{R}}_H-\boldsymbol{\mu}_H\right). 
\end{align}
When $\hat{\mathcal{O}}$ is a displacement operator we have
\begin{align}
    W_{\mathcal{O}}(\boldsymbol{r}_C) &= \frac{1}{(2\pi)^{({N}-{M})}} \text{exp}\left(i\boldsymbol{J}^T \boldsymbol{r}_C\right).
\end{align}
Now 
\begin{align}
    \nonumber \int d\boldsymbol{r}_C G(\boldsymbol{r}_C, \tilde{\boldsymbol{R}}_H) W_{\mathcal{O}}(\boldsymbol{r}_C) &= \frac{1}{(2\pi)^{({N}-{M})}} \frac{1}{\sqrt{\text{det}(2\pi \boldsymbol{\gamma})}} \text{exp}\left( -\frac{1}{2} (\tilde{\boldsymbol{R}}_H-\boldsymbol{\mu}_{H})^T \boldsymbol{\gamma
    }_{HH}^{-1} (\tilde{\boldsymbol{R}}_H-\boldsymbol{\mu}_{H})\right) \\ &\times \int d\boldsymbol{r}_C \text{exp}\left( -\frac{1}{2} \left(\boldsymbol{r}_C-\boldsymbol{\xi}\right)^T \boldsymbol{C
    }^{-1} \left(\boldsymbol{r}_C-\boldsymbol{\xi}\right)\right) \text{exp}\left(i\boldsymbol{J}^T \boldsymbol{r}_C\right).
\end{align}
We define $\boldsymbol{r}'=\boldsymbol{r}_C - \boldsymbol{\xi}$, so
\begin{align}
    \int d\boldsymbol{r}_C G(\boldsymbol{r}_C, \tilde{\boldsymbol{R}}_H) W_{\mathcal{O}}(\boldsymbol{r}_C) &= \frac{1}{(2\pi)^{({N}-{M})}} \frac{1}{\sqrt{\text{det}(2\pi \boldsymbol{\gamma})}} \text{exp}\left( -\frac{1}{2} (\tilde{\boldsymbol{R}}_H-\boldsymbol{\mu}_{H})^T \boldsymbol{\gamma
    }_{HH}^{-1} (\tilde{\boldsymbol{R}}_H-\boldsymbol{\mu}_{H})\right) \nonumber \\ &\times \text{exp}\left(i\boldsymbol{J}^T \boldsymbol{\xi}\right) \int d\boldsymbol{r}' \text{exp}\left( -\frac{1}{2} \boldsymbol{r}'^T \boldsymbol{C
    }^{-1} \boldsymbol{r}'\right) \text{exp}\left(i\boldsymbol{J}^T \boldsymbol{r}'\right).
\end{align}
Using
\begin{align}
    \int d\boldsymbol{r} \text{exp}\left( -\frac{1}{2} \boldsymbol{r}^T \boldsymbol{\gamma}^{-1} \boldsymbol{r} \right) \text{exp}\left( i \boldsymbol{J}^T \boldsymbol{r} \right) &= \sqrt{\text{det}(2\pi\boldsymbol{\gamma})} \text{exp}\left(-\frac{1}{2} \boldsymbol{J}^T \boldsymbol{\gamma} \boldsymbol{J} \right),
\end{align}
and 
\begin{align}
    \frac{\text{det}(2\pi \boldsymbol{C})}{\text{det}(2\pi \boldsymbol{\gamma})} &= \frac{1}{\text{det}(2\pi \boldsymbol{\gamma}_{HH})},
\end{align}
we have
\begin{align}
    \int d\boldsymbol{r}_C \tilde{G}(\boldsymbol{r}_C, \tilde{\boldsymbol{R}}_H) W_{\mathcal{O}}(\boldsymbol{r}_C) &= \frac{1}{(2\pi)^{({N}-{M})}} \frac{1}{\sqrt{\text{det}(2\pi \tilde{\boldsymbol{\gamma}}_{HH})}} \text{exp}\left( -\frac{1}{2} (\tilde{\boldsymbol{R}}_H-\tilde{\boldsymbol{\mu}}_{H})^T \boldsymbol{\tilde{\gamma
    }}_{HH}^{-1} (\tilde{\boldsymbol{R}}_H-\tilde{\boldsymbol{\mu}}_{H})\right) \\ &\times \text{exp}\left(i\boldsymbol{J}^T \boldsymbol{\xi}\right) \text{exp}\left(-\frac{1}{2} \boldsymbol{J}^T \boldsymbol{C} \boldsymbol{J} \right) \nonumber,
\end{align}
where we have restored the tildes from Eq. \eqref{eq:with_tilde}. 

We now introduce the $\boldsymbol{P}_H$ and $\boldsymbol{P}_C$ such that 
\begin{align}
    \tilde{\boldsymbol{\mu}}_{H} &= \boldsymbol{P}_H \tilde{\boldsymbol{\mu}} = \boldsymbol{P}_H \boldsymbol{P} \boldsymbol{\mu}\\
    \tilde{\boldsymbol{\gamma}}_{HH} &= \boldsymbol{P}_H \tilde{\boldsymbol{\gamma}} \boldsymbol{P}_H^T\\ 
    \tilde{\boldsymbol{\gamma}}_{CH} &= \boldsymbol{P}_C \tilde{\boldsymbol{\gamma}} \boldsymbol{P}_H^T, 
\end{align}
and likewise for $\tilde{\boldsymbol{\mu}}_C$ and $\tilde{\boldsymbol{\gamma}}_C$. Similar to $\boldsymbol{P}$ defined in Eq. \ref{eq:P_def}, $\boldsymbol{P}_{H}$ picks out the matrix components associated with modes that are measured; $\boldsymbol{P}_C$ picks out matrix components associated with unmeasured modes comprising the cluster state. We have
\begin{align}
    \mathcal{I} &=  \frac{1}{(2\pi)^{{N}}} \sum_{\boldsymbol{m}} c_{\boldsymbol{m}} \frac{1}{\sqrt{\text{det}(2\pi \boldsymbol{P}_H \tilde{\boldsymbol{\gamma}}_{\boldsymbol{m}}\boldsymbol{P}_H^T)}} \text{exp}\left( -\frac{1}{2} (\tilde{\boldsymbol{R}}_H-\boldsymbol{P}_H \tilde{\boldsymbol{\mu}}_{\boldsymbol{m}})^T (\boldsymbol{P}_H \tilde{\boldsymbol{\gamma}}_{\boldsymbol{m}}\boldsymbol{P}_H^T)^{-1} (\tilde{\boldsymbol{R}}_H-\boldsymbol{P}_H \tilde{\boldsymbol{\mu}}_{\boldsymbol{m}})\right) \nonumber \\ &\times \text{exp}\left(i\boldsymbol{J}^T \boldsymbol{\xi}_{\boldsymbol{m}}\right) \text{exp}\left(-\frac{1}{2} \boldsymbol{J}^T \boldsymbol{C}_{\boldsymbol{m}} \boldsymbol{J} \right). \label{eq:numerator2}
\end{align}
The denominator can be obtained by setting $\boldsymbol{J} = \boldsymbol{0}$ in Eq. \eqref{eq:numerator2}. This can be seen by recognizing that the identity operator is equivalent to the displacement operator at zero displacement.

\subsection{Cosmesis}

We now have 
\begin{align}
    \langle \hat{\mathcal{S}} \rangle &= \frac{\sum_{\boldsymbol{m}} c_{\boldsymbol{m}} g_{\boldsymbol{m}}(\tilde{\boldsymbol{R}}_H)\text{exp}\left(i\boldsymbol{J}^T \boldsymbol{\xi}_{\boldsymbol{m}}-\frac{1}{2} \boldsymbol{J}^T \boldsymbol{C}_{\boldsymbol{m}} \boldsymbol{J} \right)}{\sum_{\boldsymbol{m}} c_{\boldsymbol{m}} g_{\boldsymbol{m}}(\tilde{\boldsymbol{R}}_H)}, \label{eq:result_appendix}
\end{align}
where 
\begin{align}
    g_{\boldsymbol{m}}(\tilde{\boldsymbol{R}}_H) &= \frac{1}{\sqrt{\text{det}(2\pi {\boldsymbol{\gamma}}_{HH \boldsymbol{m}})}} \text{exp}\left( -\frac{1}{2} (\tilde{\boldsymbol{R}}_H-{\boldsymbol{\mu}}_{H 
    \boldsymbol{m}})^T ({\boldsymbol{\gamma}}_{HH \boldsymbol{m}})^{-1} (\tilde{\boldsymbol{R}}_H-{\boldsymbol{\mu}}_{H \boldsymbol{m}})\right), \\\label{eq:schur1}
    \boldsymbol{\xi} &= \boldsymbol{\mu}_C + \boldsymbol{\gamma}_{CH} \boldsymbol{\gamma}_{HH}^{-1} \left(\tilde{\boldsymbol{R}}_H-\boldsymbol{\mu}_H\right),\\
    \boldsymbol{C} &= \boldsymbol{\gamma}_{CC} - \boldsymbol{\gamma}_{CH} \boldsymbol{\gamma}_{HH}^{-1} \boldsymbol{\gamma}_{HC}. \label{eq:schur2}
\end{align}

We use Eqs. \eqref{eq:schur1} and \eqref{eq:schur2} to write 
\begin{align}
    \text{exp}\left(i\boldsymbol{J}^T \boldsymbol{\xi}_{\boldsymbol{m}}-\frac{1}{2} \boldsymbol{J}^T \boldsymbol{C}_{\boldsymbol{m}} \boldsymbol{J} \right) &= \text{exp}\left(i\boldsymbol{J}^T \boldsymbol{\mu}_C -\frac{1}{2} \boldsymbol{J}^T \boldsymbol{\gamma}_{CC} \boldsymbol{J}\right) \text{exp} \left( i \boldsymbol{J}^T \boldsymbol{\gamma}_{CH} \boldsymbol{\gamma}_{HH}^{-1}\left(\tilde{\boldsymbol{X}} - \boldsymbol{\mu}_H\right)  + \frac{1}{2} \boldsymbol{J}^T \boldsymbol{\gamma}_{CH} \boldsymbol{\gamma}_{HH}^{-1} \boldsymbol{\gamma}_{HC} \boldsymbol{J} \right).
\end{align}
Using \cite{matrix_cookbook}
\begin{align}
-\frac{1}{2} \boldsymbol{x}^T \boldsymbol{M} \boldsymbol{x} + (\boldsymbol{x}^T \boldsymbol{b})^T =  -\frac{1}{2} (\boldsymbol{x} - \boldsymbol{M}^{-1} \boldsymbol{b})^T \boldsymbol{M}   (\boldsymbol{x} - \boldsymbol{M}^{-1} \boldsymbol{b}) + \frac{1}{2} \boldsymbol{b}^T \boldsymbol{M}^{-1} \boldsymbol{b}
\end{align}
(with  $\boldsymbol{x} = i\boldsymbol{\gamma}_{HC} \boldsymbol{J}$, $\boldsymbol{b} = \boldsymbol{\gamma}_{HH}^{-1}\left( \tilde{\boldsymbol{R}}_H - \boldsymbol{\mu}_H \right), \boldsymbol{M} = \boldsymbol{\gamma}_{HH}^{-1}$)  we write
\begin{align}
     &i \boldsymbol{J}^T \boldsymbol{\gamma}_{CH} \boldsymbol{\gamma}_{HH}^{-1}\left(\tilde{\boldsymbol{R}}_H - \boldsymbol{\mu}_H\right)  + \frac{1}{2} \boldsymbol{J}^T \boldsymbol{\gamma}_{CH} \boldsymbol{\gamma}_{HH}^{-1} \boldsymbol{\gamma}_{HC} \boldsymbol{J}\\ &= - \frac{1}{2} \left( i \boldsymbol{\gamma}_{HC} \boldsymbol{J} - \boldsymbol{\gamma}_{HH} \boldsymbol{\gamma}_{HH}^{-1}\left( \tilde{\boldsymbol{R}}_H - \boldsymbol{\mu}_H \right) \right)^T \boldsymbol{\gamma}_{HH}^{-1} \left( i \boldsymbol{\gamma}_{HC} \boldsymbol{J} - \boldsymbol{\gamma}_{HH} \boldsymbol{\gamma}_{HH}^{-1}\left( \tilde{\boldsymbol{R}}_H - \boldsymbol{\mu}_H \right) \right) \nonumber\\&+ \frac{1}{2} \left( \tilde{\boldsymbol{R}}_H - \boldsymbol{\mu}_H \right)^T \left(\boldsymbol{\gamma}_{HH}^{-1}\right)^T \boldsymbol{\gamma}_{HH} \boldsymbol{\gamma}_{HH}^{-1}\left( \tilde{\boldsymbol{R}}_H - \boldsymbol{\mu}_H \right)\\
     &= - \frac{1}{2} \left( i \boldsymbol{\gamma}_{HC} \boldsymbol{J} - \left( \tilde{\boldsymbol{R}}_H - \boldsymbol{\mu}_H \right) \right)^T \boldsymbol{\gamma}_{HH}^{-1} \left( i \boldsymbol{\gamma}_{HC} \boldsymbol{J} - \left( \tilde{\boldsymbol{R}}_H - \boldsymbol{\mu}_H \right) \right)+\frac{1}{2} \left( \tilde{\boldsymbol{R}}_H - \boldsymbol{\mu}_H \right)^T \left(\boldsymbol{\gamma}_{HH}^{-1}\right) \left( \tilde{\boldsymbol{R}}_H - \boldsymbol{\mu}_H \right),
\end{align}
so 
\begin{align}
    \text{exp}\left(i\boldsymbol{J}^T \boldsymbol{\xi}_{\boldsymbol{m}}-\frac{1}{2} \boldsymbol{J}^T \boldsymbol{C}_{\boldsymbol{m}} \boldsymbol{J} \right) &= \nonumber \text{exp} \left( - \frac{1}{2} \left( \tilde{\boldsymbol{R}}_H - \boldsymbol{\mu}_H - i \boldsymbol{\gamma}_{HC} \boldsymbol{J} \right) ^T \boldsymbol{\gamma}_{HH}^{-1} \left( \tilde{\boldsymbol{R}}_H - \boldsymbol{\mu}_H - i \boldsymbol{\gamma}_{HC} \boldsymbol{J} \right)\right)\\ &\times \text{exp}\left(i\boldsymbol{J}^T \boldsymbol{\mu}_C -\frac{1}{2} \boldsymbol{J}^T \boldsymbol{\gamma}_{CC} \boldsymbol{J}\right)\text{exp}\left( \frac{1}{2} \left( \tilde{\boldsymbol{R}}_H - \boldsymbol{\mu}_H \right)^T \left(\boldsymbol{\gamma}_{HH}^{-1}\right) \left( \tilde{\boldsymbol{R}}_H - \boldsymbol{\mu}_H \right)\right). \label{eq:arg_cosmesed}
\end{align}
Putting this into Eq. \eqref{eq:result_appendix} it is clear that the final term in Eq.  \eqref{eq:arg_cosmesed} cancels the exponential term in  $g_{\boldsymbol{m}}(\tilde{\boldsymbol{R}}_H)$. Now adopting the notation in the main text ($ \tilde{\boldsymbol{\eta}} \equiv \tilde{\boldsymbol{R}}_H$), the result is 
\begin{align}
    \langle \hat{\mathcal{S}} \rangle &= \frac{\sum_{\boldsymbol{m}} c_{\boldsymbol{m}} g_{\boldsymbol{m}}(\tilde{\boldsymbol{\eta}};\boldsymbol{J})\text{exp}\left(i\boldsymbol{J}^T \boldsymbol{\mu}_C -\frac{1}{2} \boldsymbol{J}^T \boldsymbol{\gamma}_{CC} \boldsymbol{J}\right)}{\sum_{\boldsymbol{m}} c_{\boldsymbol{m}} g_{\boldsymbol{m}}(\tilde{\boldsymbol{\eta}}, \boldsymbol{0})},\\
    &= \frac{\sum_{\boldsymbol{m}} c_{\boldsymbol{m}} g_{\boldsymbol{m}}(\tilde{\boldsymbol{\eta}};\boldsymbol{J})\text{exp}\left(i\boldsymbol{J}^T \boldsymbol{P}_C \boldsymbol{A}\boldsymbol{\mu}_{\boldsymbol{m}} -\frac{1}{2} \boldsymbol{J}^T \boldsymbol{P}_C \boldsymbol{A} \boldsymbol{\gamma}_{\boldsymbol{m}} \boldsymbol{A}^T \boldsymbol{P}_C^T \boldsymbol{J}\right)}{\sum_{\boldsymbol{m}} c_{\boldsymbol{m}} g_{\boldsymbol{m}}(\tilde{\boldsymbol{\eta}}, \boldsymbol{0})},
\end{align}
where
\begin{align}
     g_{\boldsymbol{m}}(\tilde{\boldsymbol{\eta}}; \boldsymbol{J})
     &= \frac{1}{\sqrt{\text{det}(2\pi {\boldsymbol{\gamma}}_{HH})}} \text{exp} \left( - \frac{1}{2} \left( \tilde{\boldsymbol{\eta}} - \boldsymbol{\mu}_H - i \boldsymbol{\gamma}_{HC} \boldsymbol{J} \right) ^T \boldsymbol{\gamma}_{HH}^{-1} \left( \tilde{\boldsymbol{\eta}} - \boldsymbol{\mu}_H - i \boldsymbol{\gamma}_{HC} \boldsymbol{J} \right)\right), \label{eq:g_XJ_def}
\end{align}
and 
\begin{align}
    \boldsymbol{\mu}_{I} &= \boldsymbol{P}_I \boldsymbol{A} \boldsymbol{\mu}_{\boldsymbol{m}},\\
    \boldsymbol{\gamma}_{IJ} &= \boldsymbol{P}_I \boldsymbol{A} \boldsymbol{\gamma}_{\boldsymbol{m}} \boldsymbol{A}^T \boldsymbol{P}_J^T, \hspace{1cm} ({I}, {J} = \{ H,C \}).
\end{align}

{Using the results of Appendix \ref{appendix:loss}, the corresponding expressions including loss are
\begin{align}
    \langle \hat{\mathcal{S}} \rangle &= \frac{\sum_{\boldsymbol{m}} c_{\boldsymbol{m}} g_{\boldsymbol{m}}(\tilde{\boldsymbol{\eta}};\boldsymbol{J})\text{exp}\left(i\boldsymbol{J}^T \boldsymbol{P}_C \boldsymbol{T} \boldsymbol{A}\boldsymbol{\mu}_{\boldsymbol{m}} -\frac{1}{2} \boldsymbol{J}^T \boldsymbol{P}_C \boldsymbol{T} \boldsymbol{A} \boldsymbol{\gamma}_{\boldsymbol{m}} \boldsymbol{A}^T \boldsymbol{T}^T \boldsymbol{P}_C^T \boldsymbol{J} - \frac{1}{4} \boldsymbol{J}^T \boldsymbol{P}_C \boldsymbol{R} \boldsymbol{R}^T \boldsymbol{P}_C^T \boldsymbol{J} \right)}{\sum_{\boldsymbol{m}} c_{\boldsymbol{m}} g_{\boldsymbol{m}}(\tilde{\boldsymbol{\eta}}, \boldsymbol{0})},
\end{align}
where $g_{\boldsymbol{m}}(\tilde{\boldsymbol{\eta}}; \boldsymbol{J})$ is defined as in Eq. \eqref{eq:g_XJ_def}, now with
\begin{align}
    \boldsymbol{\mu}_{I} &= \boldsymbol{P}_I \boldsymbol{T}\boldsymbol{A} \boldsymbol{\mu}_{\boldsymbol{m}},\\
    \boldsymbol{\gamma}_{IJ} &= \boldsymbol{P}_I \boldsymbol{T} \boldsymbol{A} \boldsymbol{\gamma}_{\boldsymbol{m}} \boldsymbol{A}^T \boldsymbol{T}^T \boldsymbol{P}_J^T + \frac{1}{2} \boldsymbol{P}_I \boldsymbol{R} \boldsymbol{R}^T \boldsymbol{P}_J^T, \hspace{1cm} ({I}, {J} = \{ H,C \}).
\end{align}}

\newpage

\section{Stabilizer expectation values of Gaussian Random Noise states}\label{app:ExpValsGRNS}

\allowdisplaybreaks
A Gaussian Random Noise state $\hat{\tilde{\rho}}$ is a mixed state resulting from the Gaussian random displacement of a pure state $\hat{\rho}=\ketbra{\psi}{\psi}$ 
\begin{equation}
    \hat{\tilde{\rho}} = \int_\mathbb{C}\mathrm{d}^2\alpha G_{\bm{\Sigma}}(\alpha)\widehat{D}(\alpha)\ketbra{\psi}{\psi}\widehat{D}^\dagger(\alpha), \label{eq:GRN_def}
\end{equation} 
where $G_{\bm{\Sigma}}(\alpha)$ is a standard complex normal distribution~\cite{andersen1995multivariate} such that
\begin{equation}
    G_{\bm{\Sigma}(\alpha)}=\frac{\exp{-\frac{\Re{\alpha}^2}{\Sigma_0}}\exp{-\frac{\Im{\alpha}^2}{\Sigma_1}}}{\pi\sqrt{\Sigma_0\Sigma_1}},
\end{equation}
with
\begin{align}
    \hat{D}(\alpha) &= \text{exp}\left( \alpha \hat{a}^{\dagger} - \alpha^* \hat{a} \right)\\
    &= \text{exp}\left( \sqrt{2}i \text{Im}\{\alpha\} \hat{x} -\sqrt{2}i \text{Re}\{\alpha\} \hat{p} \right).
 \end{align}
{An ideal GKP state on a rectangular lattice is stabilized by (recall Section \ref{section:GRN_intro})
\begin{align}
    \hat{\mathcal{S}}_x(s) &= \exp(is\hat{x}) = \hat{D}\left(\frac{is}{\sqrt{2}}\right)\\
    \hat{\mathcal{S}}_p(s') &= \exp(-is'\hat{p}) = \hat{D}\left(\frac{s'}{\sqrt{2}}\right),
\end{align}
where $s$ and $s'$ depend on the lattice spacing.}

{We envision modelling an approximate GKP sensor state $\hat{\rho}$ using a Gaussian Random noise state $\hat{\tilde{\rho}}$ obtained by applying Eq. \eqref{eq:GRN_def} to} an ideal GKP sensor state $\ket{\varnothing}$, which is stabilized by ${\hat{\mathcal{S}}_x(\sqrt{2\pi})}=\widehat{D}(i\sqrt{\pi})$ and ${\hat{\mathcal{S}}_p(\sqrt{2\pi})}=\widehat{D}(\sqrt{\pi})$. {The GRN state should be chosen to satisfy}
\begin{equation}\label{eq:equalitiesGRNS}
\expval{{\hat{\mathcal{S}}_x(\sqrt{2\pi})}}_{{\tilde{\rho}}}=\expval{{\hat{\mathcal{S}}_x(\sqrt{2\pi})}}_{\rho}, \quad \text{and} \quad  \expval{{\hat{\mathcal{S}}_p(\sqrt{2\pi})}}_{{\tilde{\rho}}}=\expval{{\hat{\mathcal{S}}_p(\sqrt{2\pi})}}_{\rho}.
\end{equation}
The equalities in Eq.~\eqref{eq:equalitiesGRNS} are satisfied by setting
\begin{equation}
    \Sigma_{i} = \frac{1}{2\pi}\log\frac{1}{\abs{\expval{\widehat{\Theta}_i}_\rho}^2},
\end{equation}
{where $\widehat{\Theta}_0=\hat{\mathcal{S}}_x(\sqrt{2\pi})$, and $\widehat{\Theta}_1=\hat{\mathcal{S}}_p(\sqrt{2\pi})$.}
\begin{proof}
\begin{align}
      \expval{\widehat{\Theta}_j}_{\hat{\tilde{\rho}}} &= \Tr\left[\widehat{\Theta}_j\hat{\tilde{\rho}}\right]  = \Tr\left[\widehat{\Theta}_j\int_\mathbb{C}\mathrm{d}^2\alpha G_{\bm{\Sigma}}(\alpha)\widehat{D}(\alpha)\ketbra{\varnothing}{\varnothing}\widehat{D}^\dagger(\alpha)\right]\\
      &=\int_\mathbb{C}\mathrm{d}^2\alpha G_{\bm{\Sigma}}(\alpha)\Tr\left[\widehat{\Theta}_j\widehat{D}(\alpha)\ketbra{\varnothing}{\varnothing}\widehat{D}^\dagger(\alpha)\right]\\
      &=
      \int_\mathbb{C}\mathrm{d}^2\alpha G_{\bm{\Sigma}}(\alpha)\Tr\left[\expval{\widehat{D}^\dagger(\alpha)\widehat{\Theta}_j\widehat{D}(\alpha)}{\varnothing}\right]\\
      &=\int_\mathbb{C}\mathrm{d}^2\alpha G_{\bm{\Sigma}}(\alpha)\Tr\left[\expval{\widehat{D}^\dagger(\alpha)\widehat{D}(e^{i\pi(1-j)}\sqrt{\pi})\widehat{D}(\alpha)}{\varnothing}\right]\\
      &=\int_\mathbb{C}\mathrm{d}^2\alpha G_{\bm{\Sigma}}(\alpha) e^{2i(1-j)\sqrt{\pi}\Re{\alpha}}e^{-2ij\sqrt{\pi}\Im{\alpha}}\Tr\left[\expval{\widehat{\Theta}_j}{\varnothing}\right]\\
      &=\int_\mathbb{C}\mathrm{d}^2\alpha G_{\bm{\Sigma}}(\alpha) e^{2i(1-j)\sqrt{\pi}\Re{\alpha}}e^{-2ij\sqrt{\pi}\Im{\alpha}}\\
      &=\int_\mathbb{C}\mathrm{d}^2\alpha\frac{1}{\pi\sqrt{\Sigma_0\Sigma_1}}\exp{-\frac{\Re{\alpha}^2}{\Sigma_0}}\exp{-\frac{\Im{\alpha}^2}{\Sigma_1}} e^{2i(1-j)\sqrt{\pi}\Re{\alpha}}e^{-2ij\sqrt{\pi}\Im{\alpha}}\\
      &=\frac{1}{\pi\sqrt{\Sigma_{0}\Sigma_{1}}}\int_{-\infty}^\infty\mathrm{d}\Re{\alpha}\exp{-\frac{\Re{\alpha}^2}{\Sigma_0}}e^{2i(1-j)\sqrt{\pi}\Re{\alpha}}\nonumber\\ 
      &\quad\quad\quad\quad\quad\quad\int_{-\infty}^\infty\mathrm{d}\Im{\alpha}\exp{-\frac{\Im{\alpha}^2}{\Sigma_1}} e^{-2ij\sqrt{\pi}\Im{\alpha}}\\
      &=\nonumber \frac{1}{\pi\sqrt{\Sigma_{0}\Sigma_{1}}}\int_{-\infty}^\infty\mathrm{d}\Re{\alpha}\exp{-\left(\frac{\Re{\alpha}}{\sqrt{\Sigma_0}}-i(1-j)\sqrt{\pi\Sigma_0}\right)^2}\exp{-(1-j)^2\pi\Sigma_0}\\
      &\quad\quad\quad\quad\quad\quad\int_{-\infty}^\infty\mathrm{d}\Im{\alpha}\exp{-\left(\frac{\Im{\alpha}}{\sqrt{\Sigma_1}}+ij\sqrt{\pi\Sigma_1}\right)^2}\exp{-j^2\pi\Sigma_1}\\
      &=\exp{-(1-j)\pi\Sigma_0}\exp{-j\pi\Sigma_1}\\
      &=\exp{-(1-j)\pi\cdot\frac{1}{2\pi}\ln\frac{1}{\abs{\expval{\widehat{\Theta}_0}_\rho}^2}}\exp{-j\pi\cdot\frac{1}{2\pi}\ln\frac{1}{\abs{\expval{\widehat{\Theta}_1}_\rho}^2}}\\
      &=\abs{\expval{\widehat{\Theta}_0}_\rho}^{(1-j)}\abs{\expval{\widehat{\Theta}_1}_\rho}^{j}=\abs{\expval{\widehat{\Theta}_j}_\rho}
\end{align}

\end{proof}

Gaussian Random Noise states can be employed in place of the approximate GKP states to calculate the stabilizer expectation values of the two mode entangled state obtained via a $CZ$ operation onto two GKP qubits. The operation is realized with the optical circuit shown in Fig~\ref{fig:dumbbell_circuit}.
Here, we verify that the stabilizer $\widehat{D}_1(\sqrt{\pi/2})\otimes\widehat{D}_2(i\sqrt{\pi/2})$ has the same expectation value {for the output two-mode GKP state} both when the input states of the circuit are two approximate GKP states $\hat{\rho}_1, \hat{\rho}_2$ and when they are the respective GRN states $\hat{\tilde{\rho}}_1, \hat{\tilde{\rho}}_2$, {provided the GRN state parameters are chosen according to Eq.~\eqref{eq:equalitiesGRNS}.}

\begin{proof}
The expectation value of the stabilizer is

\begin{equation}
    \expval{\widehat{D}_1(\sqrt{\pi/2})\otimes\widehat{D}_2(i\sqrt{\pi/2})}_{\hat{\tilde{\rho}}_{\text{out}}}=\Tr\left\{\widehat{D}_1(\sqrt{\pi/2})\otimes\widehat{D}_2(i\sqrt{\pi/2})\left[\widehat{R}_1\left(\frac{\pi}{2}\right)\widehat{BS}\widehat{R}_1\left(-\frac{\pi}{2}\right)\hat{\tilde{\rho}}_1\otimes\hat{\tilde{\rho}}_2\widehat{R}_1\left(-\frac{\pi}{2}\right)\widehat{BS}^\dagger\widehat{R}_1\left(\frac{\pi}{2}\right)\right]\right\}
\end{equation}

Where $\widehat{BS}$ refers to the operator describing the action of the symmetric beam splitter and $\hat{R}_i(\theta)$ marks the rotation gate of angle $\theta$ acting on the $i$-th mode. The action of the beam splitter operation onto the annihilation operators on the first and second mode $(\hat{a}_1,\hat{a}_2)^T$ is described by the following unitary matrix
\begin{equation}
        \widehat{BS}=\frac{1}{\sqrt{2}}\begin{pmatrix}
        1 & -1 \\
        1 & 1
    \end{pmatrix}.
\end{equation}
We can then define the matrix for the operator $\widehat{BS}'$ such that
\begin{equation}
    \widehat{BS}'\equiv \widehat{R}_1\left(\frac{\pi}{2}\right)\widehat{BS}\widehat{R}_1\left(-\frac{\pi}{2}\right) = \frac{1}{\sqrt{2}} \begin{pmatrix}
        i & 0\\
    0 & 1

    \end{pmatrix}
    \begin{pmatrix}
        1 & -1\\
    1 & 1

    \end{pmatrix}
    \begin{pmatrix}
        -i & 0\\
    0 & 1

    \end{pmatrix} = 
    \frac{1}{\sqrt{2}}
    \begin{pmatrix}
        1 & -i \\
        -i & 1
    \end{pmatrix}
\end{equation}
As a consequence

\begin{align}
     \expval{\widehat{D}_1(\sqrt{\pi/2})\otimes\widehat{D}_2(i\sqrt{\pi/2})}_{\hat{\tilde{\rho}}_{\text{out}}}&=\Tr\left[ \widehat{BS}'^\dagger \widehat{D}_1(\sqrt{\pi/2})\otimes\widehat{D}_2(i\sqrt{\pi/2})\widehat{BS}'\hat{\tilde{\rho}}_1\otimes\hat{\tilde{\rho}}_2\right] \\
     &=\Tr\left[ \widehat{D}_1{\left(\frac{\sqrt{\pi}}{2}-\frac{\sqrt{\pi}}{2}\right)}\otimes\widehat{D}_2\left(i\frac{\sqrt{\pi}}{2}+i\frac{\sqrt{\pi}}{2}\right)\hat{\tilde{\rho}}_1\otimes\hat{\tilde{\rho}}_2\right] \\
    &=\Tr\left[\widehat{D}_2\left(i\sqrt{\pi}\right)\hat{\tilde{\rho}}_1\otimes\hat{\tilde{\rho}}_2\right]\\
     &=\Tr\left[{\hat{\mathcal{S}}_{x,2}(\sqrt{2\pi})}\hat{\tilde{\rho}}_2\right]\\
     &= \Tr\left[{\hat{\mathcal{S}}_{x,2}(\sqrt{2\pi})}\rho_2\right]\\
     &=\expval{\widehat{D}_1(\sqrt{\pi/2})\otimes\widehat{D}_2(i\sqrt{\pi/2})}_{\rho_{\text{out}}}
\end{align}
\end{proof}

We can now consider the stabilizer expectation values of the state heralded by the homodyne detection of the two-mode entangled state $\rho_\text{out}$. The scheme that prepares the heralded state is shown in Fig.~\ref{fig:DBhomocircuit}. The heralded state has two stabilizers {$\hat{\mathcal{S}}_p(2\sqrt{\pi})=\widehat{D}(i\sqrt{2\pi})$ and $\hat{\mathcal{S}}_x(2\sqrt{\pi})=\widehat{D}(\sqrt{2\pi})$.}

By postselecting on the measure of $p_2$ in the second mode, the stabilizer expectation value of $\hat{\mathcal{S}}_x(2\sqrt{\pi})$ in the first mode is 

\begin{equation}\label{eq:Zexpval}
    \expval{\hat{\mathcal{S}}_{x}(2\sqrt{\pi})}=\frac{\Tr\left[\left(\ketbra{p_2}{p_2}\widehat{BS}'^\dagger\hat{\rho}_1\otimes\hat{\rho}_2\widehat{BS}'\right)\hat{\mathcal{S}}_{x,1}(2\sqrt{\pi})\right]}{\Tr\left[\ketbra{p_2}{p_2}\widehat{BS}'^\dagger\hat{\rho}_1\otimes\hat{\rho}_2\widehat{BS}'\right]}
\end{equation}

The denominator in Eq.~\eqref{eq:Zexpval} can be rewritten as 
\begin{align}
    \mathcal{N}_{p_2}\equiv&\Tr\left[\ketbra{p_2}{p_2}\widehat{BS}'^\dagger\hat{\rho}_1\otimes\hat{\rho}_2\widehat{BS}'\right]\\
     =&\Tr\left[\widehat{BS}'\ketbra{p_2}{p_2}\widehat{BS}'^\dagger\hat{\rho}_1\otimes\hat{\rho}_2\right]\\
     =&\Tr\left[\widehat{BS}'\underbrace{\int_{-\infty}^\infty\mathrm{d}l\ket{l}_x\bra{l}_x}_{\mathbb{1}_1}\otimes\ket{p_2}_p\bra{p_2}_p\widehat{BS}'^\dagger\hat{\rho}_1\otimes\hat{\rho}_2\right]\\
     =&\int_{-\infty}^\infty\mathrm{d}l\Tr\left[\widehat{BS}'\ket{l}_x\ket{p_2}_p\bra{l}_x\bra{p_2}_p\widehat{BS}'^\dagger\hat{\rho}_1\otimes\hat{\rho}_2\right]\\
     =&\int_{-\infty}^\infty\mathrm{d}l\Tr\left[\ket{\frac{l+p_2}{\sqrt{2}}}_x\ket{-\frac{l-p_2}{\sqrt{2}}}_p\bra{\frac{l+p_2}{\sqrt{2}}}_x\bra{-\frac{l-p_2}{\sqrt{2}}}_p\hat{\rho}_1\otimes\hat{\rho}_2\right]\\
     =&\int_{-\infty}^\infty\mathrm{d}l\bra{\frac{l+p_2}{\sqrt{2}}}_x\bra{-\frac{l-p_2}{\sqrt{2}}}_p\hat{\rho}_1\otimes\hat{\rho}_2\ket{\frac{l+p_2}{\sqrt{2}}}_x\ket{-\frac{l-p_2}{\sqrt{2}}}_p\\
     =&\int_{-\infty}^\infty\mathrm{d}l\bra{\frac{l+p_2}{\sqrt{2}}}_x\hat{\rho}_1\ket{\frac{l+p_2}{\sqrt{2}}}_x\bra{-\frac{l-p_2}{\sqrt{2}}}_p\hat{\rho}_2\ket{-\frac{l-p_2}{\sqrt{2}}}_p
     \label{eq:nominator_Z}
\end{align}
Analogously, the numerator in Eq.~\eqref{eq:Zexpval} is equal to
\begin{align}
    &\Tr\left[\left(\ketbra{p_2}{p_2}\widehat{BS}'^\dagger\hat{\tilde{\rho}}_1\otimes\hat{\tilde{\rho}}_2\widehat{BS}'\right) {\hat{\mathcal{S}}_{x,1}}(2\sqrt{\pi})\right]\\
     =&\Tr\left[\widehat{BS}'{\hat{\mathcal{S}}_{x,1}}(2\sqrt{\pi})\ketbra{p_2}{p_2}\widehat{BS}'^\dagger\hat{\tilde{\rho}}_1\otimes\hat{\tilde{\rho}}_2\right]\\
     =&\Tr\left[\widehat{BS}'{\hat{\mathcal{S}}_{x,1}}(2\sqrt{\pi})\widehat{BS}'^\dagger\widehat{BS}'\ketbra{p_2}{p_2}\widehat{BS}'^\dagger\hat{\tilde{\rho}}_1\otimes\hat{\tilde{\rho}}_2\right]\\
     =&\Tr\left[{\hat{\mathcal{S}}_{x,1}}(\sqrt{2\pi}){\hat{\mathcal{S}}_{p,2}}(\sqrt{2\pi})\widehat{BS}'\ketbra{p_2}{p_2}\widehat{BS}'^\dagger\hat{\tilde{\rho}}_1\otimes\hat{\tilde{\rho}}_2\right]\\
     =&\int_{-\infty}^\infty\mathrm{d}l\bra{\frac{l+p_2}{\sqrt{2}}}_x\bra{-\frac{l-p_2}{\sqrt{2}}}_p\hat{\tilde{\rho}}_1\otimes\hat{\tilde{\rho}}_2{\hat{\mathcal{S}}_{x,1}}(\sqrt{2\pi}){\hat{\mathcal{S}}_{p,2}}(\sqrt{2\pi})\ket{\frac{l+p_2}{\sqrt{2}}}_x\ket{-\frac{l-p_2}{\sqrt{2}}}_p\\
     =&\int_{-\infty}^\infty\mathrm{d}l\bra{\frac{l+p_2}{\sqrt{2}}}_x\hat{\tilde{\rho}}_1{\hat{\mathcal{S}}_{x,1}}(\sqrt{2\pi})\ket{\frac{l+p_2}{\sqrt{2}}}_x\bra{-\frac{l-p_2}{\sqrt{2}}}_p\hat{\tilde{\rho}}_2{\hat{\mathcal{S}}_{p,2}}(\sqrt{2\pi})\ket{-\frac{l-p_2}{\sqrt{2}}}_p\\
     =&\int_{-\infty}^\infty\mathrm{d}l\bra{\frac{l+p_2}{\sqrt{2}}}_x\hat{\tilde{\rho}}_1\ket{\frac{l+p_2}{\sqrt{2}}}_xe^{i\sqrt{2\pi}\left(\frac{l+p_2}{\sqrt{2}}\right)}\bra{-\frac{l-p_2}{\sqrt{2}}}_p\hat{\tilde{\rho}}_2\ket{-\frac{l-p_2}{\sqrt{2}}}_pe^{i\sqrt{2\pi}\left(\frac{l-p_2}{\sqrt{2}}\right)}\\
     =&\int_{-\infty}^\infty\mathrm{d}le^{i2\sqrt{\pi}l}\bra{\frac{l+p_2}{\sqrt{2}}}_x\hat{\tilde{\rho}}_1\ket{\frac{l+p_2}{\sqrt{2}}}_x\bra{-\frac{l-p_2}{\sqrt{2}}}_p\hat{\tilde{\rho}}_2\ket{-\frac{l-p_2}{\sqrt{2}}}_p
     \label{eq:denominator}
\end{align}
We observe that 
\begin{align}
    \bra{q}_x\hat{\tilde{\rho}}\ket{q}_x=&\\
    =&\bra{q}_x\int_\mathbb{C}\mathrm{d}^2\alpha G_{\Sigma}(\alpha)\widehat{D}(\alpha)\ketbra{\varnothing}{\varnothing}\widehat{D}^\dagger(\alpha)\ket{q}_x\\
    =&\int_\mathbb{C}\mathrm{d}^2\alpha G_{\Sigma}(\alpha)\bra{q}_x\widehat{D}(\alpha)\ketbra{\varnothing}{\varnothing}\widehat{D}^\dagger(\alpha)\ket{q}_x\\
    =&\int_\mathbb{C}\mathrm{d}^2\alpha G_{\Sigma}(\alpha)\abs{\bra{\varnothing}\widehat{D}^\dagger(\alpha)\ket{q}_x}^2\\
     =&\int_\mathbb{C}\mathrm{d}^2\alpha G_{\Sigma}(\alpha)\abs{\bra{\varnothing}\ket{q-\sqrt{2}\Re{\alpha}}_x}^2\\
     =&\frac{1}{\pi\sqrt{\Sigma_{0}\Sigma_{1}}}\int_{-\infty}^\infty\mathrm{d}\Re{\alpha}e^{-\frac{\Re{\alpha}^2}{\Sigma_{0}}}\abs{\bra{\varnothing}\ket{q-\sqrt{2}\Re{\alpha}}_x}^2\int_{-\infty}^\infty\mathrm{d}\Im{\alpha}e^{-\frac{\Im{\alpha}^2}{\Sigma_{1}}}\\
     =&\frac{1}{\sqrt{\pi\Sigma_{0}}}\int_{-\infty}^\infty\mathrm{d}\Re{\alpha}e^{-\frac{\Re{\alpha}^2}{\Sigma_{0}}}\abs{\bra{\varnothing}\ket{q-\sqrt{2}\Re{\alpha}}_x}^2\\
     =&\frac{1}{\sqrt{\pi\Sigma_{0}}}\int_{-\infty}^\infty\mathrm{d}\Re{\alpha}e^{-\frac{\Re{\alpha}^2}{\Sigma_{0}}}\abs{M\sum_{n=-\infty}^\infty\bra{\sqrt{2\pi}n}\ket{q-\sqrt{2}\Re{\alpha}}_x}^2\\
     =&\frac{M^2}{\sqrt{\pi\Sigma_{0}}}\int_{-\infty}^\infty\mathrm{d}\Re{\alpha}e^{-\frac{\Re{\alpha}^2}{\Sigma_{0}}}\sum_{n=-\infty}^\infty\abs{\bra{\sqrt{2\pi}n}\ket{q-\sqrt{2}\Re{\alpha}}_x}^2\\
     =&\frac{M^2}{\sqrt{\pi\Sigma_{0}}}\int_{-\infty}^\infty\mathrm{d}\Re{\alpha}e^{-\frac{\Re{\alpha}^2}{\Sigma_{0}}}\sum_{n=-\infty}^\infty\delta^2\left(q-\sqrt{2}\Re{\alpha}-\sqrt{2\pi}n\right)\\
    =&\frac{M^2}{\sqrt{\pi\Sigma_{0}}}\sum_{n=-\infty}^\infty\int_{-\infty}^\infty\mathrm{d}\Re{\alpha}e^{-\frac{\Re{\alpha}^2}{\Sigma_{0}}}\delta^2\left(q-\sqrt{2}\Re{\alpha}-\sqrt{2\pi}n\right)\\
    =&\frac{M^2}{\sqrt{\pi\Sigma_{0}}}\sum_{n=-\infty}^\infty\delta(0)\exp{-\frac{\left(\frac{q}{\sqrt{2}}-\sqrt{\pi}n\right)^2}{\Sigma_{0}}}
    \label{eq:rho_on_x}
\end{align}

and

\begin{align}
    \bra{q}_p\hat{\tilde{\rho}}\ket{q}_p=&\\
    =&\bra{q}_p\int_\mathbb{C}\mathrm{d}^2\alpha G_{\Sigma,2}(\alpha)\widehat{D}(\alpha)\ketbra{\varnothing}{\varnothing}\widehat{D}^\dagger(\alpha)\ket{q}_p\\
    =&\int_\mathbb{C}\mathrm{d}^2\alpha G_{\Sigma,2}(\alpha)\bra{q}_p\widehat{D}(\alpha)\ketbra{\varnothing}{\varnothing}\widehat{D}^\dagger(\alpha)\ket{q}_p\\
    =&\int_\mathbb{C}\mathrm{d}^2\alpha G_{\Sigma,2}(\alpha)\abs{\bra{\varnothing}\widehat{D}^\dagger(\alpha)\ket{q}_p}^2\\
     =&\int_\mathbb{C}\mathrm{d}^2\alpha G_{\Sigma,2}(\alpha)\abs{\bra{\varnothing}\ket{q-\sqrt{2}\Im{\alpha}}_p}^2\\
     =&\frac{1}{\pi\sqrt{\Sigma_{0}\Sigma_{1}}}\int_{-\infty}^\infty\mathrm{d}\Re{\alpha}e^{-\frac{\Re{\alpha}^2}{\Sigma_{0}}}\int_{-\infty}^\infty\mathrm{d}\Im{\alpha}e^{-\frac{\Im{\alpha}^2}{\Sigma_{1}}}\abs{\bra{\varnothing}\ket{q-\sqrt{2}\Im{\alpha}}_p}^2\\
     =&\frac{1}{\sqrt{\pi\Sigma_{1}}}\int_{-\infty}^\infty\mathrm{d}\Im{\alpha}e^{-\frac{\Im{\alpha}^2}{\Sigma_{1}}}\abs{\bra{\varnothing}\ket{q-\sqrt{2}\Im{\alpha}}_p}^2\\
     =&\frac{1}{\sqrt{\pi\Sigma_{1}}}\int_{-\infty}^\infty\mathrm{d}\Im{\alpha}e^{-\frac{\Im{\alpha}^2}{\Sigma_{1}}}\abs{M\sum_{m=-\infty}^\infty\bra{\sqrt{2\pi}n}\ket{q-\sqrt{2}\Im{\alpha}}_p}^2\\
     =&\frac{M^2}{\sqrt{\pi\Sigma_{1}}}\int_{-\infty}^\infty\mathrm{d}\Im{\alpha}e^{-\frac{\Im{\alpha}^2}{\Sigma_{1}}}\sum_{m=-\infty}^\infty\abs{\bra{\sqrt{2\pi}m}\ket{q-\sqrt{2}\Im{\alpha}}_p}^2\\
     =&\frac{M^2}{\sqrt{\pi\Sigma_{1}}}\int_{-\infty}^\infty\mathrm{d}\Im{\alpha}e^{-\frac{\Im{\alpha}^2}{\Sigma_{1}}}\sum_{m=-\infty}^\infty\delta^2\left(q-\sqrt{2}\Im{\alpha}-\sqrt{2\pi}m\right)\\
    =&\frac{M^2}{\sqrt{\pi\Sigma_{1}}}\sum_{m=-\infty}^\infty\int_{-\infty}^\infty\mathrm{d}\Im{\alpha}e^{-\frac{\Im{\alpha}^2}{\Sigma_{1}}}\delta^2\left(q-\sqrt{2}\Im{\alpha}-\sqrt{2\pi}m\right)\\
    =&\frac{M^2}{\sqrt{\pi\Sigma_{1}}}\sum_{m=-\infty}^\infty\delta(0)\exp{-\frac{\left(\frac{p}{\sqrt{2}}-\sqrt{\pi}m\right)^2}{\Sigma_{1}}}
    \label{eq:rho_on_p}
\end{align}
By inserting the results of Eqs.~\eqref{eq:rho_on_x} and \eqref{eq:rho_on_p} into Eqs.~\eqref{eq:nominator_Z} and \eqref{eq:denominator} we can rewrite Eq.~\eqref{eq:Zexpval} as
\begin{align}
    &\Rightarrow\expval{{\hat{\mathcal{S}}_{x}}}=\frac{1}{{N}_{p_2}}\int_{-\infty}^\infty\mathrm{d}le^{i2\sqrt{\pi}l}\bra{\frac{l+p_2}{\sqrt{2}}}_x\hat{\tilde{\rho}}_1\ket{\frac{l+p_2}{\sqrt{2}}}_x\bra{-\frac{l-p_2}{\sqrt{2}}}_p\hat{\tilde{\rho}}_2\ket{-\frac{l-p_2}{\sqrt{2}}}_p\\
       &=\frac{1}{{N}_{p_2}}\int_{-\infty}^\infty\mathrm{d}le^{i2\sqrt{\pi}l}\bra{\frac{l+p_2}{\sqrt{2}}}_x\hat{\tilde{\rho}}_1\ket{\frac{l+p_2}{\sqrt{2}}}_x\bra{-\frac{l-p_2}{\sqrt{2}}}_p\hat{\tilde{\rho}}_2\ket{-\frac{l-p_2}{\sqrt{2}}}_p\\
    &=\frac{\int_{-\infty}^\infty\mathrm{d}le^{i2\sqrt{\pi}l}\bra{\frac{l+p_2}{\sqrt{2}}}_x\hat{\tilde{\rho}}_1\ket{\frac{l+p_2}{\sqrt{2}}}_x\bra{-\frac{l-p_2}{\sqrt{2}}}_p\hat{\tilde{\rho}}_2\ket{-\frac{l-p_2}{\sqrt{2}}}_p}{\int_{-\infty}^\infty\mathrm{d}l\bra{\frac{l+p_2}{\sqrt{2}}}_x\hat{\tilde{\rho}}_1\ket{\frac{l+p_2}{\sqrt{2}}}_x\bra{-\frac{l-p_2}{\sqrt{2}}}_p\hat{\tilde{\rho}}_2\ket{-\frac{l-p_2}{\sqrt{2}}}_p}\\
     &=\frac{\int_{-\infty}^\infty\mathrm{d}le^{i2\sqrt{\pi}l}\frac{M^2}{\sqrt{\pi\Sigma_{0,1}}}\sum_{n=-\infty}^\infty\delta(0)\exp{-\frac{\left(\frac{l}{2}+\frac{p_2}{2}-\sqrt{\pi}n\right)^2}{\Sigma_{0,1}}}\frac{M^2}{\sqrt{\pi\Sigma_{1,2}}}\sum_{m=-\infty}^\infty\delta(0)\exp{-\frac{\left(\frac{l}{2}-\frac{p_2}{2}-\sqrt{\pi}m\right)^2}{\Sigma_{1,2}}}}{\int_{-\infty}^\infty\mathrm{d}l\frac{M^2}{\sqrt{\pi\Sigma_{0,1}}}\sum_{n=-\infty}^\infty\delta(0)\exp{-\frac{\left(\frac{l}{2}+\frac{p_2}{2}-\sqrt{\pi}n\right)^2}{\Sigma_{0,1}}}\frac{M^2}{\sqrt{\pi\Sigma_{1,2}}}\sum_{m=-\infty}^\infty\delta(0)\exp{-\frac{\left(\frac{l}{2}-\frac{p_2}{2}-\sqrt{\pi}m\right)^2}{\Sigma_{1,2}}}}\\
     &=\frac{\sum_{n,m=-\infty}^\infty\int_{-\infty}^\infty\mathrm{d}le^{i2\sqrt{\pi}l}\exp{-\frac{\left(\frac{l}{2}+\frac{p_2}{2}-\sqrt{\pi}n\right)^2}{\Sigma_{0,1}}}\exp{-\frac{\left(\frac{l}{2}-\frac{p_2}{2}-\sqrt{\pi}m\right)^2}{\Sigma_{1,2}}}}{\sum_{n,m=-\infty}^\infty\int_{-\infty}^\infty\mathrm{d}l\exp{-\frac{\left(\frac{l}{2}+\frac{p_2}{2}-\sqrt{\pi}n\right)^2}{\Sigma_{0,1}}}\exp{-\frac{\left(\frac{l}{2}-\frac{p_2}{2}-\sqrt{\pi}m\right)^2}{\Sigma_{1,2}}}}\\
     &=\frac{\sum_{n,m=-\infty}^\infty2\sqrt{\frac{\Sigma_{0,1}\Sigma_{1,2}}{\Sigma_{0,1}+\Sigma_{1,2}}\pi}\exp\left(-\frac{\left[\sqrt{\pi}(n-m)-p_2\right]^2}{\Sigma_{0,1}+\Sigma_{1,2}}\right)\exp\left(-4\pi\frac{\Sigma_{0,1}\Sigma_{1,2}}{\Sigma_{0,1}+\Sigma_{1,2}}\right)\exp\left(i4\pi\frac{\Sigma_{1,2}n+\Sigma_{0,1}m}{\Sigma_{1,2}+\Sigma_{0,1}}\right)\exp\left(-i2\sqrt{\pi}\frac{\Sigma_{1,2}-\Sigma_{0,1}}{\Sigma_{1,2}+\Sigma_{0,1}}p_2\right)}{\sum_{n,m=-\infty}^\infty2\sqrt{\frac{\Sigma_{0,1}\Sigma_{1,2}}{\Sigma_{0,1}+\Sigma_{1,2}}\pi}\exp\left(-\frac{\left[\sqrt{\pi}(n-m)-p_2\right]^2}{\Sigma_{0,1}+\Sigma_{1,2}}\right)}\\
    &=\exp\left(-4\pi\frac{\Sigma_{0,1}\Sigma_{1,2}}{\Sigma_{0,1}+\Sigma_{1,2}}\right)\exp\left(-i2\sqrt{\pi}\frac{\Sigma_{1,2}-\Sigma_{0,1}}{\Sigma_{1,2}+\Sigma_{0,1}}p_2\right) \nonumber\\
    &\quad\quad\quad\quad\quad\quad\quad\quad\frac{\sum_{n=-\infty}^\infty\sum_{m=-\infty}^\infty\exp\left(-\frac{\left[\sqrt{\pi}(n-m)-p_2\right]^2}{\Sigma_{0,1}+\Sigma_{1,2}}\right)\exp\left(i4\pi\frac{\Sigma_{1,2}n+\Sigma_{0,1}m}{\Sigma_{1,2}+\Sigma_{0,1}}\right)}{\sum_{n=-\infty}^\infty\sum_{m=-\infty}^\infty\exp\left(-\frac{\left[\sqrt{\pi}(n-m)-p_2\right]^2}{\Sigma_{0,1}+\Sigma_{1,2}}\right)}\\
    &\stackrel{m'=m-n}{=}\exp\left(\frac{-4\pi\Sigma_{0,1}\Sigma_{1,2}-2i\sqrt{\pi}p_2\left(\Sigma_{1,2}-\Sigma_{0,1}\right)}{\Sigma_{0,1}+\Sigma_{1,2}}\right) \nonumber\\
    &\quad\quad\quad\quad\quad\quad\quad\frac{\sum_{n=-\infty}^\infty\sum_{m'=-\infty}^\infty\exp\left(-\frac{\left[\sqrt{\pi}m'+p_2\right]^2}{\Sigma_{0,1}+\Sigma_{1,2}}\right)\exp\left(i4\pi\frac{\Sigma_{1,2}+\Sigma_{0,1}}{\Sigma_{1,2}+\Sigma_{0,1}}n+i4\pi\frac{\Sigma_{0,1}m'}{\Sigma_{1,2}+\Sigma_{0,1}}\right)}{\sum_{n=-\infty}^\infty\sum_{m'=-\infty}^\infty\exp\left(-\frac{\left[\sqrt{\pi}m'+p_2\right]^2}{\Sigma_{0,1}+\Sigma_{1,2}}\right)}\\
    &=\exp\left(\frac{-4\pi\Sigma_{0,1}\Sigma_{1,2}-2i\sqrt{\pi}p_2\left(\Sigma_{1,2}-\Sigma_{0,1}\right)}{\Sigma_{0,1}+\Sigma_{1,2}}\right)\frac{\sum_{m'=-\infty}^\infty\exp\left(-\frac{\left[\sqrt{\pi}m'+p_2\right]^2}{\Sigma_{0,1}+\Sigma_{1,2}}\right)\exp\left(i4\pi\frac{\Sigma_{0,1}m'}{\Sigma_{1,2}+\Sigma_{0,1}}\right)}{\sum_{m'=-\infty}^\infty\exp\left(-\frac{\left[\sqrt{\pi}m'+p_2\right]^2}{\Sigma_{0,1}+\Sigma_{1,2}}\right)}\\
    &=\exp\left(\frac{-4\pi\Sigma_{0,1}\Sigma_{1,2}-2i\sqrt{\pi}p_2\left(\Sigma_{1,2}-\Sigma_{0,1}\right)}{\Sigma_{0,1}+\Sigma_{1,2}}\right) \nonumber\\
    &\quad\quad\frac{\sqrt{\Sigma_{0,1}+\Sigma_{1,2}}\exp\left(-4\pi\frac{\Sigma_{0,1}^2}{\Sigma_{0,1}+\Sigma_{1,2}}\right)\exp\left(-4i\sqrt{\pi}\frac{\Sigma_{0,1}}{\Sigma_{0,1}+\Sigma_{1,2}}p_2\right)\theta_3\left(-\sqrt{\pi}p_2+2i\pi\Sigma_{0,1},e^{-(\Sigma_{0,1}+\Sigma_{1,2})\pi}\right)}{\sqrt{\Sigma_{0,1}+\Sigma_{1,2}}\theta_3\left(-\sqrt{\pi}p_2,e^{-(\Sigma_{0,1}+\Sigma_{1,2})\pi}\right)}\\
    &=\exp\left(-4\pi\Sigma_{0,1}-2i\sqrt{\pi}p_2\right)\frac{\theta_3\left(\sqrt{\pi}p_2-2i\pi\Sigma_{0,1},e^{-(\Sigma_{0,1}+\Sigma_{1,2})\pi}\right)}{\theta_3\left(\sqrt{\pi}p_2,e^{-(\Sigma_{0,1}+\Sigma_{1,2})\pi}\right)},
\end{align}
where $\Sigma_{i,j}$ is the $\Sigma_i$ of the state in the $j$-th mode, while the Jacobi Theta function $\theta_3(z,q)=\sum_{n=-\infty}^\infty q^{n^2}e^{2inz}$.

The expectation value of the second operator ${\hat{\mathcal{S}}_{p}}$ will be given instead by

\begin{equation}\label{eq:Xexpval}
    \expval{{\hat{\mathcal{S}}_{p}}}=\frac{\Tr\left[\left(\ketbra{p_2}{p_2}\widehat{BS}'^\dagger\hat{\rho}_1\otimes\hat{\rho}_2\widehat{BS}'\right){\hat{\mathcal{S}}_{p,1}}(2\sqrt{\pi})\right]}{\Tr\left[\ketbra{p_2}{p_2}\widehat{BS}'^\dagger\hat{\rho}_1\otimes\hat{\rho}_2\widehat{BS}'\right]}
\end{equation}
The numerator of Eq.~\eqref{eq:Xexpval} is 

\begin{align}
    &\Tr\left[\left(\ketbra{p_2}{p_2}\widehat{BS}'^\dagger\hat{\rho}_1\otimes\hat{\rho}_2\widehat{BS}'\right){\hat{\mathcal{S}}_{p,1}}(2\sqrt{\pi})\right]\\
     =&\Tr\left[\widehat{BS}'{\hat{\mathcal{S}}_{p,1}}(2\sqrt{\pi})\ketbra{p_2}{p_2}\widehat{BS}'^\dagger\hat{\rho}_1\otimes\hat{\rho}_2\right]\\
     =&\Tr\left[\widehat{BS}'{\hat{\mathcal{S}}_{p,1}}(2\sqrt{\pi})\widehat{BS}'^\dagger\widehat{BS}'\ketbra{p_2}{p_2}\widehat{BS}'^\dagger\hat{\rho}_1\otimes\hat{\rho}_2\right]\\
     =&\Tr\left[{\hat{\mathcal{S}}_{p,1}}(\sqrt{2\pi}){\hat{\mathcal{S}}_{x,2}}(-\sqrt{2\pi})\widehat{BS}'\ketbra{p_2}{p_2}\widehat{BS}'^\dagger\hat{\rho}_1\otimes\hat{\rho}_2\right]\\
     =&\int_{-\infty}^\infty\mathrm{d}l\bra{\frac{l+p_2}{\sqrt{2}}}_x\bra{-\frac{l-p_2}{\sqrt{2}}}_p\hat{\rho}_1\otimes\hat{\rho}_2{\hat{\mathcal{S}}_{p,1}}(\sqrt{2\pi}){\hat{\mathcal{S}}_{x,2}}(-\sqrt{2\pi})\ket{\frac{l+p_2}{\sqrt{2}}}_x\ket{-\frac{l-p_2}{\sqrt{2}}}_p\\
     =&\int_{-\infty}^\infty\mathrm{d}l\bra{\frac{l+p_2}{\sqrt{2}}}_x\rho_1{\hat{\mathcal{S}}_{p,1}}(\sqrt{2\pi})\ket{\frac{l+p_2}{\sqrt{2}}}_x\bra{-\frac{l-p_2}{\sqrt{2}}}_p\rho_2{\hat{\mathcal{S}}_{x,2}}(-\sqrt{2\pi})\ket{-\frac{l-p_2}{\sqrt{2}}}_p.
     \label{eq:numeratorX}
\end{align}
The left term in the integral of Eq.~\ref{eq:numeratorX} is
\begin{align}
    &\bra{q}_x\hat{\tilde{\rho}}{\hat{\mathcal{S}}_{p}}(\sqrt{2\pi})\ket{q}_x\\
    =&\bra{q}_x\int_\mathbb{C}\mathrm{d}^2\alpha G_{\Sigma}(\alpha)\widehat{D}(\alpha)\ketbra{\varnothing}{\varnothing}\widehat{D}^\dagger(\alpha){\hat{\mathcal{S}}_{p}}(\sqrt{2\pi})\ket{q}_x\\
    =&\int_\mathbb{C}\mathrm{d}^2\alpha G_{\Sigma}(\alpha)\bra{q}_x\widehat{D}(\alpha)\ketbra{\varnothing}{\varnothing}\widehat{D}^\dagger(\alpha){\hat{\mathcal{S}}_{p}}(\sqrt{2\pi})\ket{q}_x\\
    =&\int_\mathbb{C}\mathrm{d}^2\alpha G_{\Sigma}(\alpha)\bra{q}_x\widehat{D}(\alpha)\ketbra{\varnothing}{\varnothing}{\hat{\mathcal{S}}_{p}}(\sqrt{2\pi})\widehat{D}^\dagger(\alpha)\ket{q}_xe^{-2i\sqrt{\pi}\Im{\alpha}}\\
    =&\int_\mathbb{C}\mathrm{d}^2\alpha G_{\Sigma}(\alpha)e^{-2i\sqrt{\pi}\Im{\alpha}}\bra{q}_x\widehat{D}(\alpha)\ketbra{\varnothing}{\varnothing}\widehat{D}^\dagger(\alpha)\ket{q}_x\\
    =&\int_\mathbb{C}\mathrm{d}^2\alpha G_{\Sigma}(\alpha)e^{-2i\sqrt{\pi}\Im{\alpha}}\abs{\bra{\varnothing}\widehat{D}(-\alpha)\ket{q}_x}^2\\
    =&\int_\mathbb{C}\mathrm{d}^2\alpha G_{\Sigma}(\alpha)e^{-2i\sqrt{\pi}\Im{\alpha}}\abs{\bra{\varnothing}\ket{q-\sqrt{2}\Re{\alpha}}_x}^2\\
    =&\int_\mathbb{C}\mathrm{d}^2\alpha\frac{1}{2\pi\sqrt{\Sigma_{0}\Sigma_{1}}}\exp{-\frac{\Re{\alpha}^2}{\Sigma_0}}\exp{-\frac{\Im{\alpha}^2}{\Sigma_1}}\exp{-2i\sqrt{\pi}\Im{\alpha}}\abs{\bra{\varnothing}\ket{q-\sqrt{2}\Re{\alpha}}_x}^2\\
    =&\frac{1}{\pi\sqrt{\Sigma_{0}\Sigma_{1}}}\int_{-\infty}^\infty\mathrm{d}\Re{\alpha}\exp{-\frac{\Re{\alpha}^2}{\Sigma_{0}}}\abs{\bra{\varnothing}\ket{q-\sqrt{2}\Re{\alpha}}_x}^2 \nonumber\\
    &\quad\quad\quad\quad\quad\quad\int_{-\infty}^\infty\mathrm{d}\Im{\alpha}\exp{-\frac{\Im{\alpha}^2}{\Sigma_{1}}} e^{-2i\sqrt{\pi}\Im{\alpha}}\\
    =&\frac{1}{\pi\sqrt{\Sigma_{0}\Sigma_{1}}}\int_{-\infty}^\infty\mathrm{d}\Re{\alpha}\exp{-\frac{\Re{\alpha}^2}{\Sigma_{0}}}\abs{\bra{\varnothing}\ket{q-\sqrt{2}\Re{\alpha}}_x}^2\nonumber\\
    &\quad\quad\quad\quad\quad\quad\int_{-\infty}^\infty\mathrm{d}\Im{\alpha}\exp{-\left(\frac{\Im{\alpha}}{\sqrt{\Sigma_{1}}}+i\sqrt{\pi\Sigma_{1}}\right)^2}\exp{-\pi\Sigma_{1}}\\
    =&\frac{\exp{-\pi\Sigma_{1}}}{\pi\sqrt{\Sigma_{0}\Sigma_{1}}}\int_{-\infty}^\infty\mathrm{d}\Re{\alpha}\exp{-\frac{\Re{\alpha}^2}{\Sigma_{0}}}\abs{\bra{\varnothing}\ket{q-\sqrt{2}\Re{\alpha}}_x}^2\nonumber\\
    &\quad\quad\quad\quad\quad\quad\int_{-\infty}^\infty\mathrm{d}\Im{\alpha}\exp{-\frac{\Im{\alpha}^2}{\Sigma_{1}}}\\
     =&\frac{\exp{-\pi\Sigma_{1}}}{\pi\sqrt{\Sigma_{0}\Sigma_{1}}}\int_{-\infty}^\infty\mathrm{d}\Re{\alpha}\exp{-\frac{\Re{\alpha}^2}{\Sigma_{0}}} \nonumber\\
    &\quad\quad\quad\quad\quad\quad\int_{-\infty}^\infty\mathrm{d}\Im{\alpha}\exp{-\frac{\Im{\alpha}^2}{\Sigma_{1}}}\abs{\bra{\varnothing}\widehat{D}^\dagger(\alpha)\ket{q}_x}^2\\
    =&\exp{-\pi\Sigma_{1}}\int_{-\infty}^\infty\mathrm{d}^2\alpha G_{\Sigma}(\alpha)\bra{q}_x\widehat{D}(\alpha)\ketbra{\varnothing}{\varnothing}\widehat{D}^\dagger(\alpha)\ket{q}_x\\
    =&\exp{-\pi\Sigma_{1}}\bra{q}_x\hat{\tilde{\rho}}\ket{q}_x\label{eq:leftTermX}
\end{align}
The right term in the integral of Eq.~\ref{eq:numeratorX} is
\begin{align}
    \bra{q}_p\hat{\tilde{\rho}}{\hat{\mathcal{S}}_{x}}(-\sqrt{2\pi})\ket{q}_p=&\\
    =&\bra{q}_p\int_\mathbb{C}\mathrm{d}^2\alpha G_{\Sigma}(\alpha)\widehat{D}(\alpha)\ketbra{\varnothing}{\varnothing}\widehat{D}^\dagger(\alpha){\hat{\mathcal{S}}_{x}}(-\sqrt{2\pi})\ket{q}_p\\
    =&\int_\mathbb{C}\mathrm{d}^2\alpha G_{\Sigma}(\alpha)\bra{q}_p\widehat{D}(\alpha)\ketbra{\varnothing}{\varnothing}\widehat{D}^\dagger(\alpha){\hat{\mathcal{S}}_{x}}(-\sqrt{2\pi})\ket{q}_p\\
    =&\int_\mathbb{C}\mathrm{d}^2\alpha G_{\Sigma}(\alpha)\bra{q}_p\widehat{D}(\alpha)\ketbra{\varnothing}{\varnothing}{\hat{\mathcal{S}}_{x}}(-\sqrt{2\pi})\widehat{D}^\dagger(\alpha)\ket{q}_pe^{-2i\sqrt{\pi}\Re{\alpha}}\\
    =&\int_\mathbb{C}\mathrm{d}^2\alpha G_{\Sigma}(\alpha)e^{-2i\sqrt{\pi}\Re{\alpha}}\bra{q}_p\widehat{D}(\alpha)\ketbra{\varnothing}{\varnothing}\widehat{D}^\dagger(\alpha)\ket{q}_p\\
    =&\int_\mathbb{C}\mathrm{d}^2\alpha G_{\Sigma}(\alpha)e^{-2i\sqrt{\pi}\Re{\alpha}}\abs{\bra{\varnothing}\widehat{D}(-\alpha)\ket{q}_p}^2\\
    =&\int_\mathbb{C}\mathrm{d}^2\alpha G_{\Sigma}(\alpha)e^{-2i\sqrt{\pi}\Re{\alpha}}\abs{\bra{\varnothing}\ket{q-\sqrt{2}\Im{\alpha}}_p}^2\\
    =&\int_\mathbb{C}\mathrm{d}^2\alpha\frac{1}{\pi\sqrt{\Sigma_{0}\Sigma_{1}}}\exp{-\frac{\Re{\alpha}^2}{\Sigma_0}}\exp{-\frac{\Im{\alpha}^2}{\Sigma_1}}\exp{-2i\sqrt{\pi}\Re{\alpha}}\abs{\bra{\varnothing}\ket{q-\sqrt{2}\Im{\alpha}}_p}^2\\
    =&\frac{1}{\pi\sqrt{\Sigma_{0}\Sigma_{1}}}\int_{-\infty}^\infty\mathrm{d}\Re{\alpha}\exp{-\frac{\Re{\alpha}^2}{\Sigma_0}}e^{-2i\sqrt{\pi}\Re{\alpha}} \nonumber\\
    &\quad\quad\quad\quad\quad\quad\int_{-\infty}^\infty\mathrm{d}\Im{\alpha}\exp{-\frac{\Im{\alpha}^2}{\Sigma_1}} \abs{\bra{\varnothing}\ket{q-\sqrt{2}\Im{\alpha}}_p}^2\\
    =&\frac{1}{\pi\sqrt{\Sigma_{0}\Sigma_{1}}}\int_{-\infty}^\infty\mathrm{d}\Re{\alpha}\exp{-\left(\frac{\Re{\alpha}}{\sqrt{\Sigma_0}}+i\sqrt{\pi\Sigma_0}\right)^2}\exp{-\pi\Sigma_0} \nonumber\\
    &\quad\quad\quad\quad\quad\quad\int_{-\infty}^\infty\mathrm{d}\Im{\alpha}\exp{-\frac{\Im{\alpha}^2}{\Sigma_1}}\abs{\bra{\varnothing}\ket{q-\sqrt{2}\Im{\alpha}}_p}^2\\
    =&\frac{\exp{-\pi\Sigma_0}}{\pi\sqrt{\Sigma_{0}\Sigma_{1}}}\int_{-\infty}^\infty\mathrm{d}\Re{\alpha}\exp{-\frac{\Re{\alpha}^2}{\Sigma_0}} \nonumber\\
    &\quad\quad\quad\quad\quad\quad\int_{-\infty}^\infty\mathrm{d}\Im{\alpha}\exp{-\frac{\Im{\alpha}^2}{\Sigma_1}}\abs{\bra{\varnothing}\ket{q-\sqrt{2}\Im{\alpha}}_p}^2\\
     =&\frac{\exp{-\pi\Sigma_0}}{\pi\sqrt{\Sigma_{0}\Sigma_{1}}}\int_{-\infty}^\infty\mathrm{d}\Re{\alpha}\exp{-\frac{\Re{\alpha}^2}{\Sigma_0}}\nonumber\\
    &\quad\quad\quad\quad\quad\quad\int_{-\infty}^\infty\mathrm{d}\Im{\alpha}\exp{-\frac{\Im{\alpha}^2}{\Sigma_1}}\abs{\bra{\varnothing}\widehat{D}^\dagger(\alpha)\ket{q}_p}^2\\
    =&\exp{-\pi\Sigma_0}\int_{-\infty}^\infty\mathrm{d}^2\alpha G_{\Sigma}(\alpha)\bra{q}_p\widehat{D}(\alpha)\ketbra{\varnothing}{\varnothing}\widehat{D}^\dagger(\alpha)\ket{q}_p\\
    =&\exp{-\pi\Sigma_0}\bra{q}_p\hat{\tilde{\rho}}\ket{q}_p
    \label{eq:rightTermX}
\end{align}

By inserting the results of Eqs.~\eqref{eq:leftTermX} and \eqref{eq:rightTermX} into Eq.~\eqref{eq:numeratorX}, we find that the stabilizer of ${\hat{\mathcal{S}}_{p}}$ introduced in~\eqref{eq:Xexpval} becomes

\begin{align}
    \expval{{\hat{\mathcal{S}}_{p}}} =&\frac{\Tr\left[\left(\ketbra{p_2}{p_2}\widehat{BS}'^\dagger\hat{\tilde{\rho}}_1\otimes\hat{\tilde{\rho}}_2\widehat{BS}'\right){\hat{\mathcal{S}}_{p,1}}(2\sqrt{\pi})\right]}{\Tr\left[\ketbra{p_2}{p_2}\widehat{BS}'^\dagger\hat{\tilde{\rho}}_1\otimes\widetilde{\rho_2}\widehat{BS}'\right]}\\
    =&\frac{1}{{N}_{p_2}}\int_{-\infty}^\infty\mathrm{d}l\bra{\frac{l+p_2}{\sqrt{2}}}_x\hat{\tilde{\rho}}_1{\hat{\mathcal{S}}_{p,1}}(\sqrt{2\pi})\ket{\frac{l+p_2}{\sqrt{2}}}_x\bra{-\frac{l-p_2}{\sqrt{2}}}_p\hat{\tilde{\rho}}_2{\hat{\mathcal{S}}_{x,2}}(-\sqrt{2\pi})\ket{-\frac{l-p_2}{\sqrt{2}}}_p\\
    =&\frac{1}{{N}_{p_2}}\int_{-\infty}^\infty\mathrm{d}l\exp{-\pi\Sigma_{1,1}}\bra{\frac{l+p_2}{\sqrt{2}}}_x\hat{\tilde{\rho}}_1\ket{\frac{l+p_2}{\sqrt{2}}}_x\exp{-\pi\Sigma_{0,2}}\bra{-\frac{l-p_2}{\sqrt{2}}}_p\hat{\tilde{\rho}}_2\ket{-\frac{l-p_2}{\sqrt{2}}}_p\\
    =&\exp{-\pi\Sigma_{1,1}}\exp{-\pi\Sigma_{0,2}}\frac{\int_{-\infty}^\infty\mathrm{d}l\bra{\frac{l+p_2}{\sqrt{2}}}_x\hat{\tilde{\rho}}_1\ket{\frac{l+p_2}{\sqrt{2}}}_x\bra{-\frac{l-p_2}{\sqrt{2}}}_p\hat{\tilde{\rho}}_2\ket{-\frac{l-p_2}{\sqrt{2}}}_p}{{N}_{p_2}}\\
    =&\exp{-\pi\Sigma_{1,1}}\exp{-\pi\Sigma_{0,2}}\\
    =&\exp{-\pi\left(\frac{1}{2\pi}\log\frac{1}{\abs{\expval{{\hat{\mathcal{S}}_{p,1}}}_{\rho_1}}^2}+\frac{1}{2\pi}\log\frac{1}{\abs{\expval{{\hat{\mathcal{S}}^\dagger_{x,2}}}_{\rho_2}}^2}\right)}\\
    =&\abs{\expval{{\hat{\mathcal{S}}_{p,1}}}_{\rho_1}}\abs{\expval{{\hat{\mathcal{S}}^\dagger_{x,2}}}_{\rho_2}}
\end{align}

\end{document}